\documentclass{article}
\pdfoutput=1 

\usepackage{jheppub} 

\usepackage[T1]{fontenc} 

\usepackage{graphicx} 
\usepackage{dcolumn} 
\usepackage{bm} 
\usepackage{amssymb} 
\usepackage{amsmath} 
\usepackage{amsfonts}
\usepackage{color}
\usepackage{comment}
\usepackage{rotating}
\usepackage{multirow}
\usepackage{colortbl}
\usepackage{afterpage}
\usepackage{slashed}
\usepackage{array}
\usepackage{subcaption}
\usepackage{bbm}

\usepackage{hyperref}

\usepackage{ctable}

\newcommand{\CP}{\textsf{CP}}

\newcommand{\tchi}{\tilde{\chi}}
\newcommand{\tpsi}{\tilde{\psi}}

\newcommand{\Lag}{\mathcal{L}}

\title{A Closer Look at $\textsf{CP}$-Violating Higgs Portal Dark Matter as a Candidate for the GCE}
\author{Katherine Fraser, Aditya Parikh, and Weishuang Linda Xu}
\affiliation{Department of Physics, Harvard University, Cambridge, MA 02138, USA}
\date{today}
\abstract{A statistically significant excess of gamma rays has been reported and robustly confirmed in the Galactic Center over the past decade. Large local dark matter densities suggest that this Galactic Center Excess (GCE) may be attributable to new physics, and indeed it has been shown that this signal is well-modelled by annihilations dominantly into $b\bar{b}$ with a WIMP-scale cross section. In this paper, we consider Majorana dark matter annihilating through a Higgs portal as a candidate source for this signal, where a large  $\textsf{CP}$-violation in the Higgs coupling may serve to severely suppress scattering rates. In particular, we explore the phenomenology of two UV completions, a singlet-doublet model and a doublet-triplet model, and map out the available parameter space which can give a viable signal while respecting current experimental constraints.}

\begin{document}
\maketitle

\section{Introduction}
Dark matter constitutes the majority of the mass in our universe, but its properties remain largely unknown. An extensive experimental program is in place aimed at understanding the nature of dark matter using its interactions with the Standard Model. This includes direct production at colliders and fixed-target experiments, scattering at direct detection experiments, and indirect detection of annihilation products. Over the years, there have been tantalizing hints in various experiments; while many of these signals have vanished due to increased statistics or a better understanding of systematic uncertainties, some signals, such as the Galactic Center Excess (GCE), have persisted for over a decade.

In astrophysical settings, the Galactic Center is expected to have some of the largest dark matter densities, and is therefore one of the most promising targets for indirect searches. The GCE is a statistically significant excess of gamma rays at energies of $\sim 2 - 3$ GeV observed in the Galactic Center by the Fermi Gamma Ray Space Telescope~\cite{TheFermi-LAT:2015kwa}. As pointed out by~\cite{Goodenough:2009gk,Hooper:2010mq,Hooper:2011ti,Gordon:2013vta,Abazajian:2014fta,Daylan:2014rsa,Calore:2014xka}, the GCE could be explained by a thermal WIMP annihilating to Standard Model particles. 
To truly confirm such a hypothesis, it is crucial to observe a signal in other indirect channels. In fact, it is possible that AMS-02 is observing an antiproton excess~\cite{Aguilar:2016kjl} at a concordant energy range~\cite{Cuoco:2017rxb,Cuoco:2019kuu,Cholis:2019ejx,Hooper:2019xss}, though the existence of this excess is not as well established~\cite{Boudaud:2019efq,Heisig:2020nse}.
While promising, it has also been suggested that the GCE signal could be generated by millisecond pulsars~\cite{Cholis:2014noa,Cholis:2014lta}. In recent years, the debate surrounding the origin of the GCE has intensified~\cite{Lee:2014mza,Bartels:2015aea,Lee:2015fea,Macias:2016nev,Haggard:2017lyq,Bartels:2017vsx,Macias:2019omb,Leane:2019xiy,Zhong:2019ycb,Leane:2020nmi,Leane:2020pfc,Buschmann:2020adf,Abazajian:2020tww,List:2020mzd,Mishra-Sharma:2020kjb,Karwin:2016tsw}. New measurements in the coming decade and a better theoretical understanding of Galactic diffuse emission models will help settle this debate, but until then, the origin of the GCE remains unknown and dark matter annihilation remains a viable explanation.

As discussed in~\cite{Daylan:2014rsa,Calore:2014xka}, the GCE can be well described by dark matter annihilations, particularly to $b\bar{b}$. This has fostered the development of many dark matter models with WIMP-like annihilation mechanisms, which are too numerous to review here (see \cite{Arcadi:2017kky} for a review).  Of these, models with pseudoscalar s-channel mediators are particularly well-motivated because they are neutral and can evade direct detection constraints. In particular, if the dark matter lives close to resonance, the annihilation cross section can be boosted enough to explain the GCE~\cite{Huang:2013apa,Boehm:2014hva,Cheung:2014lqa,Guo:2014gra,Cao:2014efa,Berlin:2015wwa,Gherghetta:2015ysa,Duerr:2015bea}. While much of the previous work relies on the introduction of a new pseudoscalar mediator, the authors of~\cite{Carena:2019pwq} proposed an interesting alternative. In their setup, the dark sector is connected to the visible sector via a $\CP$-violating coupling to the Higgs, which allows annihilation and spin-independent scattering to be governed by different parameters. In principle, the $\CP$-violating coupling can generate a viable thermal relic candidate even away from the resonance, by suppressing the scattering rather than enhancing the annihilation. 
However, in~\cite{Carena:2019pwq}, the authors consider specific model realizations within the context of supersymmetry where the benchmark best fit model still has the dark matter mass very close to half the Higgs mass.

In this work, we extract the key ingredients of their model, namely a Majorana dark matter candidate with $\CP$-violating coupling to the Higgs, and explore the extent of freedom away from the mass resonance that can be achieved with larger $\CP$-violating couplings. We see that for large enough coupling in the dark matter EFT, there is $\mathcal{O}(10)$ GeV flexibility for the dark matter mass when the phase is approximately $\pi/2$. 

We also consider and explore the phenomenology  of two different minimal UV realizations of this scenario: singlet-doublet dark matter~\cite{Mahbubani:2005pt,DEramo:2007anh,Enberg_2007,Cohen:2011ec,Cheung:2013dua,Abe:2014gua,Calibbi:2015nha,Freitas:2015hsa,Banerjee:2016hsk,Cai_2017,Lopez-Honorez:2017ora} and doublet-triplet dark matter~\cite{Dedes:2014hga,Abe:2014gua,Freitas:2015hsa,Lopez-Honorez:2017ora}. We study both how these models translate to EFT parameters, and constraints governing these UV realizations, including contributions to the electron electric dipole moment (EDM), the Peskin-Takeuchi parameters, as well as possible collider signatures. We find that while the dark matter mass and {\CP}-violating phase are independent parameters in the EFT, their dependence in the UV completion is quite nonlinear since the Yukawa coupling directly affects the dark matter mass. Specifically, it is difficult to achieve the phase tuning scenario without also tuning the mass in the UV completion, because the large couplings that are required to generate the annihilation cross section when away from resonance also change the dark matter mass. Additionally, we find that the amount of {\CP}-violation in the UV may not be reflective of that observed in the EFT. In the singlet-doublet case, we find two different types of viable parameter space. When the UV couplings are small, both the singlet mass in the UV and the dark matter mass must be very close to $m_h/2$, but the phase is flexible. When the UV couplings are larger, parameters must be chosen such that both the phase of the dark matter-Higgs coupling and the dark matter mass must be somewhat tuned, but there is more flexibility in the dark matter and singlet masses than in the small coupling case. In the doublet-triplet model, we find that EDM, spin-independent direct detection, and charged fermion collider search constraints are sufficient to rule out any WIMP-scale annihilation signal. 

The rest of this paper is organized as follows. In Section~\ref{sec:EFT}, we discuss the effective field theory of Majorana dark matter interacting with the Standard Model through a $\CP$-violating Higgs coupling. The EFT parameters dictate the annihilation and scattering cross sections which are broadly applicable independent of specific UV completions. In Section~\ref{sec:singlet_doublet}, we UV complete the EFT by introducing a singlet Majorana fermion and a doublet Dirac fermion. In Section~\ref{sec:doublet_triplet}, we consider another UV completion by introducing a doublet Dirac fermion and a triplet Majorana fermion. We discuss the strong constraints placed on each of these models by a variety of complementary experimental probes such as the electron EDM, precision electroweak parameters, and collider searches. Finally, we offer concluding remarks in Section~\ref{sec:conclusion}.

\section{Model Independent Constraints in the Effective Theory}
\label{sec:EFT}

In this section we take an effective field theory approach and focus on the phenomenology of a single species of Majorana dark matter which couples to the visible sector via a $\CP$-violating Higgs portal. After spontaneous symmetry breaking (SSB), the corresponding terms in the Lagrangian are given by  

\begin{equation}
    \Lag \ni  \frac{y_{h\chi}}{2 \sqrt{2}} h \bar \chi P_L \chi + \frac{y^*_{h\chi}}{2 \sqrt{2}} h \bar \chi P_R \chi +  \frac{g_{Z\chi}}{2} Z_\mu \bar \chi \gamma^\mu \gamma^5  \chi 
\label{eq:eft_lag}
\end{equation}
where the $\CP$-violation manifests in the complex nature of dark matter-Higgs coupling $y_{h\chi}$. Furthermore, we have also allowed for a coupling $g_{Z\chi}$ to the $Z$ boson.\footnote{$\chi$ does not have a vector current coupling because $\bar \chi \gamma^\mu \chi$ vanishes identically for Majorana fermions.}

As in all WIMP-type solutions to the GCE, the burden of the model is to reconcile the $\mathcal{O}(1)$ pb annihilation cross section necessary to achieve both the observed gamma-ray excess and the dark matter relic density, with the $\mathcal{O}(10^{-10})$ pb bounds on spin-independent scattering with nucleons from direct detection experiments. Traditionally, this is achieved for Higgs-portal dark matter by tuning the dark matter mass to the s-channel resonance $2 m_\chi \sim m_h$, but an additional avenue is available in the case of our model.


In the non-relativistic limit, two Majorana fermions form a $\CP$-odd state, so annihilation into the $\CP$-even Higgs through a $\CP$-conserving coupling is p-wave suppressed.  It then follows that if the coupling is complex, the annihilation in this limit is dominantly set by the imaginary part of $y_{h\chi}$, which is reflected in the result we obtain in Equation \ref{eqn:Annihilation_Cross_Section}.
 Conversely, the dark matter scattering off of the nucleon (or quark) does not require any $\CP$-violation since the initial and final states have the same $\CP$ properties, and thus we expect the spin-independent scattering cross section to be proportional to the real part of $y_{h\chi}$. This is reflected in the result we obtain in Equation~\ref{eqn:spin_indep_cross_section}.
 Therefore, the phase of the Higgs coupling can also contribute to a large hierarchy between the scattering and annihilation cross sections.  With this intuition, we describe the details and corresponding phenomenology of this theory in the remainder of this section.

\subsection{Annihilation}
\label{sec:anni}
Annihilation is mediated by both the Higgs and the $Z$ boson through an s-channel diagram. The dark sector couplings contributing to dark matter annihilation into SM fermions are given in Equation~\ref{eq:eft_lag}, and the visible sector couplings have the form

\begin{align}
    \Lag \ni \sum_{f} \frac{y_{hf}}{\sqrt{2}} h \bar ff  +  g_{Zf} Z_\mu \bar f \gamma^\mu (v_{f} - a_{f} \gamma^5)  f.
    \label{eq:sm_lag}
\end{align}
The couplings are given by their SM values
\begin{equation}
     y_{hf} = -\frac{\sqrt{2}m_f}{v}, \quad  g_{Zf} = \frac{e}{2\cos \theta_w \sin \theta_w},  \quad v_{f}= I_3 - 2Q\sin^2\theta_w,  \quad a_{f} = I_3,
\end{equation}
where $v$ is the Higgs vev, $\theta_w$ is the Weinberg angle, and $m_f$, $I_3$, and $Q$ are the mass, weak isospin, and electric charge of the fermion respectively.  In the non-relativistic limit, the total spin-averaged amplitude squared for annihilation can be written as

\begin{equation}
\label{eqn:Annihilation_Cross_Section}
    |\mathcal{M}|_{\chi \chi \to f\bar{f} }^2 = 4 m^2_\chi \left[ \frac{m_f^2 }{m_Z^4}g^2_{Zf} g_{Z\chi}^2 a_f^2    + y^2_{hf} \mathrm{Im}[y_{h\chi}]^2\frac{ (m_\chi^2 - m_f^2 )}{(m_h^2 - 4m_\chi^2)^2 + m_h^2 \Gamma_h^2} \right],
\end{equation}
where $\Gamma_h$ denotes the width of the Higgs. The Higgs mediated piece depends only on the imaginary part of the coupling as expected. The cross section is correspondingly given by 

\begin{equation}\langle \sigma v \rangle  = \sum_{m_f \leq m_\chi} \frac{N_c \sqrt{m_\chi^2 - m_f^2}}{64 \pi m_\chi^3} |\mathcal{M}|_{\chi \chi \to f\bar{f} }^2. \end{equation}

If the dark matter is a thermal relic, then the present-day dark matter abundance, $\Omega_\chi h^2 =0.11$, sets the annihilation cross section at the time of freeze-out, which is the well-known  $\mathcal{O}(1)$ pb weak-scale cross section~\cite{Zeldovich:1965gev,Chiu:1966kg,Lee:1977ua,Hut:1977zn,Wolfram:1978gp,Steigman:1979kw,Scherrer:1985zt,Bertstein:1985,Srednicki:1988ce,Griest:1990kh,Gondolo:1990dk,Steigman:2012nb}. Recent work~\cite{Binder:2017rgn,Abe:2020obo} has shown that for models with a hierarchy between annihilation and scattering strengths, early kinetic decoupling before freeze-out alters this number, requiring a larger cross section to achieve the observed abundance. At most extreme, a $\sim 20$ pb annihilation cross section may be needed for a $\sim 57$ GeV dark matter with purely imaginary couplings, though this is quite sensitive to the details of the QCD phase transition. However, this effect is significantly weaker for masses $\gtrsim m_h/2$, so we do not take our annihilation cross section to be this large.   

At present, dark matter annihilation is expected to produce a distribution of gamma-rays whose flux is given by 
\begin{equation}
    \frac{d^2 \Phi_\gamma}{d\Omega dE_\gamma} = \frac{1}{2} \langle \sigma v \rangle \left( \sum_f \frac{dN_\gamma}{dE\gamma } \mathrm{Br}_{\chi\chi \to f\bar{f} }\right) \int_{\rm los} \frac{\rho_\chi^2(r) d\ell }{4\pi m_\chi^2},
\end{equation}
where $\mathrm{Br}_{\chi\chi \to f\bar{f}}$ denotes the branching ratio to the $f\bar{f}$ final state, and $dN_\gamma/ dE_\gamma$ its corresponding injection spectrum. $\rho_\chi(r)$ denotes the dark matter halo profile and is integrated over the line-of-sight to the Galactic Center. It has been shown that the Fermi GCE data is well-modeled by a Higgs portal dark matter with a cross section $\langle \sigma v \rangle \sim 3$ pb, assuming a modified NFW profile~\cite{Hooper:2010mq}. 
As the precise best fit depends on many details, including the galactic profile and background modeling~\cite{DiMauro:2021raz}, in conjunction with the modeling uncertainties of the thermal relic argument, we will consider here a range of cross sections $\langle \sigma v \rangle$ from 1 to 10 pb to be in concordance with both the GCE and the relic abundance.

\subsection{Direct Detection}
\label{sec:didt}

In contrast with annihilation, processes relevant for direct detection occur below the weak scale and should be  considered  in  terms  of  effective  interactions  with  target  nuclei. Much of the subsequent discussion follows~\cite{Lin:2019uvt}. At momentum transfers  $t \ll m_Z^2$, the interactions in Equations~\ref{eq:eft_lag} -- \ref{eq:sm_lag} are rewritten as the following dimension-6 operators

\begin{align} \mathcal{L}  \ni & \frac{C_{S}}{m_h^2} \bar \chi \chi \bar f f  + \frac{C_{PS}}{m_h^2}  \bar \chi \gamma^5 \chi \bar f f    
+ \frac{C_{V}}{m_Z^2} \bar \chi \gamma^\mu \gamma^5    \chi \bar f \gamma_\mu f   + \frac{C_{PV}}{m_Z^2} \bar \chi \gamma^\mu  \gamma^5 \chi \bar f \gamma_\mu \gamma^5 f  
\end{align}
with $C_S$, $C_{PS}$, $C_V$, and $C_{PV}$ denoting the scalar, pseudo-scalar, vector, and  pseudo-vector pieces of the quark-gauge couplings respectively. The contributions governed by $C_{PS}$ and $C_V$ are velocity-suppressed and we neglect them in the following. After matching to the UV theory, the coefficients are given by 
\begin{equation} 
C_S = \frac{1}{2} \mathrm{Re}[y_{h\chi}] y_{hf} \qquad C_{PV} = g_{Z\chi} g_{Zf} a_f.
\end{equation}
In the zero momentum transfer limit, the nucleon-level operators are matched to the quark-level ones via form factors
\begin{align}
\langle N(p) |\bar f \gamma^\mu \gamma^5 f| N(p') \rangle  & =  \bar u_N (p)  \left[  \Delta_1^{f,N} (q^2) \gamma^\mu \gamma^5 \right] u_N(p') \\
\langle N(p) |\bar f  f| N(p') \rangle  & =  \frac{m_N}{m_f} f^N_{f} \bar u_N (p)  u_N(p') 
\end{align} 
where $N$ represents a nucleon (a proton or neutron), $q = p'-p$ denotes the momentum transfer, and the form factors are listed in Table~\ref{tab:form_fac}. We have neglected higher order terms in $q^{2}$. For the scalar term specifically, the heavy quarks also contribute via a gluon loop. After integrating out heavy quarks, the relevant operator for each flavor appears as 
\begin{equation}
- \frac{C_S}{m_h^2} \frac{\alpha_s}{12\pi m_f} \bar \chi  \chi G^{\mu\nu} G_{\mu\nu}. 
\end{equation} 
To match to the nucleon-level picture the following matrix element is taken into account
\begin{equation}
\langle N(p) |G^{\mu\nu} G_{\mu\nu} | N(p') \rangle   =  -  \frac{8\pi}{9 \alpha_s} m_N f^N_{g} \bar u_N (p)  u_N(p'). \end{equation} 
In terms of the quark-level couplings, the nucleon-level spin-independent cross section is given by

\begin{equation}
    \sigma_{SI} = \frac{m_N^2 m_\chi^2}{4\pi (m_\chi + m_N)^2} \left[  \frac{ \text{Re}[y_{h\chi}]}{m_h^2}\left[ \sum_{f \in u,d,s} y_{hf}  \frac{m_N}{m_f} f^N_{f} +  \sum_{f \in c,b,t}  y_{hf} \frac{2}{27} \frac{m_N}{m_f}  f^N_{g}   \right] \right]^2.   
    \label{eqn:spin_indep_cross_section}
\end{equation}

As discussed earlier, the cross section only depends on the real part of the Higgs coupling. Furthermore, the dependence on the coupling to the $Z$ boson vanishes in the $q\to 0$ limit.
Likewise the spin-dependent cross section is given by 
\begin{align}
    \sigma_{SD} = & \frac{3 m_N^2 m_\chi^2 }{\pi (m_\chi + m_N)^2} \left[\frac{g_{Z\chi}}{4m_Z^2} \sum_{f \in u,d,s} g_{Zf}  a_f   \Delta_1^{N,f}\right]^2. 
\end{align}

\begin{table}
\begin{center}
\begin{tabular}{ c | c c c | c c c c}
\toprule
     & $\Delta_1^{N, u}$ & $\Delta_1^{N, d}$ & $\Delta_1^{N, s}$  & $f^N_{u}$ & $f^N_{d}$ & $f^N_{s}$  & $f^N_{g}$ \\
     \hline
     \hline
     Protons & 0.80 & -0.46 & -0.12 & 0.018 & 0.027 & 0.037 & 0.917\\
     Neutrons & -0.46 & 0.80 & -0.12 &  0.013 & 0.040 & 0.037 & 0.910 \\
     \bottomrule
\end{tabular}
\end{center}
\caption{Here we show the light quark and gluon form factors for the proton and neutron. These values come from~\cite{DelNobile:2013sia,Hill:2014yxa,Bishara:2017pfq,Ellis:2018dmb} and are summarized in~\cite{Lin:2019uvt}.}
\label{tab:form_fac}
\end{table}

\begin{figure}
\begin{subfigure}{0.5\linewidth}
    \centering
    \includegraphics[width = \textwidth]{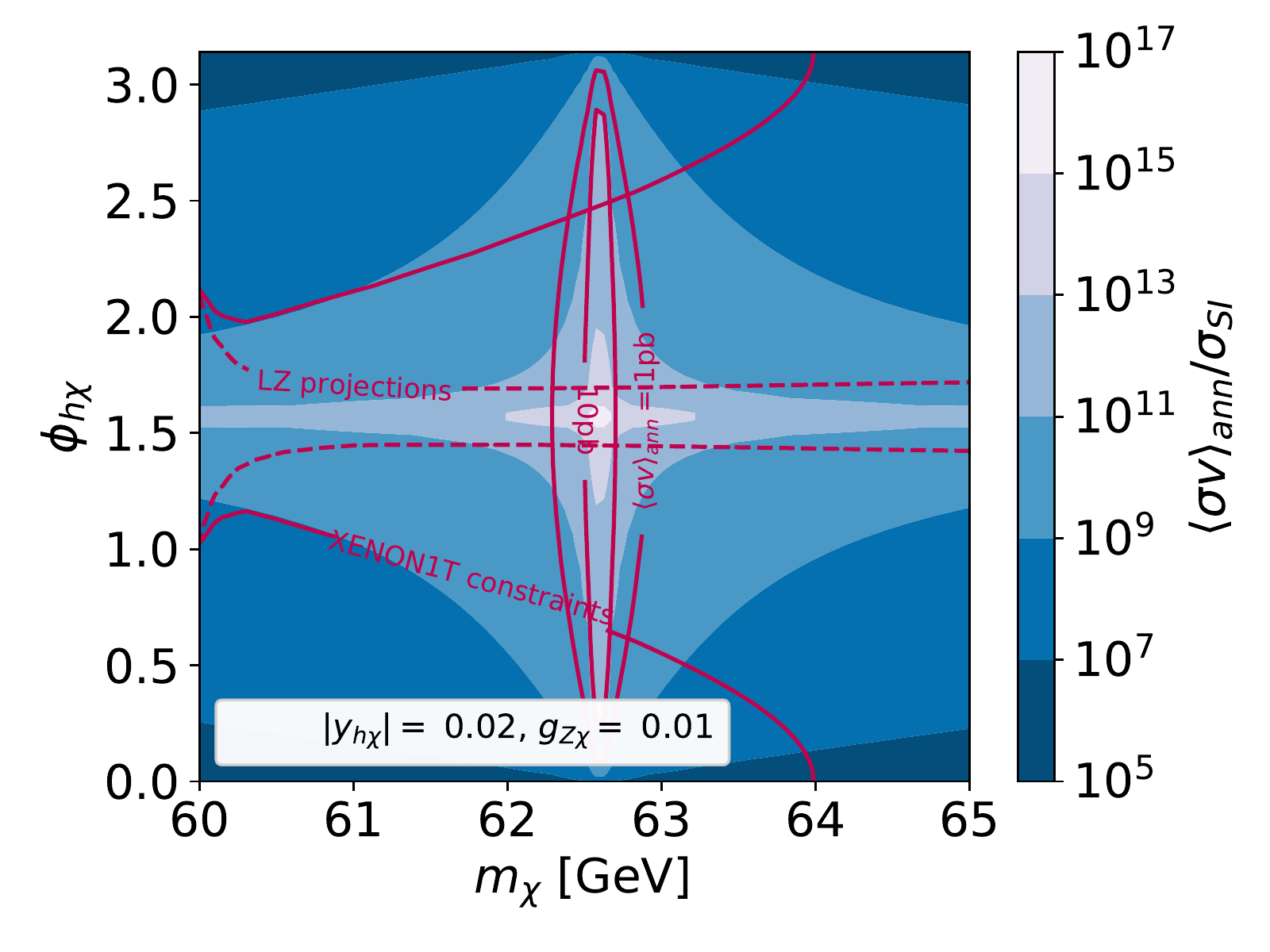}
    \end{subfigure}
    \begin{subfigure}{0.5\linewidth}
    \centering
    \includegraphics[width = \textwidth]{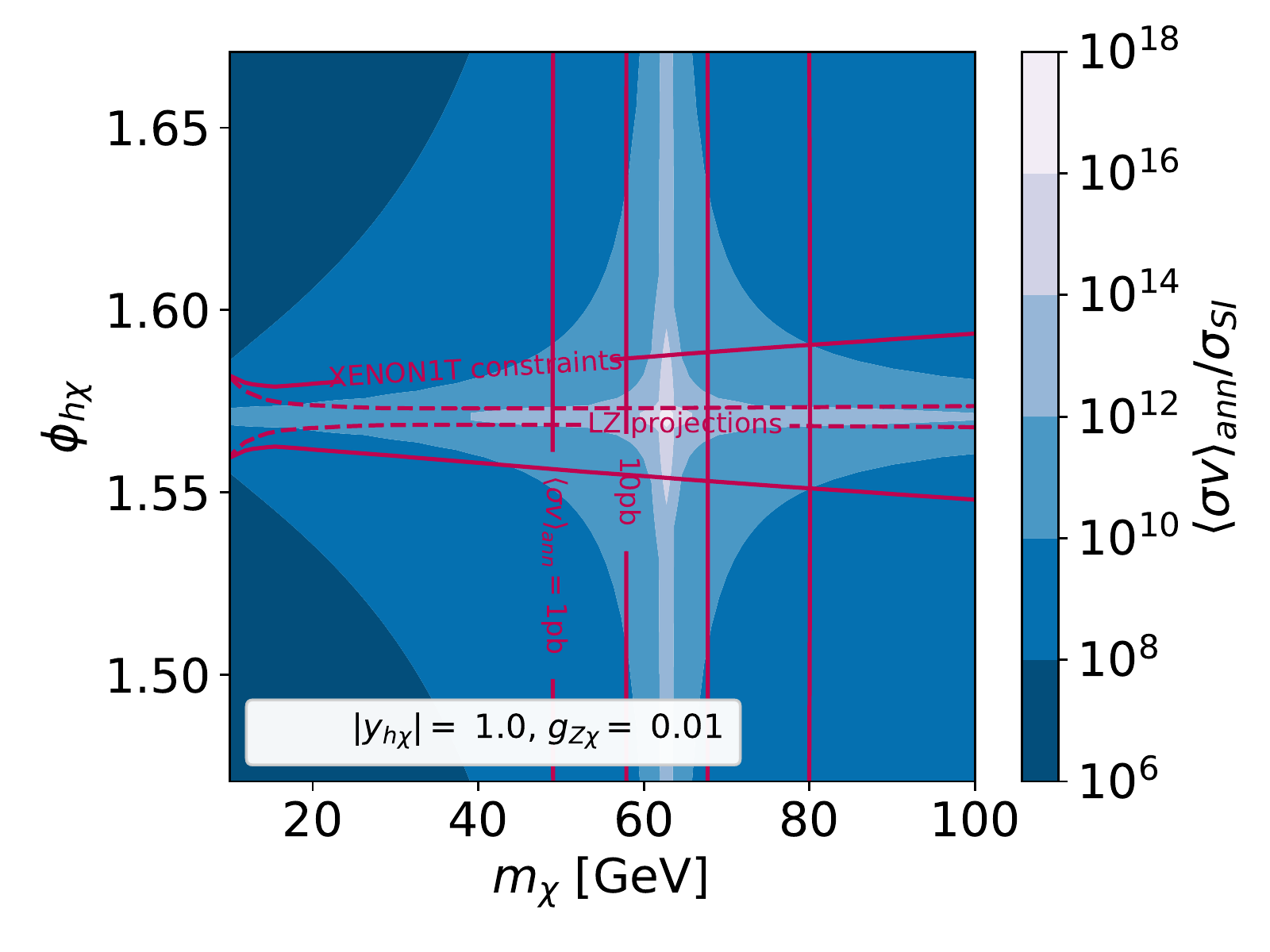}
    \end{subfigure}
    \caption{The ratio between annihilation and spin-independent direct detection cross sections on the $m_\chi-\phi_{h\chi}$ plane for different values of $|y_{h \chi}|$. The region allowed by direct detection is inside the solid XENON1T~\cite{Aprile:2017iyp,Aprile:2018dbl} constraint line, while the region allowed by annihilation is between the solid 1 pb and 10 pb lines. We also show projected limits from LZ~\cite{Akerib:2018lyp} as dashed lines. Note that the axis scales on the two plots are different. We assume $m_h = 125.2$ GeV here and throughout this paper. The left plot shows the mass resonance with small $y_{h\chi}$, for which the dark matter mass must be tuned to within less than a GeV of the pole, but there is some flexibility in the phase. The right plot shows the phase tuning: away from $m_h = 2 m_\chi$ a large coupling is required to achieve a sufficient annihilation cross section, but tuning the phase near $\pi/2$ avoids direct detection limits despite the large coupling. In this case, the flexibility of the allowed mass range changes to $\mathcal{O}(10)$ GeV. Both of these plots include a small non-zero $Z$ coupling that is consistent with spin-dependent direct detection constraints. The limits are similar for vanishing $Z$ coupling.}
    \label{fig:Linda_EFT_plot}
\end{figure}

\begin{figure}
    \centering
    \includegraphics[width = 0.5\textwidth]{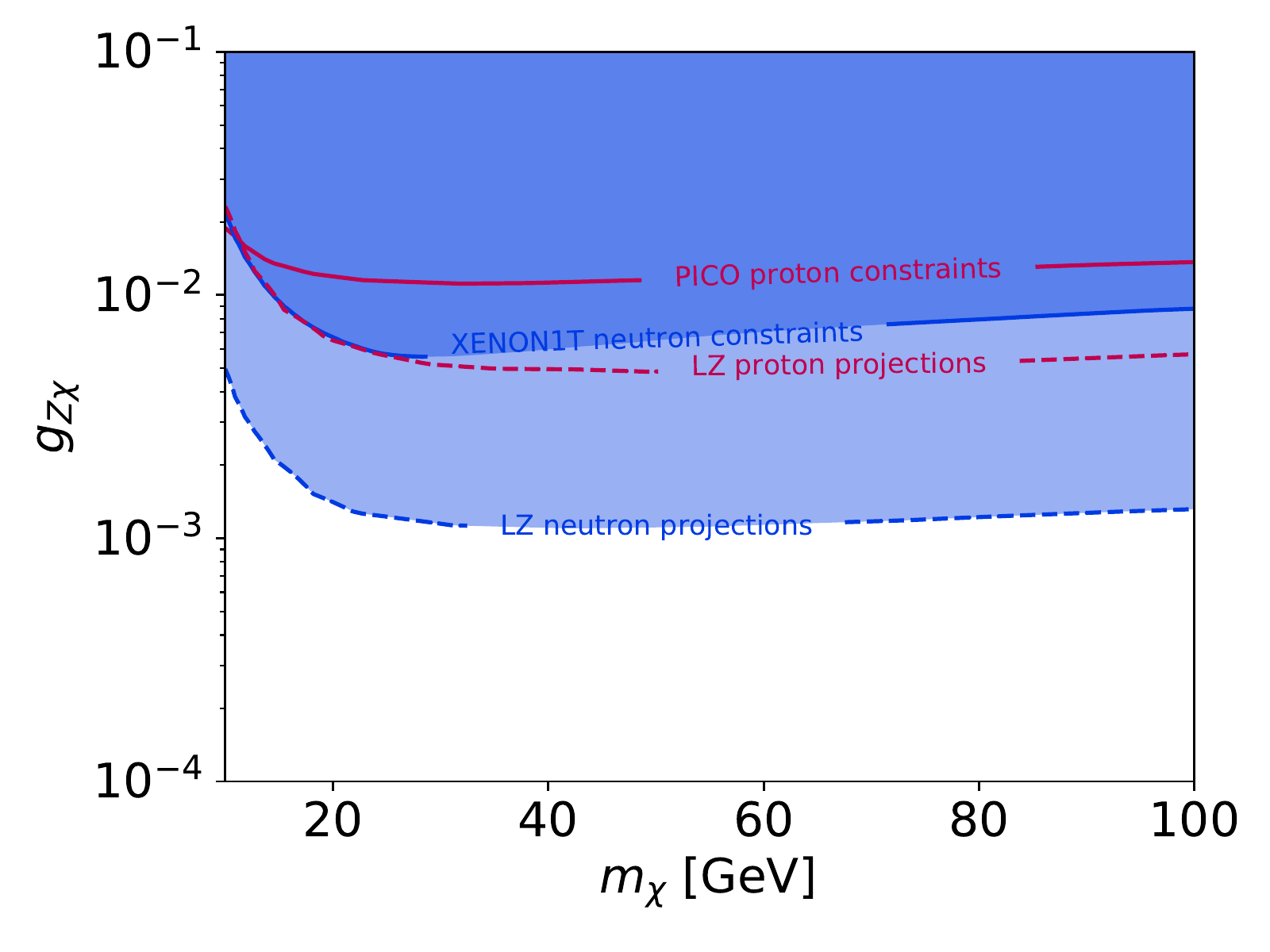}
    \caption{Spin-dependent direct detection limits as a function of dark matter mass and dark matter-$Z$ coupling. Constraints are close to horizontal because the spin-dependent cross section depends on the reduced mass. For neutrons, XENON1T~\cite{Aprile:2019dbj} is the strongest model independent constraint. For protons, PICO~\cite{PhysRevLett.118.251301,PhysRevD.100.022001} provides a slightly stronger constraint. Additionally, we show projected limits from LZ~\cite{Akerib:2018lyp}.}
    \label{fig:EFT_Z_coupling}
\end{figure}

\subsection{Results and Discussion}

In this subsection we examine the phenomenology of the effective theory, and discuss the regions of parameter space where a high annihilation and low scattering cross section can be achieved -- specifically we are interested in an annihilation cross section between approximately 1 and 10 pb to fit the GCE and a scattering cross section consistent with direct detection experiments. For spin-independent scattering, the strongest limits come from  XENON1T~\cite{Aprile:2017iyp,Aprile:2018dbl}, while for spin-dependent scattering, the strongest limits come from both XENON1T~\cite{Aprile:2019dbj} and PICO~\cite{PhysRevLett.118.251301,PhysRevD.100.022001}. LZ~\cite{Akerib:2018lyp} and XENONnT~\cite{Aprile:2020vtw} are projected to improve on current limits within the parameter space of interest. The projected limits are comparable, so we show only one in our figures for clarity. We omit limits from IceCube~\cite{Aartsen:2016zhm}, LUX~\cite{Akerib:2016lao,Akerib:2016vxi} and PandaX-II~\cite{Cui:2017nnn} because they are slightly weaker than those we've shown for $\mathcal{O}(60)$ GeV dark matter. For the spin-independent constraints, we consider only dark matter-proton scattering because in this case the difference between proton and neutron cross sections is negligible. 

First we review which masses and coupling magnitudes are in general concordance with scattering constraints and annihilation requirements. Typical couplings that can generate an annihilation cross section of $\sim 1$ pb are shown in Equation~\ref{approx_annihilation} for two different dark matter masses. In Equation~\ref{approx_DD}, we show approximate couplings and masses that are consistent with direct detection constraints.
\begin{equation}
\begin{split}
\left[\frac{ \langle \sigma v \rangle}{1\text{ pb}} \right] &= \left[\frac{m_\chi}{80\text{ GeV}} \right]^2 \left[ \frac{4m_\chi^2 - m_h^2}{10^4 \text{ GeV}^2} \right]^{-2} \left[ \frac{y_{h\chi} \sin \phi_{h\chi}}{1.0} \right]^2 \\
&= \left[\frac{m_\chi}{62.5\text{ GeV}} \right]^2 \left[ \frac{4m_\chi^2 - m_h^2}{50 \text{ GeV}^2} \right]^{-2} \left[ \frac{y_{h\chi} \sin \phi_{h\chi}}{0.007} \right]^2
\end{split}
\label{approx_annihilation}
\end{equation}
\begin{equation}
\begin{split}
\left[ \frac{\sigma_{SI}}{10^{-10} \text{ pb}}\right] & =  \left[ \frac{y_{h\chi} \cos \phi_{h\chi}}{0.02}\right]^2  \\
\left[ \frac{\sigma_{SD}}{10^{-5} \text{ pb}} \right] & =  \left[ \frac{g_{Z\chi}}{0.01} \right]^2 
\end{split}
\label{approx_DD}
\end{equation}

We remind the reader that the free parameters of the theory are $m_\chi$, $g_{Z\chi}$, and the complex coupling $y_{h\chi}$ with phase $\phi_{h\chi}$. While $g_{Z\chi}$ and Im[$y_{h\chi}$] set the annihilation cross section, only Re[$y_{h\chi}$] sets the magnitude of scattering. In order to generate a large enough annihilation cross section while avoiding direct detection constraints, Higgs portal dark matter models typically tune the dark matter mass close to half the Higgs mass~\cite{Huang:2013apa,Boehm:2014hva,Cheung:2014lqa,Guo:2014gra,Cao:2014efa,Berlin:2015wwa,Gherghetta:2015ysa,Carena:2018nlf,Carena:2019pwq}. While tuning the mass is one way to generate the correct ratio in this model, we emphasize that in the EFT, the correct ratio can also be obtained for a wider mass range by increasing the magnitude of the Higgs coupling while tuning the phase, $\phi_{h\chi}$, of the Higgs coupling close to $\pi/2$ to suppress direct detection constraints. This is illustrated in Figure~\ref{fig:Linda_EFT_plot}, which plots annihilation and spin independent direct detection constraints in the $m_\chi - \phi_{h\chi}$ plane for different magnitudes of the Higgs couplings. We can see that near the mass resonance, a small Higgs coupling ($\sim 0.02$) is sufficient to generate the annihilation cross section and the phase does not need to be near $\pi/2$ to avoid direct detection constraints. However, with phase tuning, the larger Higgs coupling required to generate the correct annihilation cross section away from resonance is allowed because direct detection only constrains the real part of $y_{h \chi}$. This widens the mass range considerably to $\mathcal{O}(10)$ GeV. Even for the mass resonance, the coupling cannot be purely real, because the leading velocity dependent term is not large enough to generate the required annihilation cross section given the finite Higgs width. See Appendix~\ref{Section:yhpsi_real} for more details. Note that while in principle a large pseudo-vector $Z$ coupling could also generate a sufficient annihilation cross section, this is constrained by spin-dependent direct detection constraints, as shown in Figure~\ref{fig:EFT_Z_coupling}. Within the range of $Z$ couplings allowed by direct detection, the effect on the allowed annihilation signal is negligible.

\section{Singlet-Doublet Model}
\label{sec:singlet_doublet}
A well-motivated way to UV complete the dark matter EFT provided in Section~\ref{sec:EFT} in a gauge invariant manner is to introduce additional particles charged under $G_{SM}$. In this section, we discuss a simple potential UV completion, where the only additional particles we introduce to the Standard Model are a  singlet Majorana fermion and a doublet Dirac fermion. This model has previously been discussed in other contexts in~\cite{Mahbubani:2005pt,DEramo:2007anh,Enberg_2007,Cohen:2011ec,Cheung:2013dua,Abe:2014gua,Calibbi:2015nha,Freitas:2015hsa,Banerjee:2016hsk,Cai_2017,Lopez-Honorez:2017ora}.

\subsection{Model in the UV\label{section:model}}
We start by establishing notation and describing the model. The model contains a singlet Majorana fermion $\psi_1$ and an additional SU(2) doublet Dirac fermion with hypercharge 1/2. We describe the SU(2) doublet with two left handed Weyl fermions $\psi_2$ (with neutral component $\psi_2^0$ and charged component $\psi_2^1$) and  $\tpsi_2$ (with neutral component $\tpsi_2^0$ and charged component $\tpsi_2^{-1}$). All new fermions are SU(3) singlets. The Lagrangian for this model is
\begin{equation}
    \Lag = \Lag_{SM} + \Lag_{\rm kinetic}  -m_2 \psi_{2} \cdot \tpsi_2  - \frac{m_1}{2} \psi_1 \psi_{1} + Y \psi_1 H^\dag \psi_{2} - \tilde Y \psi_{1} H \cdot  \tpsi_2     + \text{h.c.} 
\end{equation}
As we introduce three new fields and four free parameters, there is one remaining physical phase. We make the choice to fix each of the Yukawa terms to the same phase, which carries the {\CP}-violation,
\begin{equation}
    Y \equiv y \, \mathrm{e}^{i\delta_{CP}/2}, \qquad
    \tilde Y \equiv \tilde y \, \mathrm{e}^{i\delta_{CP}/2}. 
\end{equation}
After SSB, the mass terms are written as

\begin{align}
    &\begin{aligned}
        \Lag_{\rm mass} =  &- m_2\left(\tpsi_{2}^{-1} \psi_{2}^1 - \tpsi_{2}^{0} \psi_{2}^0 \right) -\frac{m_1}{2} \psi_1 \psi_{1} +\frac{v}{2} y \, \mathrm{e}^{i\delta_{CP}/2}  \psi_1\psi_{2}^0
     + \frac{v}{2} \tilde y\, \mathrm{e}^{i\delta_{CP}/2} \psi_1\tpsi_{2}^0  + \text{h.c.} \\ 
    \end{aligned}
\end{align}

Let us define $\psi_2^s \equiv \frac{1}{\sqrt{2}} (\psi_2^0 + \tpsi_2^0)$ and $\psi_2^d \equiv \frac{1}{\sqrt{2}} (\psi_2^0 - \tpsi_2^0)$ to be the two Majorana fermions that constitute the neutral Dirac fermion $\{\psi_2^0, \tpsi_2^0\}$. The mass eigenstates thus result from the mixing of the doublet and singlet Majorana fermions,  $ \psi_i = (\psi_2^s, \psi_2^d, \psi_1)_i$. We will denote the mass eigenstates $ \chi_i =( \chi, \chi_1, \chi_2 )_i$, the lightest of which, $\chi$, is the dark matter candidate. Then

\begin{equation}
        \Lag_{\rm mass} = - m_2\tpsi_{2}^{-1} \psi_{2}^1 -  \frac{1}{2}\psi_i M_{ij} \psi_j, \\
\end{equation}
where $M$ is the mass matrix. This basis change is governed by $J$, the matrix of eigenvectors that diagonalizes both $M^\dag M$ and $M$, phase rotated such that $J^T M J$ has real eigenvalues. After diagonalizing, the Higgs Yukawa couplings are
\begin{equation}
    \Lag_{\rm Higgs} = \frac{1}{2} h \chi_i[J^T U_h J]_{ij}\chi_j + \text{h.c.}
\end{equation}
where 
\begin{equation} U_h =
    \begin{pmatrix}
   0 & 0 & \frac{(Y + \tilde{Y})}{2} \\
    0 & 0 & \frac{(Y - \tilde{Y})}{2} \\
    \frac{(Y + \tilde{Y})}{2} & \frac{(Y - \tilde{Y})}{2} & 0 \\
    \end{pmatrix}.
    \end{equation}

\begin{figure}[b]
    \begin{subfigure}{0.5\linewidth}
    \centering
    \includegraphics[width = \textwidth]{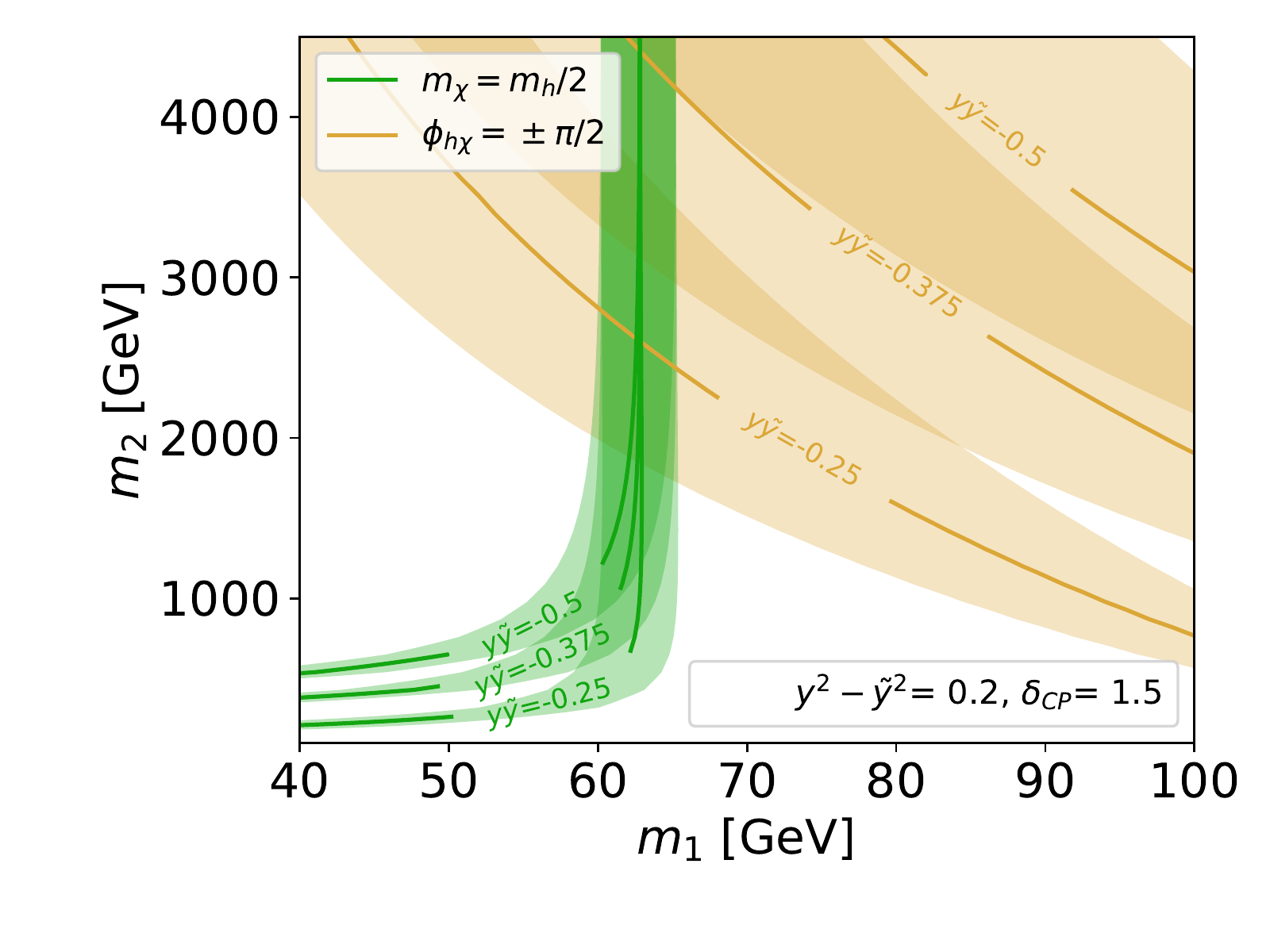} 
    \end{subfigure}
    \begin{subfigure}{0.5\linewidth}
    \includegraphics[width = \textwidth]{ 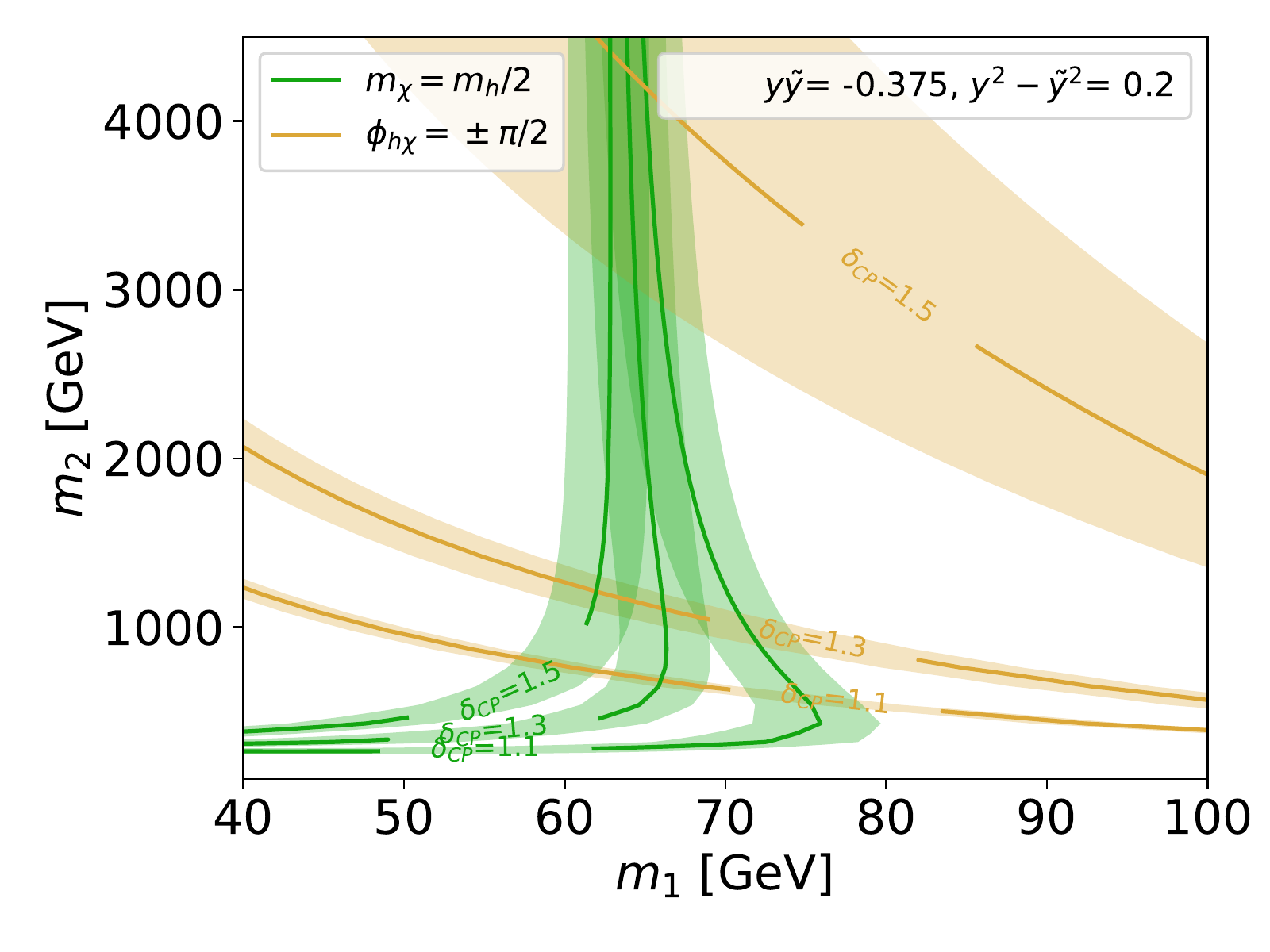}
    \end{subfigure}
    \caption{Plots show EFT coupling phase and dark matter mass as a function of $m_1$ and $m_2$ for different values of $y, \tilde{y},$ and $\delta_{CP}$. Left: $\delta_{CP}$ is fixed to 1.5 while $y\tilde{y}$ is varied. Right: $y\tilde{y}$ is fixed to $-0.375$ while $\delta_{CP}$ is varied. The shaded regions give a sense of the width of the regions of interest: $60 \, \mathrm{GeV} \leq m_\chi \leq 65 \, \mathrm{GeV}$ and $1.55 \leq \phi_{h \chi} \leq 1.60$. We can see that changing $y\tilde{y}$ has a minimal effect on $m_\chi$ at large $m_2$ but strongly affects which masses correspond to the central value of $\phi_{h\chi} = \pi/2$. $y\tilde{y}$ also affects the smallest value of $m_2$ that can lead to a mass resonance. Changing $\delta_{CP}$ has a larger effect on which $m_1$ is required to get the mass resonance, and also affects the width of the $\phi_{h\chi} = \pi/2$ band in addition to the position of its central value.}
    \label{fig:m1_vs_m2_EFT_params}
\end{figure} 

Since one of the new fermions is an SU(2) doublet, the new fermions also couple to the electroweak gauge bosons. The $Z$ couplings are
    \begin{equation}
    \Lag_{Z}= \chi^\dagger_{i} [J^\dagger U_{Z} J]_{ij} Z \overline{\sigma}  \chi_{j} + g_Z(\cos^2 \theta_W-\sin^2 \theta_W) (\psi_{2}^{1\dagger}Z\overline{\sigma}\psi_{2}^1 -\tilde{\psi}_{2}^{-1\dagger}Z\overline{\sigma}\tilde{\psi}_{2}^{-1})
    \end{equation}  
while the $W$ couplings are
    \begin{equation}
    \Lag_{W}= \chi^\dagger_{i}  [b J^*]_{i} W^+ \overline{\sigma}\tilde{\psi}_{2}^{-1} + {\psi}_{2}^{1\dagger} W^+ \overline{\sigma} [a J]_j \psi_{j} + \text{h.c.} 
    \label{singlet_doublet_Wcouplings}
    \end{equation}
    where $g$ is the SU(2) gauge coupling, $g'$ is the U(1) hypercharge gauge coupling, $\theta_W$ is the Weinberg angle, and $g_Z \equiv \sqrt{g^2 + g'^2}/2$. Here, $a_i = (g/2, g/2,0)_i, b_i = ( g/2, -g/2,0)_i$ and
    \begin{equation}
    U_Z = \begin{pmatrix}
    0 & -g_Z & 0 \\
    -g_Z & 0 & 0 \\
    0 & 0 & 0
    \end{pmatrix}.
    \end{equation}

The dark matter candidate $\chi$ obtains the couplings seen in the EFT via mixing between the singlet and doublet. The strength of these couplings can be adjusted by altering the makeup of the lightest Majorana fermion. The theory at this level is fully specified by five degrees of freedom: the singlet mass $m_1$, the doublet mass $m_2$, the doublet Yukawa coupling magnitudes $\{y, \tilde y\}$ and the associated {\CP}-violating phase $\delta_{CP}$.

\subsection{Translating to the EFT \label{subsec:UV_to_EFT}}

Now we discuss how the EFT parameters
$g_{Z\chi}, m_\chi,$ and $y_{h\chi}$ depend on the UV parameters $y, \tilde{y}, m_1, m_2$, and $\delta_{CP}$. We focus mostly on the region where $m_2$ is large, but also comment on the more general case.\footnote{We also omit the case where both $m_1$ and $m_2$ are large. In this case, extremely large couplings are required in order to get dark matter with mass near $m_h/2$. This means the $\delta_{CP}$ must be small to avoid EDM constraints, which leaves us with $\phi_{h\chi}$ mostly real and prevents us from simultaneously evading spin-independent constraints.}  Since the theory has a charged fermion with mass $m_2$, parameter space with small $m_2$ will generically be ruled out by collider constraints~\cite{LEPSUSYWG1,LEPSUSYWG2}. EDM and electroweak constraints are likewise more stringent in this regime.

Figure~\ref{fig:m1_vs_m2_EFT_params} shows the EFT mass and phase as a function of $m_1$ and $m_2$ for different values of the UV coupling magnitudes and phase. On the left we show multiple values of $y\tilde{y}$ for fixed $\delta_{CP}$ while on the right we show multiple values of $\delta_{CP}$ for fixed $y\tilde{y}$. In both cases, we can see that only a narrow range in $m_1$ translates to dark matter with mass near the mass resonance. When $m_2$ is large, the lightest fermion is mostly $m_1$. In this limit, mixing is small, so to have the dark matter mass near the mass resonance, $m_1$ must be fairly close to half the Higgs mass. We can see that changing $y\tilde{y}$ changes where $\phi_{h\chi} = \pi/2$ is located but only has a minimal effect on which $m_1$ value translates to the mass resonance. We can also see that for the same $m_1$, smaller $y\tilde{y}$ requires a correspondingly smaller $m_2$ to get dark matter with $m_\chi \approx m_h/2$. Changing $\delta_{CP}$ also changes the location of $\phi_{h\chi} = \pi/2$ contour, but additionally affects the $m_1$ required to get the mass resonance and the width of the $\phi_{h\chi} \approx \pi/2$ band.

\begin{figure}[h!]
    \centering
    \includegraphics[width = 0.86 \textwidth]{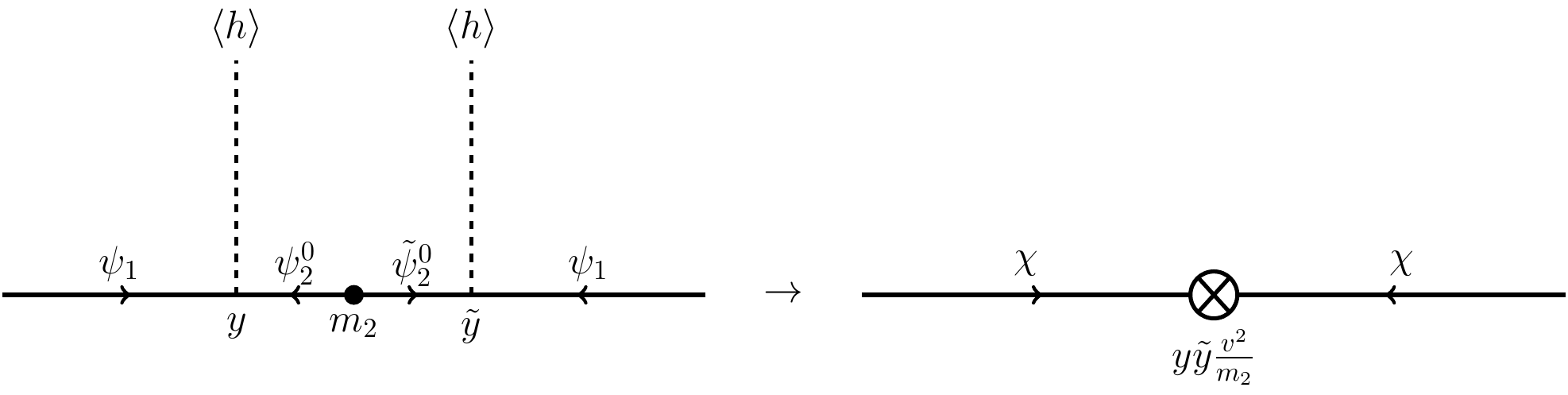}
    \caption{Diagram generating dark matter mass in the limit where $m_2$ is large.}
    \label{fig: mass m2 gg m1}
\end{figure}

\begin{figure}[h!]
    \centering
    \includegraphics[width = 74mm]{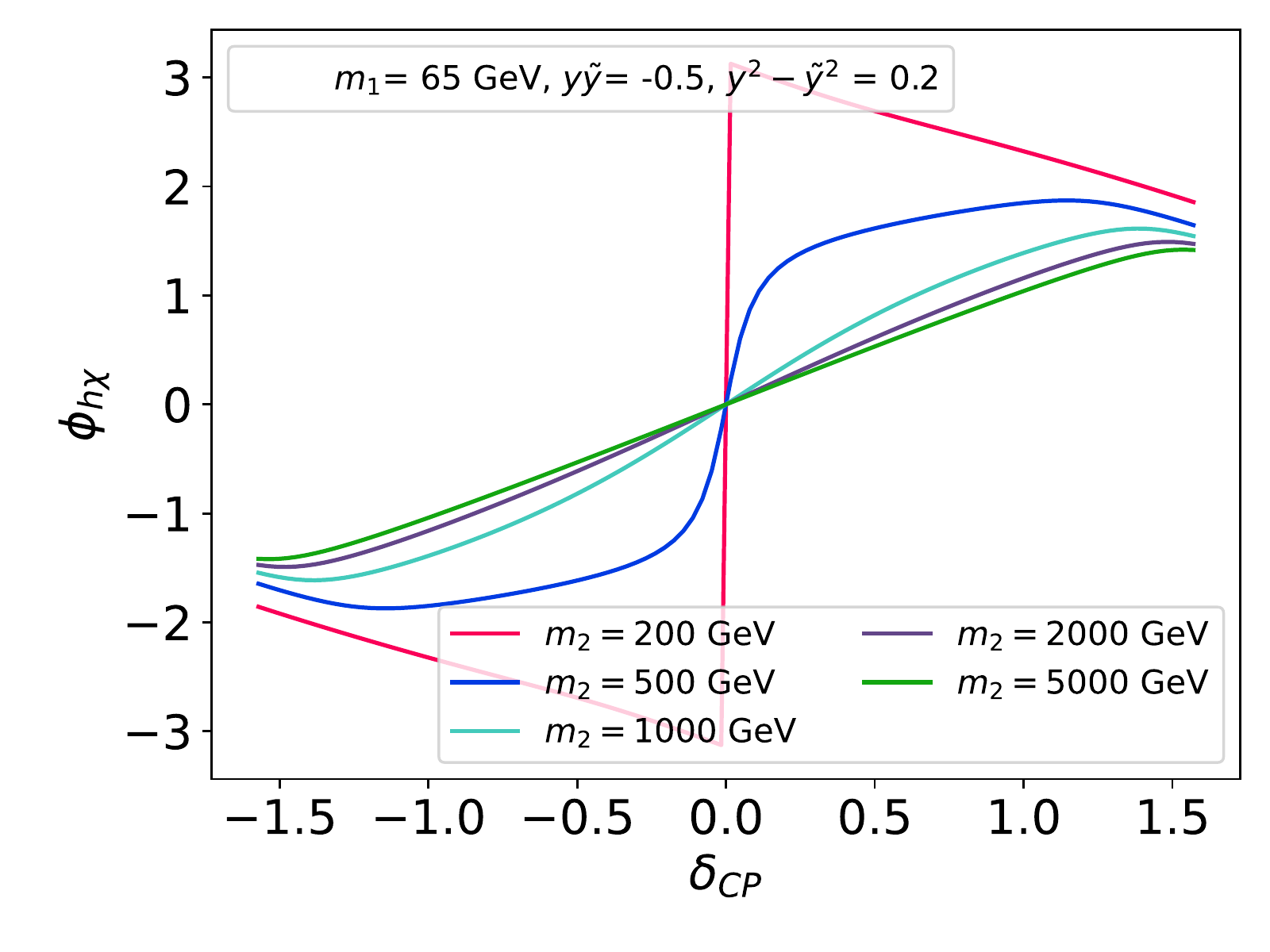}
    \caption{Plot of $\phi_{h\chi}$ as a function of $\delta_{CP}$ for different values of $m_2$. We note that as $m_2$ increases, the IR phase maps directly to the UV phase and $\phi_{h\chi} \sim \delta_{CP}$.}
    \label{fig:phi_EFT_params}
\end{figure}

\begin{figure}[h!]
    \begin{subfigure}{0.5\linewidth}
    \centering
    \includegraphics[width = \textwidth]{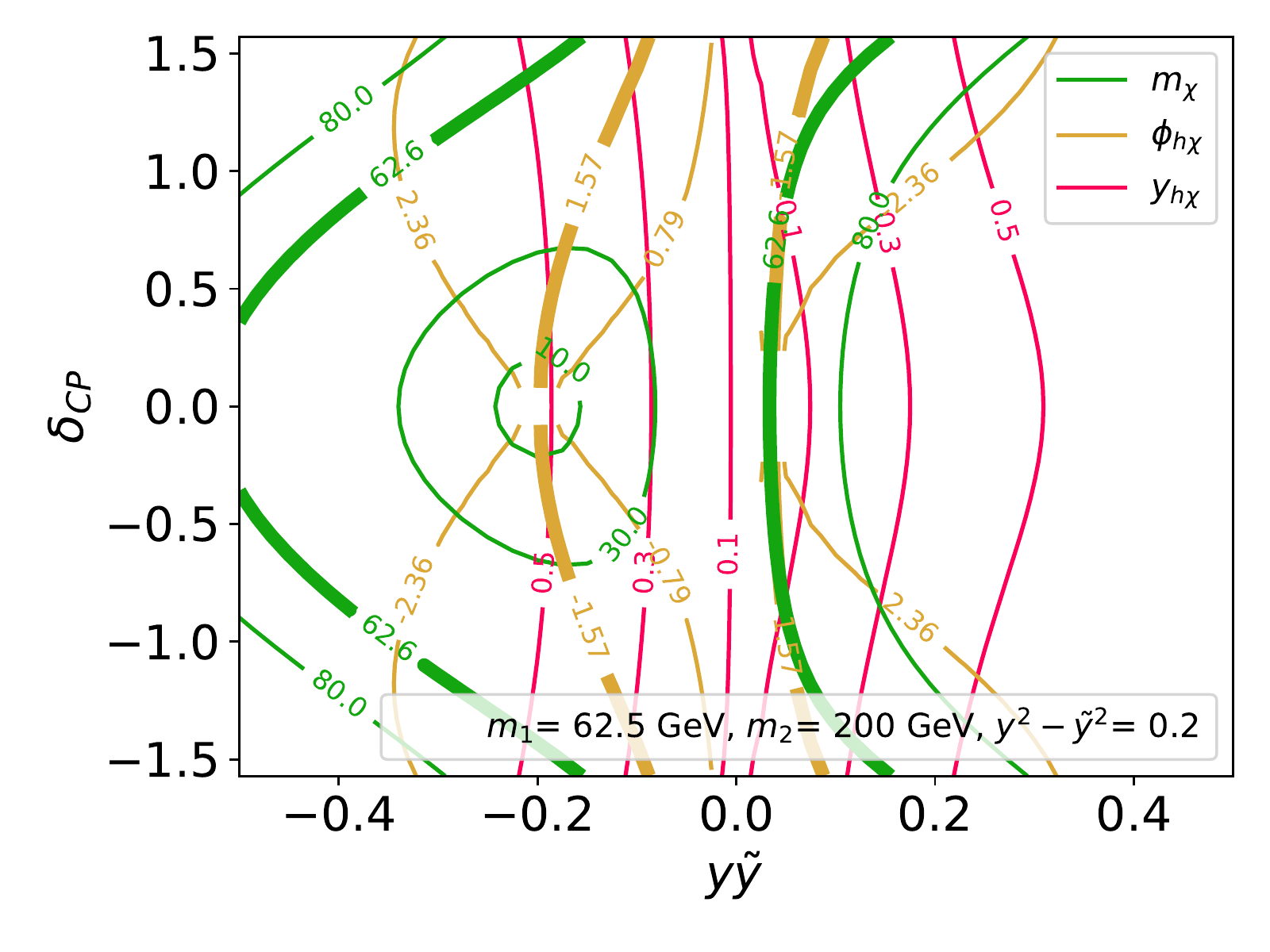} 
    \end{subfigure}
    \begin{subfigure}{0.5\linewidth}
    \includegraphics[width = \textwidth]{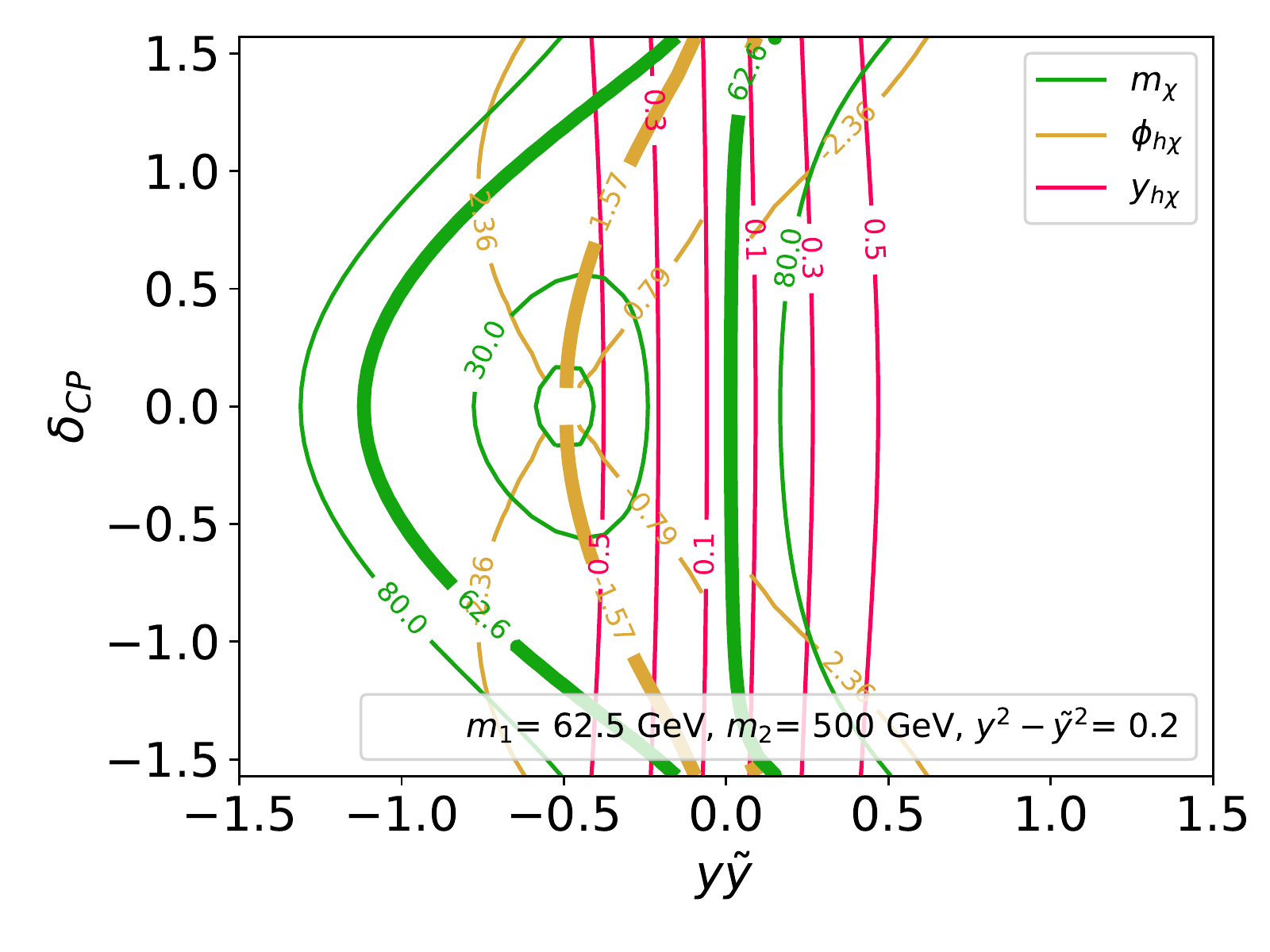}
    \end{subfigure}
    \begin{subfigure}{0.5\linewidth}
    \centering
    \includegraphics[width = \textwidth]{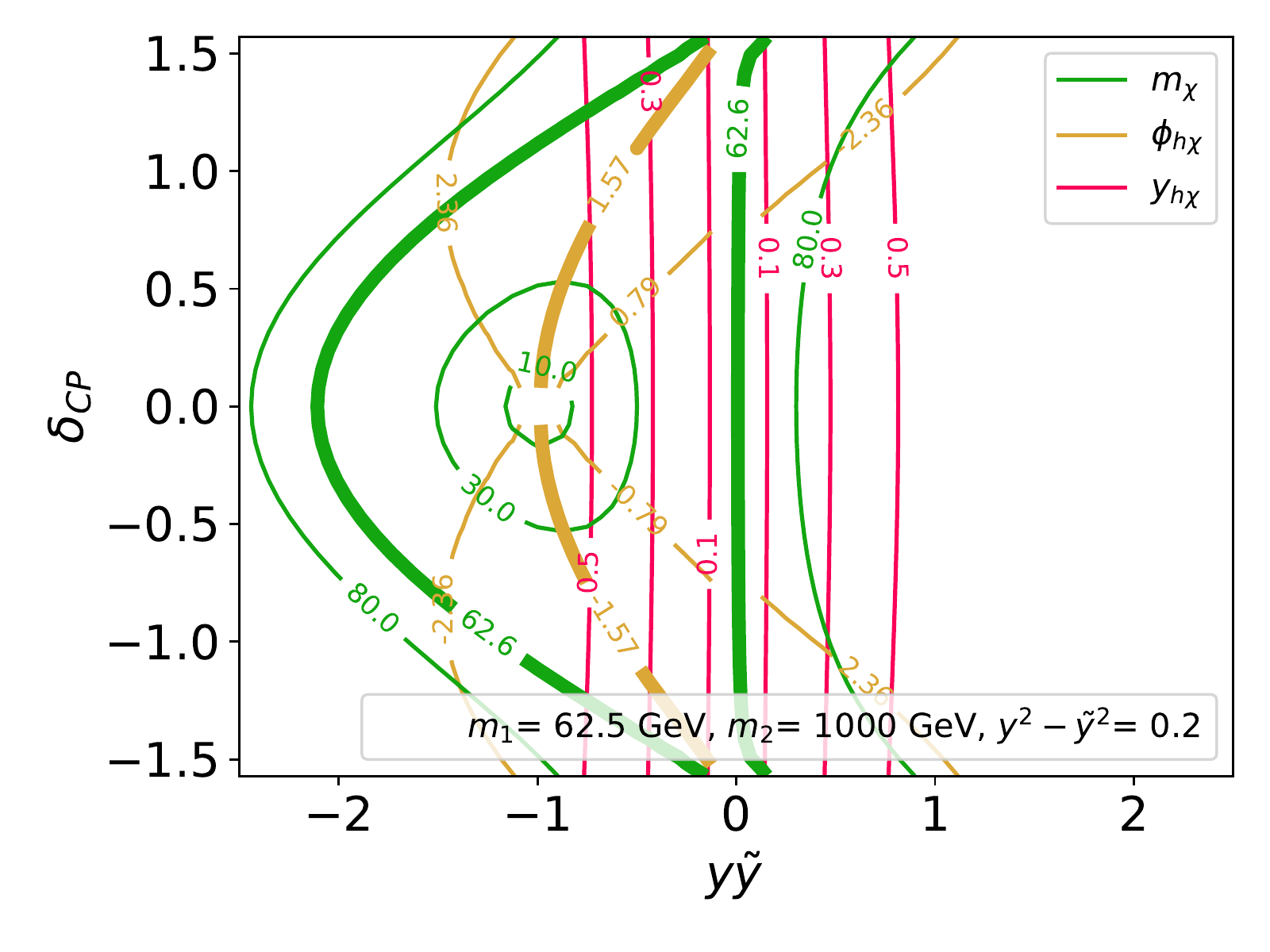} 
    \end{subfigure}
    \begin{subfigure}{0.5\linewidth}
    \includegraphics[width = \textwidth]{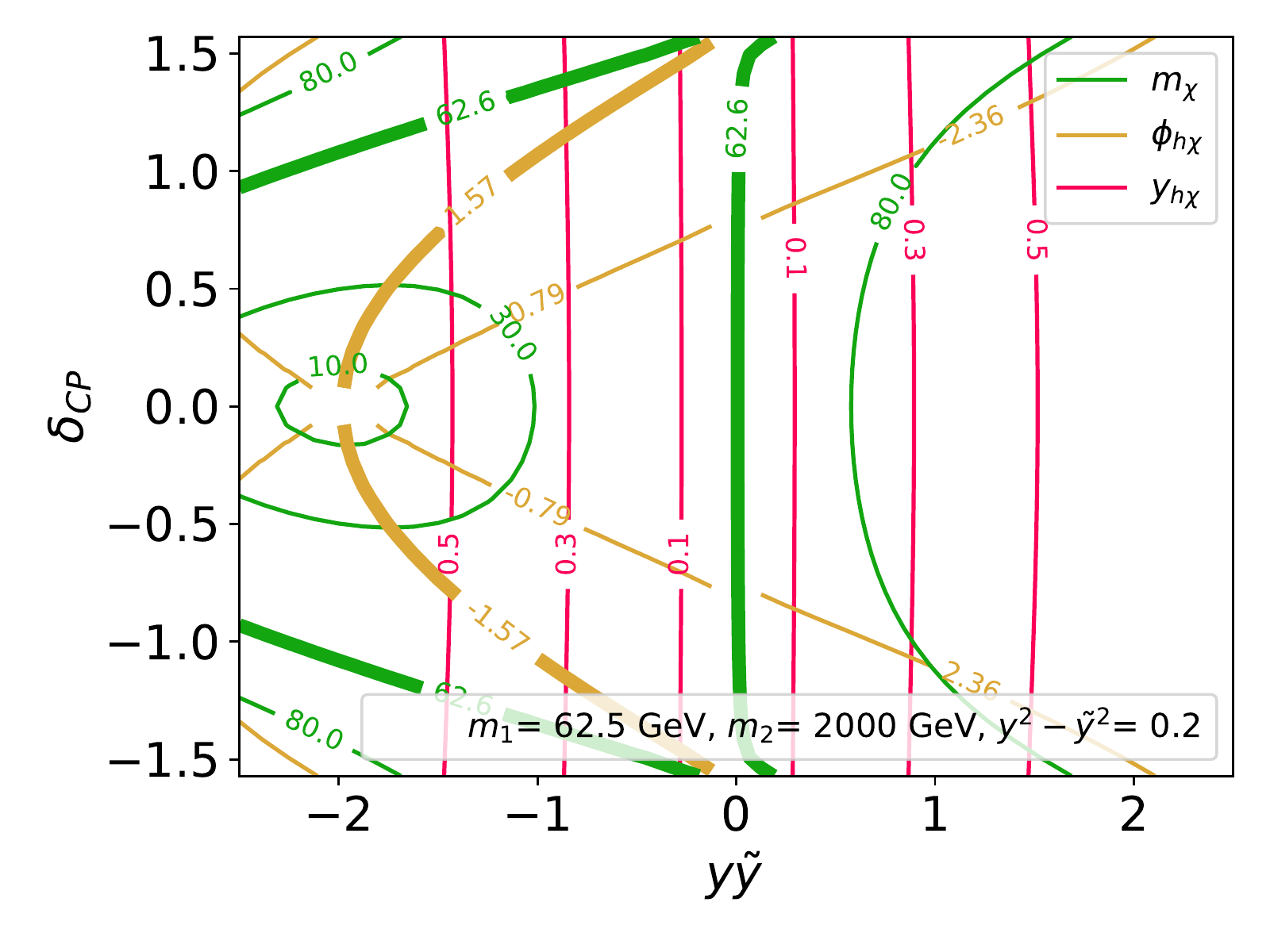}
    \end{subfigure}
    \caption{Dark matter mass, EFT phase, and dark matter-Higgs Yukawa coupling as a function of the UV parameters $y \tilde{y}$ and $\delta_{CP}$ for different values of $m_2$. In each plot we see a similar mass structure: we see a massless state when $y\tilde{y}$ and $m_1$ have opposite signs, and have a lightest fermion near 60 GeV for both larger and smaller $y \tilde{y}$ than this value. We can also see the scaling of both the EFT mass and Higgs coupling with $y \tilde{y}$ and $m_2$. Note the different values on the $y \tilde{y}$ axis in each of the plots. }
    \label{fig:alpha_vs_delta_CP_EFT_params}
\end{figure}

Figure~\ref{fig: mass m2 gg m1} shows that the corrections to the mass scale as $y \tilde{y} v^2/m_2$.\footnote{Although we need to phase rotate $\psi_{2}^s$, the phase rotations in the couplings and mass insertion cancel out.} This diagram also tells us that $\phi_{h\chi} = \delta_{CP}$ in the large $m_2$ limit, as long as mixing is small and the dark matter mass comes mostly from $m_1$ rather than the Higgs vev. This can also be seen in Figure~\ref{fig:phi_EFT_params}. When the dark matter mass gets a large contribution from the Higgs vev the story is more complicated: when $y \tilde{y}$ and $m_1$ have opposite signs, the Higgs contribution can cancel with $m_1$ at $y \tilde{y} = -m_2 m_1/v^2$ to get a massless state. There is a mass resonance contour for $y \tilde{y}$ both larger and smaller than this value, which can be seen in Figure~\ref{fig:alpha_vs_delta_CP_EFT_params}. We might also ask whether a small $\delta_{CP}$ in the UV can translate to $\phi_{h\chi} \approx \pi/2$ in the IR and produce an annihilation signal that evades both direct detection and EDM constraints. However, from the same figure, we can see that although there is a point where small $\delta_{CP}$ translates to $\phi_{h\chi} \approx \pi/2$, it corresponds precisely to the massless state mentioned above and cannot generate our annihilation signal. This is evidenced by all the phase contours converging at the massless point, because when $m_\chi$ is zero, we can freely rotate $m_1$ to absorb the phase in $y$ since the phase is no longer physical.

\begin{figure}
    \centering
    \includegraphics[width = 100 mm]{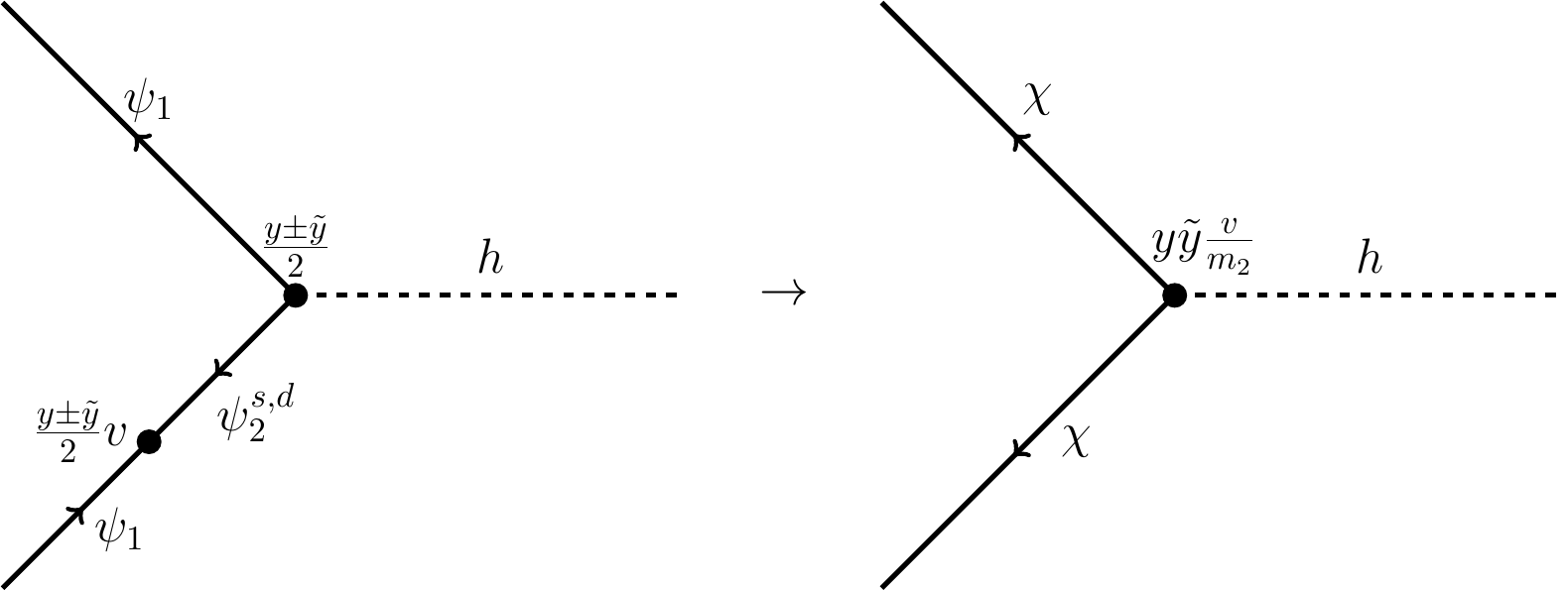}
    \caption{Diagram that generates the dark matter-Higgs coupling in the limit where $m_2$ is large.}
\label{fig: higgs coupling m2 gg m1}
\end{figure}

In the small mixing and large $m_2$ limit, there are two contributions to the Higgs coupling: one where $\psi_1$ mixes into $\psi_{2}^s$ and one where it mixes into  $\psi_{2}^d$, as shown in Figure~\ref{fig: higgs coupling m2 gg m1}. Each of these contributes $(y \pm \tilde{y})^2v/m_2$, with a relative minus sign between the two contributions because we need to phase rotate $\psi_{2}^s$ to have positive mass. This means the Higgs coupling scales as $y\tilde{y}v/m_2$, which determines the scaling of the annihilation signal. This can also be seen from the pink lines in Figure~\ref{fig:alpha_vs_delta_CP_EFT_params}. Note that this scaling breaks down once the Yukawa contributions become the dominant contribution to the mass. 

\begin{figure}
    \centering
    \includegraphics[width = 100 mm]{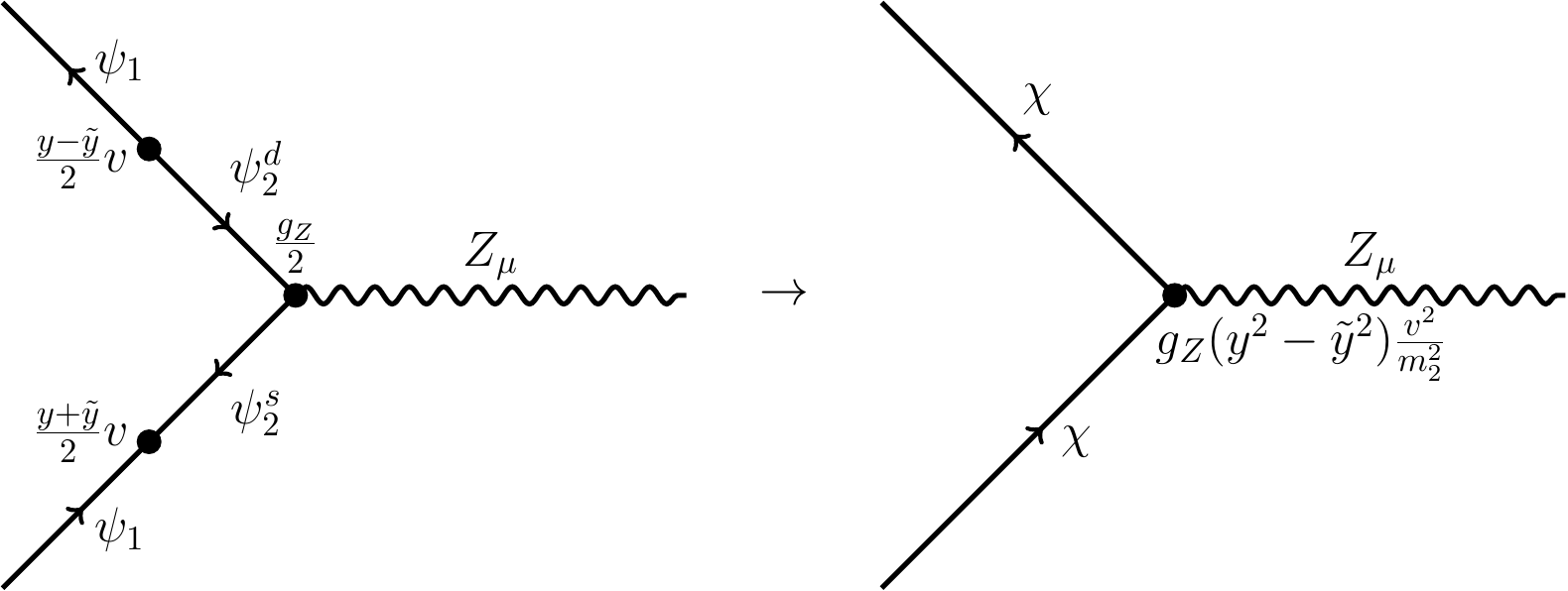}
    \caption{Diagram that generates the dark matter-$Z$ coupling in the limit where $m_2$ is large.}
\label{fig: Z coupling m2 gg m1}
\end{figure}

\begin{figure}
    \centering
    \includegraphics[width = 80 mm]{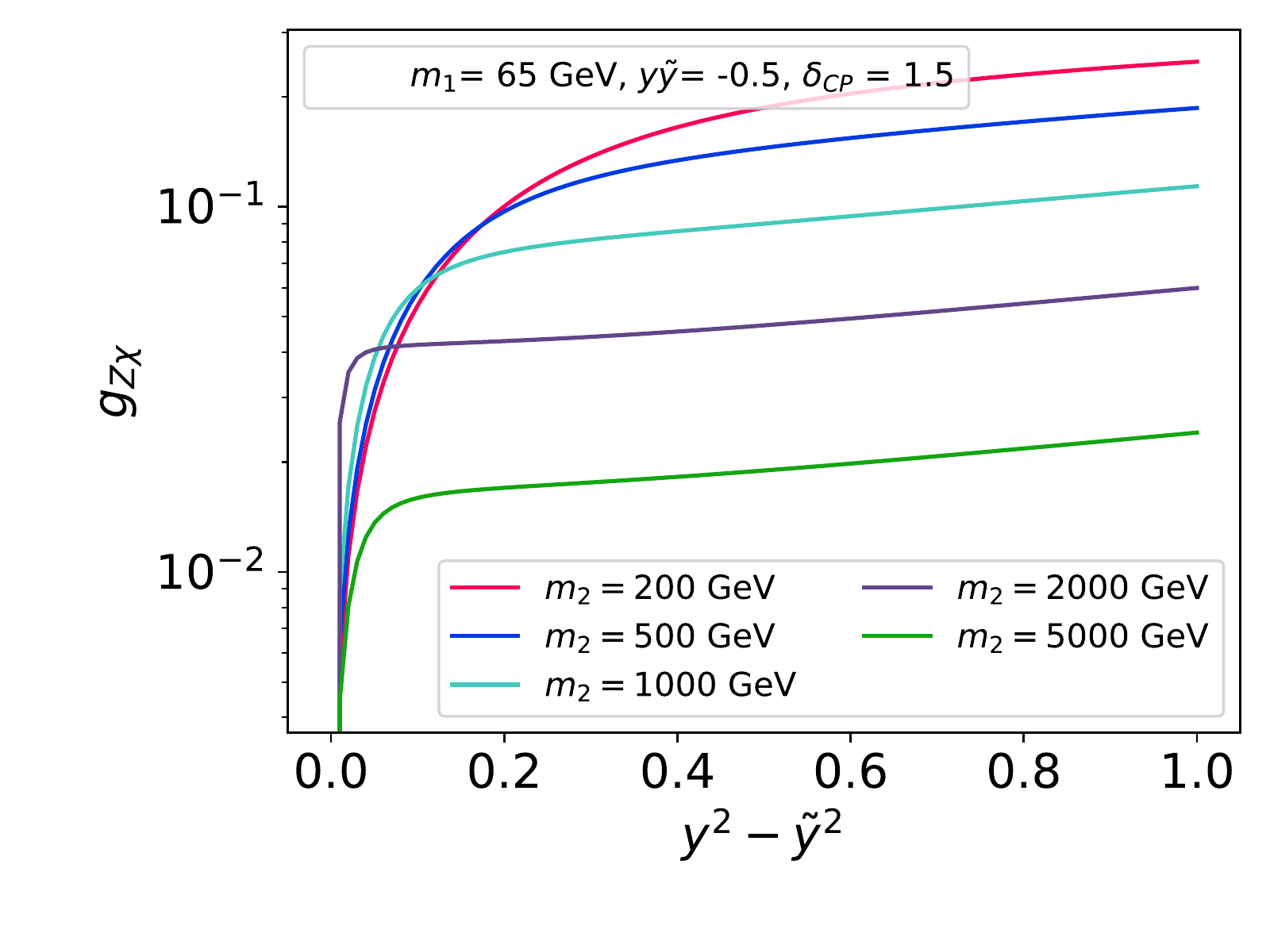}
    \caption{Plot of $g_{Z\chi}$ as a function of $y^2 - \tilde{y}^2$ for different values of $m_2$. $g_{Z\chi}$ increases with increasing $y^2 - \tilde{y}^2$ and decreases with increasing $m_2$, corroborating the scaling derived from the diagrams in Figure~\ref{fig: Z coupling m2 gg m1}.}
    \label{fig:gz_EFT_params}
\end{figure}

In the same limit, the dominant contribution to the $Z$ coupling comes from Figure~\ref{fig: Z coupling m2 gg m1}, which scales as
$g_Z(y^2 - \tilde{y}^2)v^2/m_2^2$. Even away from this limit, we still get a vanishing $Z$ coupling for $y = \tilde{y}$, because only one of the doublet states mixes with the singlet when $y = \tilde{y}$. For small $m_2$, spin-dependent direct detection constraints require $y \approx \tilde{y}$, but for $m_2 \gtrsim 500$ GeV this constraint becomes irrelevant, since the Higgs coupling (which determines the annihilation signal) scales as $m_2^{-1}$ while the $Z$ coupling scales as $m_2^{-2}$. This can be seen in Figure~\ref{fig:gz_EFT_params}. 

\subsection{Constraints}
In this section, we discuss the experimental constraints that apply to the singlet-doublet model. We focus on constraints that apply directly to the parameters in the UV theory, including discussing their scaling in the large $m_2$ limit.

\subsubsection{Electric Dipole Moment}

Any new source of $\CP$-violation in a given model can lead to additional contributions to electric dipole moments. Since our model contains new $\CP$-violating couplings to the Higgs, we expect electron EDM constraints to be relevant for our model. For small $m_2$, the EDM limit will be one of the strongest on our model, since the EDM is precisely constrained to be below $1.1 \times 10^{-29}$ \textit{e} cm \cite{Andreev:2018ayy}.

\begin{figure}
    \centering
    \includegraphics[width = 120 mm]{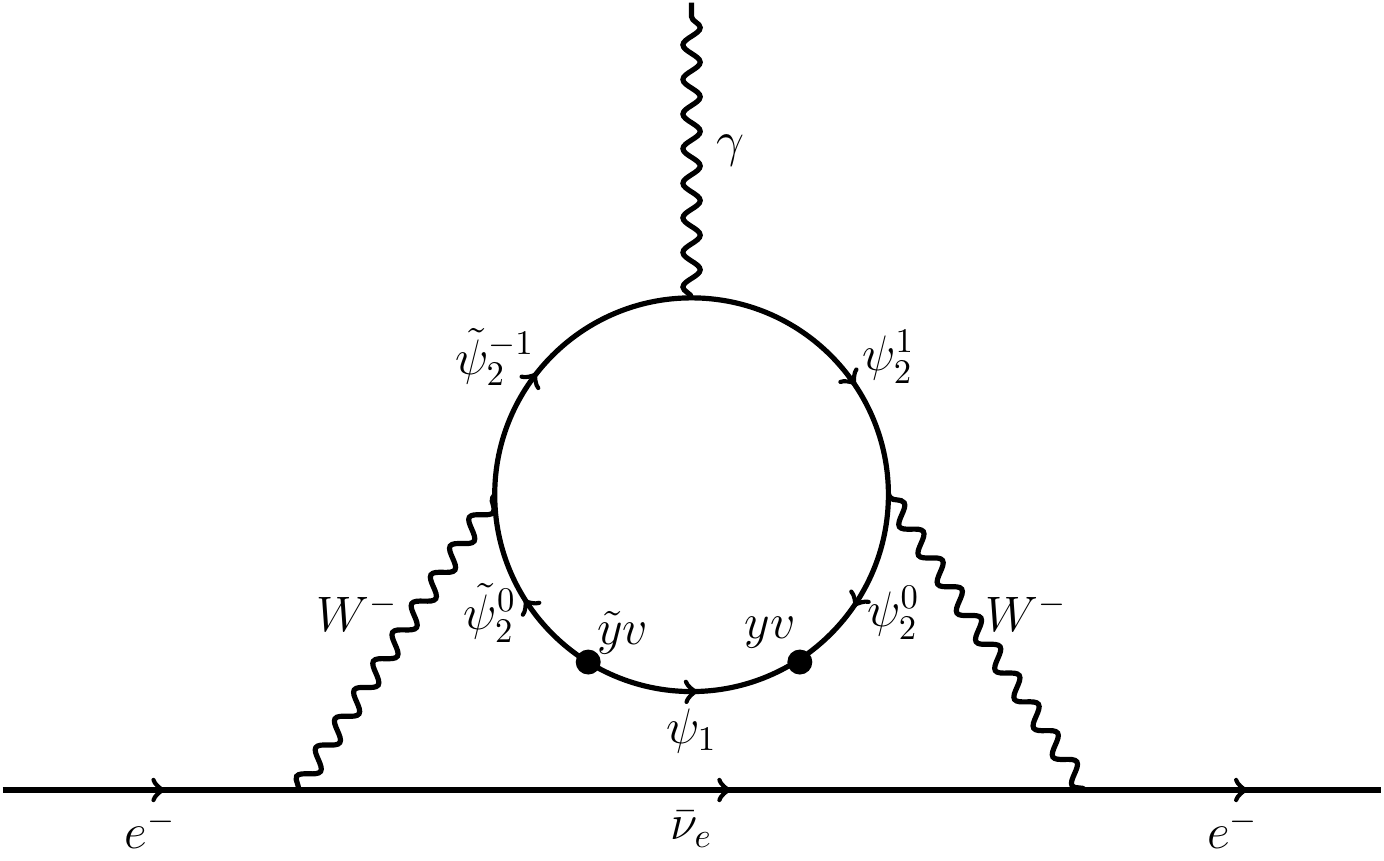}
    \caption{Leading contribution to the non-zero Barr-Zee diagram in the large $m_2$ limit. In this limit, we can work perturbatively in the gauge basis. The relevant $W$ couplings are the coefficients of $\chi_{21}^\dag \frac{\sqrt{g}}{2} W^- \chi_{20}$ and $\tchi_{20}^\dag \frac{\sqrt{g}}{2} W^+ \tchi_{2-1}$.}
\label{fig:BarrZeeW}
\end{figure}

For the singlet-doublet model above, the only relevant diagram is the Barr-Zee diagram with $W$ bosons in the outer loop \cite{Barr:1990vd}, displayed in Figure~\ref{fig:BarrZeeW}. There are no other Barr-Zee diagrams with Higgs or $Z$ legs; since $\CP$-violation is only in the neutral sector of this model and a charged particle is necessary to radiate a photon, the inner loop must contain both a neutral and charged particle.
Additionally, there are no other non-Barr-Zee diagrams that contribute to the EDM at 2 or fewer loops. For any non-Barr-Zee diagrams to contribute, there would have to be a $\CP$-odd correction to a gauge boson or Higgs propagator. With only a single external momentum, it is impossible to contract with an epsilon tensor and make a non-vanishing $\CP$-odd Lorentz invariant.

To compute the value of the relevant Barr-Zee diagram, we use a simplified version of Equation 21 in~\cite{Atwood:1990cm}, where we have neglected the neutrino mass, approximated lepton couplings as flavor diagonal, and used the fact that one of the fermions in the loop is neutral:
\begin{equation}
    \begin{split}
    \frac{d_e}{e} = -\frac{g^2}{(4 \pi)^4}\sum_{i} \text{Im}([aJ]_{i}^* [bJ^*]_{i})\left(\frac{m_c m_{n,i} m_e}{M_W^4}\right)G(x_c, x_i, 0).
    \end{split}
\end{equation}
Here, $x_\alpha = m_\alpha^2/M_W^2$ and $G(a,b,c)$ is defined as
\begin{equation}
   G(a,b,c) = \frac{1}{1-c}\int_0^1 \frac{dx}{1 - x} \left( \frac{c}{z - c} \log\left(\frac{c}{z}\right) + \frac{1}{1 - z} \log \left( \frac{1}{z}\right)\right)
\end{equation}
with
\begin{equation}
    z(x,a,b) = \frac{b}{x} + \frac{a}{1 - x}.
\end{equation}
Recall from Section~\ref{section:model} that couplings $a_{i}$ and $b_{i}$ parameterize the $W$ boson couplings to the inner loop fermions in the gauge basis, which are given in Equation~\ref{singlet_doublet_Wcouplings}, and $J$ is the change of basis matrix.

When $m_2$ is large enough that we can integrate out the doublet and mixing is small, the dominant contribution to the EDM comes from Figure~\ref{fig:BarrZeeW}, since each helicity of charged fermion couples to a different neutral doublet component and mixing with the singlet is necessary to generate $\CP$-violation. This contribution scales as $y\tilde{y}v^{2}/m_2^2$. The $m_2$ scaling follows from dimensional analysis: three factors of $m_2$ from the integral measure cancel with three of the five factors of $m_2$ from the propagators.\footnote{The $\psi_1$ propagator also scales as $m_2^{-1}$ since $p = m_2 \gg m_1$.}

\subsubsection{Electroweak parameters}

\begin{figure}
    \centering
    \includegraphics[width = 120 mm]{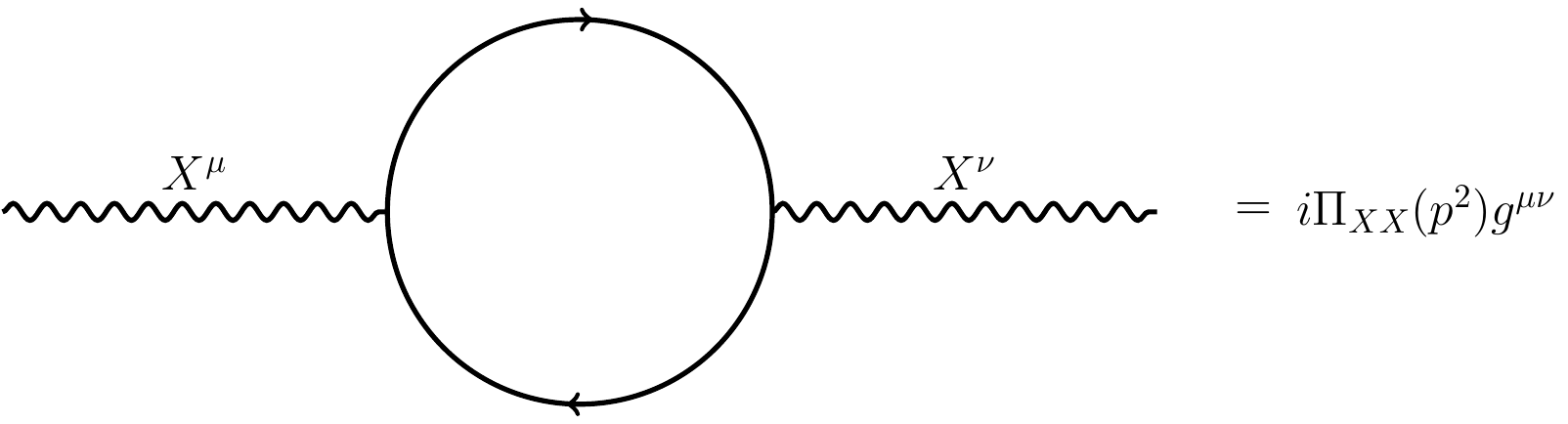}
    \caption{New particles that couple to the Standard Model gauge bosons contribute to the vacuum polarization at 1-loop through this diagram. The $X^{\mu}$ represents an electroweak gauge boson. We ignore the $p^{\mu}p^{\nu}$ terms since they aren't relevant for Equations~\ref{eqns:STU1} - \ref{eqns:STU2}.}
\label{fig:Oblique_Parameter}
\end{figure}

Here we consider constraints from electroweak precision measurements, where deviations from the SM are parametrized by oblique parameters $S, T , U, W,$ and $Y$~\cite{Peskin:1991sw,Cacciapaglia:2006pk}, defined in Equations~\ref{eqns:STU1}-\ref{eqns:STU2}.

\begin{align}
    S & \equiv \frac{4c^2 s^2}{\alpha_e} \left[ \Pi'_{ZZ}(0) -\frac{c^2 -s^2}{cs}  \Pi_{Z\gamma}'(0) -\Pi_{\gamma\gamma}'(0) \right]
    \label{eqns:STU1}\\
    T & \equiv \frac{1}{\alpha_e} \left[ \frac{\Pi_{WW}(0)}{m_W^2} -\frac{\Pi_{ZZ}(0)}{m_Z^2} \right]\\
    U & \equiv \frac{4s^2}{\alpha_e} \left[ \Pi'_{WW}(0) -\frac{c}{s}  \Pi_{Z\gamma}'(0) -\Pi_{\gamma\gamma}'(0) \right] -S \\
    W & \equiv \frac{m_W^2 s^2 c^2}{8\pi\alpha_e} \left[ \Pi''_{ZZ}(0) +\frac{2s}{c}\Pi''_{Z\gamma}(0) +\frac{s^2}{c^2}\Pi''_{\gamma\gamma}(0) \right]\\
    Y &= \frac{m_W^2 s^2}{8 \pi \alpha_e}\Bigg[c^2 \Pi_{\gamma \gamma}''(0) + s^2 \Pi_{ZZ}''(0) - 2sc \Pi_{\gamma Z}''(0)\Bigg]
    \label{eqns:STU2}
\end{align}
The masses and couplings are evaluated at $m_Z^2$ and $c$ and $s$ are $\cos\theta_{W}$ and $\sin\theta_{W}$ respectively. $\Pi_{XX}$ represents the new particles' contribution to the vacuum polarization of the gauge boson $X$ at 1-loop, computed in $\overline{\mathrm{MS}}$ scheme under the convention shown in Figure~\ref{fig:Oblique_Parameter}.

To lend intuition, we note that $T$ parametrizes custodial SU(2) breaking inherent in the asymmetry within the doublet terms; in our theory this manifests in the difference in Yukawa couplings $y$ and $\tilde y$. $U$ is the derivative of $T$, and thus is typically smaller. All these parameters fall off with increasing $m_2$.

The most recent constraints, at 95\% CL, from the LHC yield 

\begin{equation}
    S = -0.01\pm 0.10 \qquad T=0.03\pm 0.12 \qquad U = -0.01 \pm 0.10
\end{equation}
with correlations +0.92 between $S$ and $T$, -0.80 between $S$ and $U$, and -0.93 between $T$ and $U$ \cite{RPP2020}. W and Y are measured to be
\begin{equation}
    W = (-2.7 \pm 2.0) \times 10^{-3} \qquad Y = (4.2 \pm 4.9) \times 10^{-3}
\end{equation} with correlation $-0.96$ \cite{Barbieri:2004qk}, though we find these to be subdominantly constraining for this theory.

\subsubsection{Collider Experiments}
Constraints from many collider searches (in particular SUSY searches) can be applied to this model. Specifically, we consider those searches included in the database of the publicly available SModelS version 1.2.4 software \cite{Khosa:2020zar,Ambrogi:2018ujg,Dutta:2018ioj,Ambrogi:2017neo,Kraml:2013mwa,ATL-PHYS-PUB-2019-029,Skands:2003cj,Alwall:2006yp,Buckley:2013jua}. To generate the necessary input, we use SARAH 4.14.3 \cite{Staub:2008uz,Staub:2013tta,Staub:2015kfa} to create modified versions of SPheno \cite{Porod:2003um,Porod:2011nf} and Madgraph \cite{Alwall:2014hca,Alwall:2011uj} which include the singlet and doublet. Then we use this version of SPheno at tree level to compute the spectrum and branching ratios for SModelS and the run card for Madgraph, which was used to obtain the production cross sections that SModelS also needs as input. These constraints are combined into a single exclusion limit labeled LHC when included in our plots. In addition to this constraint, we also show the constraint from invisible Higgs decay. We do not include the constraint from invisible $Z$ decay, since it is not kinematically allowed in the parameter space of interest.

\subsection{Full Exclusion Limits and Discussion}
\begin{figure}
    \centering
    \begin{subfigure}{0.8\linewidth}
    \centering
    \includegraphics[width = \textwidth]{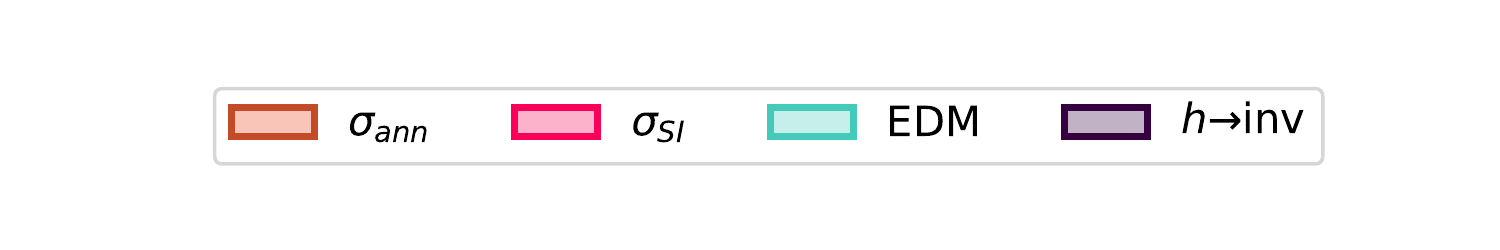} 
    \end{subfigure}
    \begin{subfigure}{0.5\textwidth}
    \includegraphics[width = \textwidth]{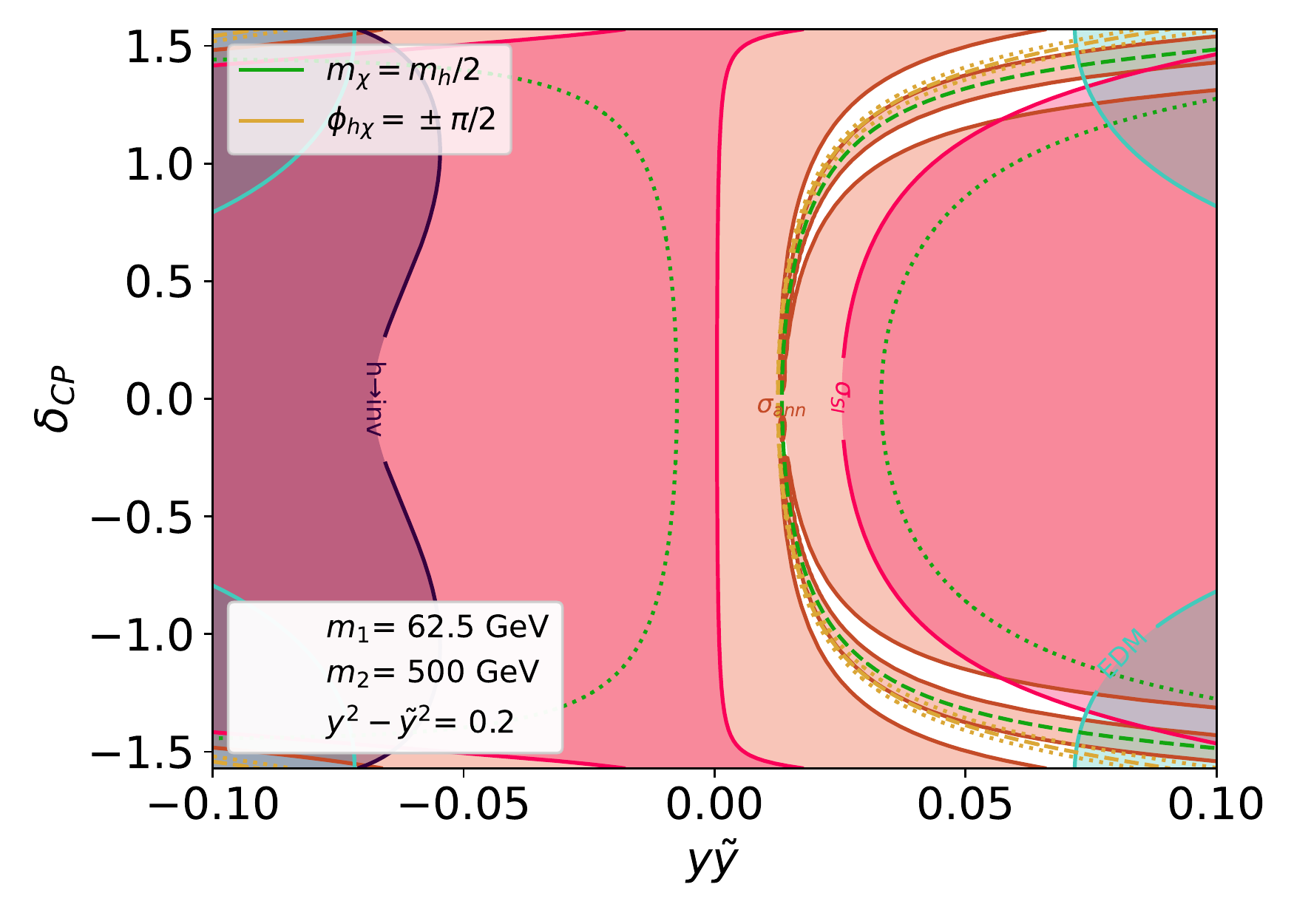} 
     \captionsetup{justification=raggedleft, singlelinecheck=false}%
    \caption{\phantom{hellothereifudgeaspace}}
    \label{Singlet_Doublet_case_a}
    \end{subfigure}%
    \begin{subfigure}{0.5\textwidth}
    \includegraphics[width = \textwidth]{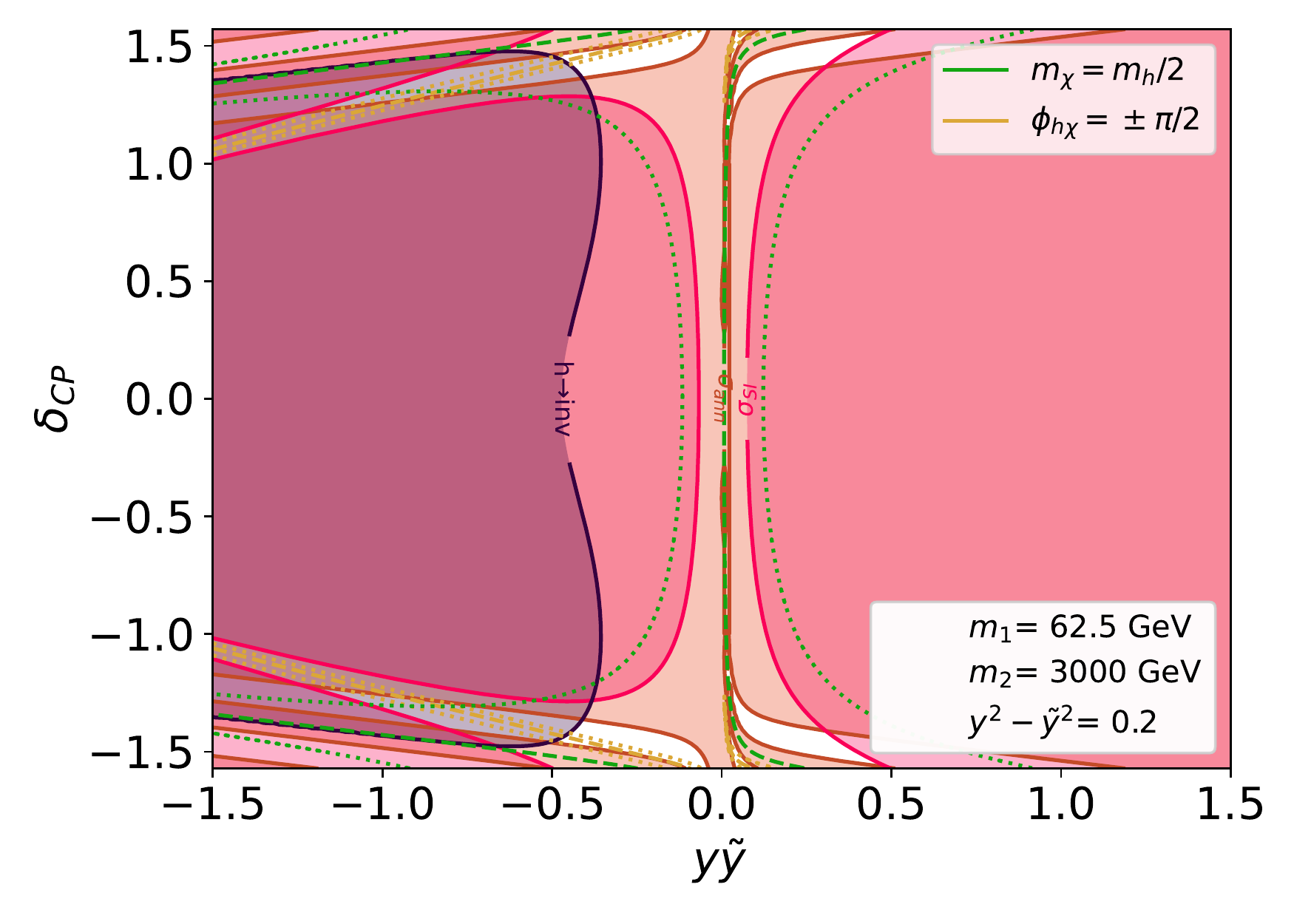}
     \captionsetup{justification=raggedleft, singlelinecheck=false}%
    \caption{\phantom{hellothereifudgeispace}}
    \label{Singlet_Doublet_case_b}
    \end{subfigure}
    \begin{subfigure}{0.5\textwidth}
    \includegraphics[width = \textwidth]{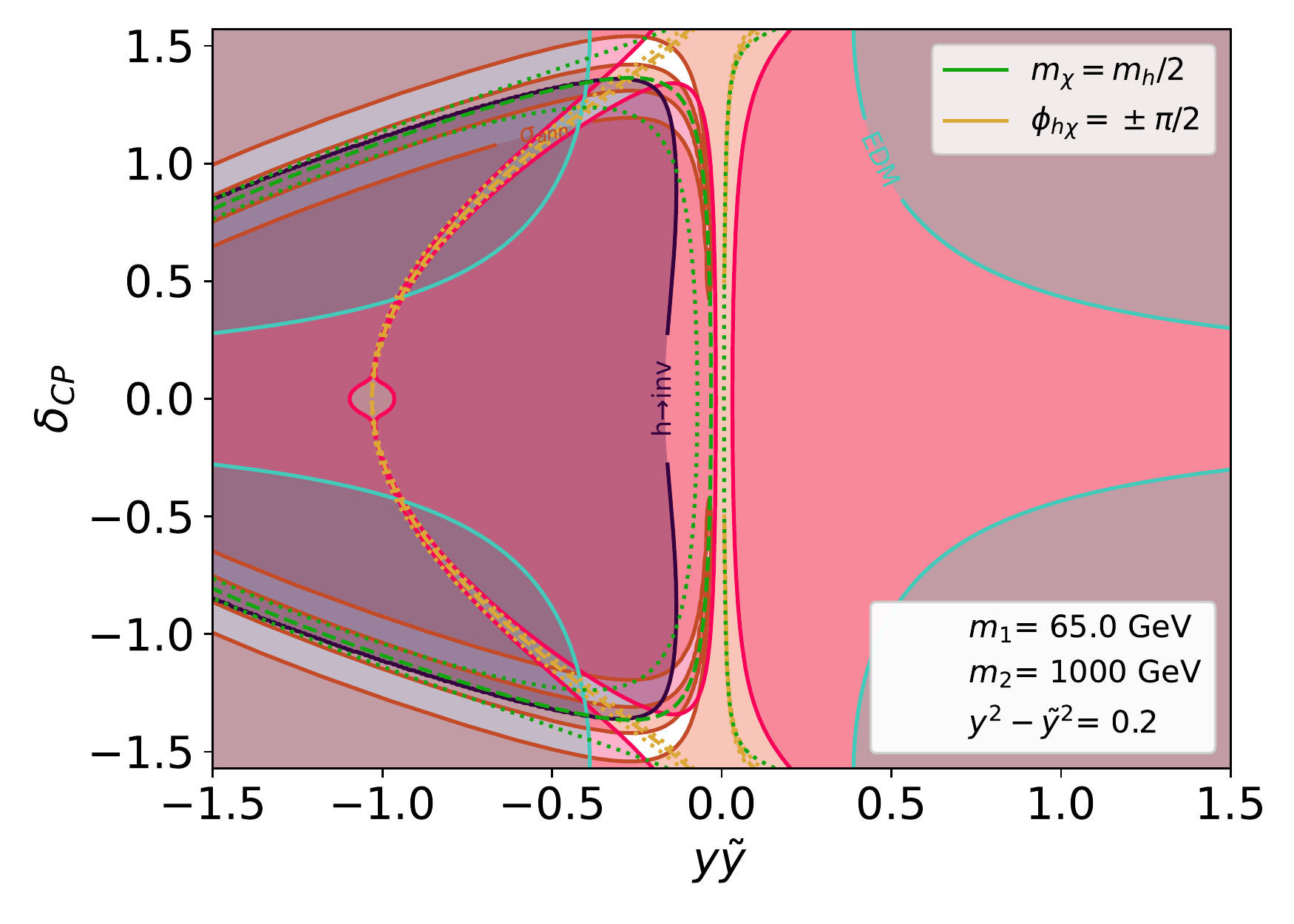}
     \captionsetup{justification=raggedleft, singlelinecheck=false}%
    \caption{\phantom{hellothereifudgeaspace}}
    \label{Singlet_Doublet_case_c}
    \end{subfigure}%
    \begin{subfigure}{0.5\textwidth}
    \includegraphics[width = \textwidth]{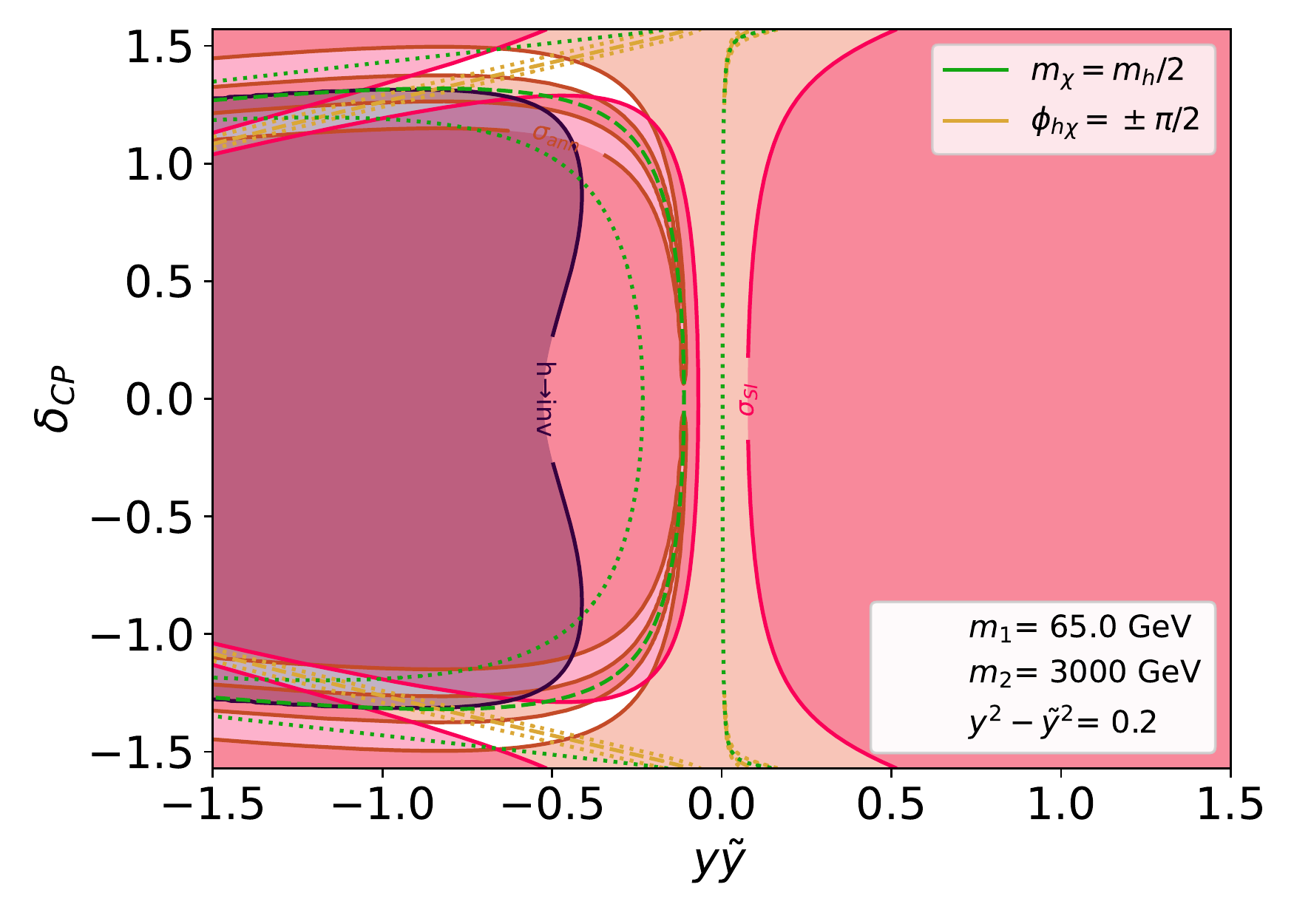}
     \captionsetup{justification=raggedleft, singlelinecheck=false}%
    \caption{\phantom{hellothereifudgeispace}}
    \label{Singlet_Doublet_case_d}
    \end{subfigure}    
    \caption{Full constraints on the model parameter space in the $y \tilde{y} - \delta_{CP}$ plane, for different values of $m_1$ and $m_2$.  In this and subsequent plots the shaded regions denote parameter space ruled out by experimental bounds~\cite{Andreev:2018ayy,Sirunyan:2019twz,Aprile:2017iyp,Aprile:2018dbl}. For annihilation, we include both an upper and lower bound. Other constraints are not relevant for these slices of parameter space. Spin-dependent direct detection constraints in particular are weak since $y^2 - \tilde{y}^2$ is small. Dotted lines indicate proximity to mass resonance and pure imaginary EFT coupling: the green dotted lines bound a region with dark matter mass $60\, \text{GeV} \leq m_\chi \leq 65 \, \text{GeV}$, the yellow with EFT phase  $1.55 \leq \phi_{h\chi} \leq 1.6$.
    In Figures (a) and (b) we show that viable parameter space can be found at small couplings, corresponding to a pure mass resonance with flexibility in $\phi_{h\chi}$. In this case, smaller values of $m_2$ are allowed but $m_1$ must be close to $m_h/2$. Figures (b) - (d) also show allowed parameter space for larger couplings: (b) shows $m_1 \approx m_h/2$; (c) and (d) show $m_1$ further away from $m_h/2$ for two different values of $m_2$. In all of (b) - (d), viable parameter space requires $m_1 \approx 60-70$ GeV, $\delta_{CP} \gtrsim 1$, and $\phi_{h \chi} \approx \pi/2$.}
    \label{fig:alpha_delta}
\end{figure}

\begin{figure}
    \centering
    \begin{subfigure}{0.8\linewidth}
    \centering
    \includegraphics[width = \textwidth]{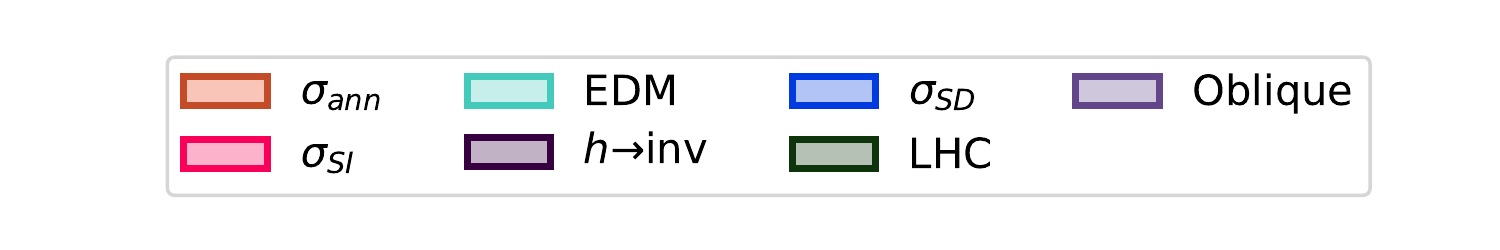} 
    \end{subfigure}
    \begin{subfigure}{0.5\linewidth}
    \includegraphics[width = \textwidth]{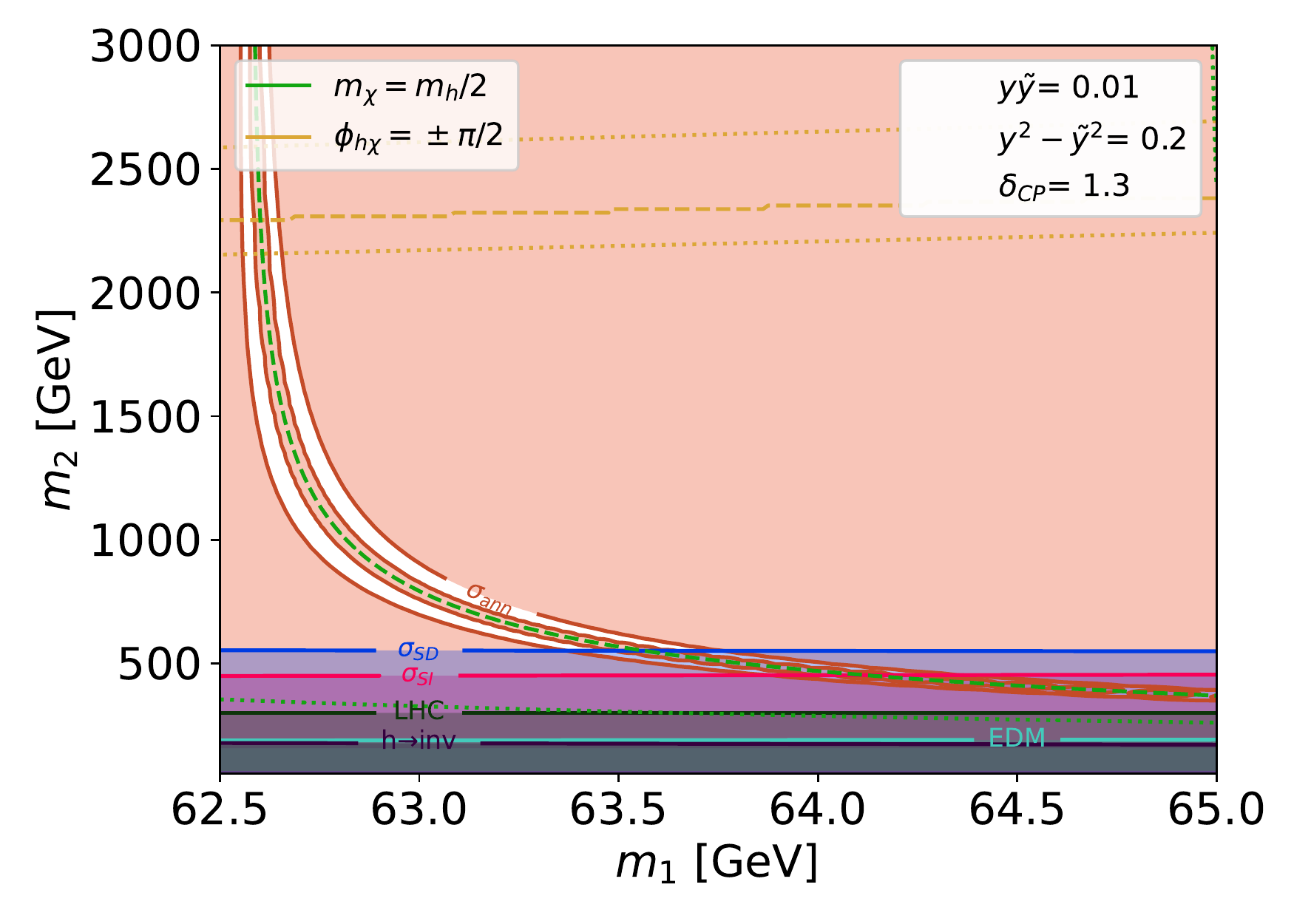}
     \captionsetup{justification=raggedleft, singlelinecheck=false}%
    \caption{\phantom{hellothereifudgeaspace}}
    \label{m1_m2_small_couplings}
    \end{subfigure}%
    \begin{subfigure}{0.5\linewidth}
    \centering
    \includegraphics[width = \textwidth]{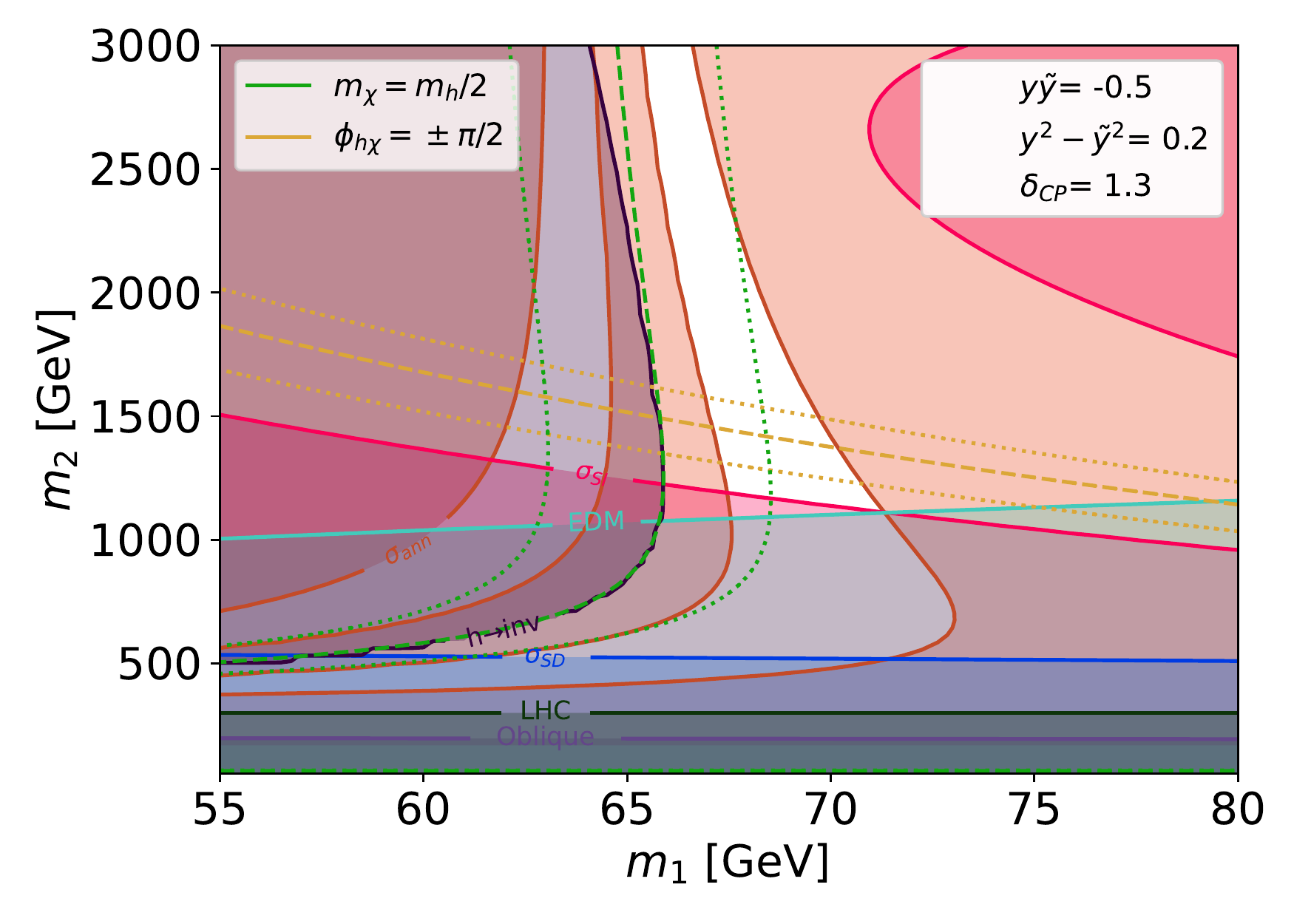} 
     \captionsetup{justification=raggedleft, singlelinecheck=false}%
    \caption{\phantom{hellothereifudgeispace}}
    \label{m1_m2_large_couplings}
    \end{subfigure}
     \caption{Similar to Figure~\ref{fig:alpha_delta}, in slices of the $m_1 - m_2$ plane and for different values of $y\tilde y$ and $\delta_{CP}$. We show the same constraints as Figure~\ref{fig:alpha_delta} and constraints from~\cite{RPP2020,Barbieri:2004qk,Aprile:2019dbj,PhysRevLett.118.251301,PhysRevD.100.022001,Khosa:2020zar,Ambrogi:2018ujg,Dutta:2018ioj,Ambrogi:2017neo,Kraml:2013mwa,ATL-PHYS-PUB-2019-029,Skands:2003cj,Alwall:2006yp,Buckley:2013jua}. Dotted lines around the critical mass and phase values give a guide towards the proximity of any viable space to mass resonance and pure imaginary EFT coupling: the green dotted lines bound a region with dark matter mass $60\, \text{GeV} \leq m_\chi \leq 65 \, \text{GeV}$, the yellow with EFT phase  $1.55 \leq \phi_{h\chi} \leq 1.6$. The left shows the case of a mass resonance with small couplings, where $m_2$ down to $\sim 500$ GeV is allowed. The right shows the case of larger couplings, where we need $m_2 \gtrsim \mathcal{O}(1)$ TeV. We omit light charged fermion constraints since small $m_2$ is already ruled out.}
    \label{fig:m1_m2}
\end{figure}

Finally, combining all of these constraints, we examine the remaining parameter space for singlet-doublet dark matter that has the desired amount of annihilation. Our results are shown in Figures~\ref{fig:alpha_delta} and~\ref{fig:m1_m2}. We find that in all cases, some tuning of the parameters is required, but that there is flexibility in which UV parameters we need to tune. 

As in the EFT, in order to achieve a pure mass resonance (and not have to tune the EFT phase) we need small couplings. This can be seen in Figures~\ref{Singlet_Doublet_case_a} and~\ref{Singlet_Doublet_case_b}.
The spin-independent constraints are weak for small couplings, regardless of $\delta_{CP}$ or the EFT phase. Other constraints are even less restrictive, except for the EDM at very large $\delta_{CP}$. Since the couplings are small, $m_1$ must be tuned near $m_h/2$ in order to achieve a sufficient annihilation signal, but there is flexibility in the value of $m_2$, as can be seen in Figure~\ref{m1_m2_small_couplings}. This is the region of parameter space that is relevant for the best fit in \cite{Carena:2019pwq}. 

If instead we choose our parameters so that we allow  the EFT phase to be tuned near $\pi/2$, there is other viable parameter space with larger couplings. Here, we have slightly more flexibility in $m_1$ (which still needs to be roughly $60-70$ GeV), but $m_2$ must be large ($m_2 \gtrsim \mathcal{O}(1)$ TeV) to avoid EDM, electroweak, and collider constraints. This can be seen in Figure~\ref{m1_m2_large_couplings}. Additionally, to achieve an EFT phase near $\pi/2$ and avoid spin-independent constraints, generally $\delta_{CP} \gtrsim 1$. Note that limits from spin-dependent scattering can be avoided, since they vanish when $y = \pm \tilde y$. This part of parameter space generally requires proximity to both the mass resonance and the phase $\pi/2$ line.  However, there is still some flexibility in both values; masses $m_\chi > 65 \, \text{GeV}$ and phases $\phi_{h\chi} < 1.5$ are allowed in these intersections, albeit not simultaneously. Unlike in the case of the EFT, it is very difficult to tune only the phase because we cannot make couplings arbitrarily large without affecting the mass spectrum, as we saw in  Section~\ref{subsec:UV_to_EFT}.

Figures ~\ref{Singlet_Doublet_case_b} - ~\ref{Singlet_Doublet_case_d} shows several examples of this. In Figure~\ref{Singlet_Doublet_case_b}, we can see the case where we still choose $m_1$ to be near $m_h/2$ but allow larger couplings. If instead we choose $m_1$ further away from $m_h/2$, the only viable parameter space requires large couplings in order to get the dark matter mass sufficiently close to resonance. This is shown in Figures~\ref{Singlet_Doublet_case_c} and~\ref{Singlet_Doublet_case_d}. Comparing these two plots, we can see that there is more flexibility in $\delta_{CP}$ and larger required coupling values for higher $m_2$, because higher $m_2$ changes the shape of the EFT phase $\pi/2$ contour. Specifically, there is more overlap between $\phi_{h \chi}$ near $\pi/2$ and the annihilation signal in the large $m_2$ case since the condition $\phi_{h \chi} =\pi/2$ becomes less dependent on $y\tilde{y}$ at larger $m_2$.\footnote{This is because the $\phi_{h \chi} =\pi/2$ contour always goes through the massless state that exists for negative $y \tilde{y}$, which occurs at larger couplings for larger $m_2$. All phase contours go through this point since the phase becomes unphysical when the lightest state is massless.}

\section{Doublet-Triplet Model \label{sec:doublet_triplet}}

In this section, we describe another potential UV completion, doublet-triplet dark matter. This model includes the addition of a doublet Dirac fermion and a triplet Majorana fermion to the Standard Model. This model has been previously discussed in other contexts in~\cite{Dedes:2014hga,Abe:2014gua,Freitas:2015hsa,Lopez-Honorez:2017ora}. 
\subsection{Model in the UV \label{sec:doublet_triplet_UV_Model}}

We begin by describing our model and establishing the notation. This model contains a Dirac doublet of two left handed Weyl fermions with hypercharge 1/2 (denoted by $\psi_2$ and $\tpsi_2$ as in the singlet-doublet case) and a triplet of Majorana fermions (with components $\psi_3^{-1}, \psi_3^0, \psi_3^1$), all of which are SU(3) singlets. The Lagrangian is given by 

\begin{equation}
\mathcal{L} = \Lag_{\text{SM}} + \Lag_{\text{kinetic}} -\frac{1}{2} m_3 \psi_3 \psi_3  - m_2 \psi_2 \cdot \tpsi_2 - Y H^\dag \psi_3 \psi_2 - \tilde{Y} (\epsilon H^*)^\dag \psi_3 \tpsi_2 + \textrm{h.c.}
\end{equation}
As in the singlet-doublet case, this theory also has a single physical phase, and we can choose the same convention as the previous section to localize $\CP$-violation to the Yukawa couplings, where

\begin{equation}
    Y \equiv y \, \mathrm{e}^{i\delta_{CP}/2}, \qquad
    \tilde Y \equiv \tilde y \, \mathrm{e}^{i\delta_{CP}/2}. 
\end{equation}

Next we describe our notation after SSB. We denote the gauge basis neutral particles by $\psi_n = \{\psi^0_3, \psi^s_2, \tilde\psi^d_2\} $ and the gauge basis charged particles by $\psi^{+}_c = \{\psi^{+1}_3, \psi^{+1}_2 \}$ and $\psi^{-}_c = \{\psi^{-1}_3, \tilde\psi^{-1}_2 \}$. We label the neutral mass eigenstates $\chi_n =\{\chi, \chi_1, \chi_2\} $ and the charged mass eigenstates $\chi^{+}_c = \{\chi^{+1}_1, \chi^{+1}_2 \}$, and  $\chi^{-}_c = \{\chi^{-1}_1, \chi^{-1}_2 \}$. Each is ordered from least to most massive, and $\chi$ again denotes the dark matter. We call the basis change matrices $J_n$ and $J^\pm_c$, which are defined by $\psi_n = J_n \chi_n$, $\psi_c^\pm = J_c^\pm \chi_c^\pm$. The phases of the eigenvectors are chosen such that the mass eigenvalues are real. Then the mass terms are given by

\begin{equation}
    \Lag_{mass} = - \frac{1}{2} \chi_n [J_n^T  M_n J_n] \chi_{n} - \chi_c^{-}[J_-^T M_c J_+ ]\chi^+_{c} + \textrm{h.c.}
\end{equation}
with
\begin{equation}
    M_n \equiv \begin{pmatrix}
    m_{3} & (y-\tilde{y}) v/2\sqrt{2} & (y+\tilde{y}) v/2\sqrt{2} \\
    (y-\tilde{y}) v/2\sqrt{2} & -m_{2} & 0 \\
    (y+\tilde{y}) v/2\sqrt{2} & 0 & m_{2}
    \end{pmatrix}, \qquad    M_{c} \equiv
    \begin{pmatrix}
    m_{3} & -yv/\sqrt{2} \\
    -\tilde{y}v/\sqrt{2} & m_{2}
    \end{pmatrix}.
\end{equation}
The Higgs Yukawa couplings are

\begin{equation}
    \mathcal{L}_{\text{Higgs}} = \frac{1}{2} h \chi_n  [J_n^T Y_n J_n]\chi_{n} + h \chi_c^{-}[J_-^T Y_c  J_+]\chi^+_{c} + \textrm{h.c.}
\end{equation}
with
\begin{equation}
    Y_n \equiv \begin{pmatrix}
     0 & -(y- \tilde{y})/2\sqrt{2} & -(y+ \tilde{y})/2\sqrt{2} \\
    -(y- \tilde{y})/2\sqrt{2} & 0 & 0  \\
    -(y+ \tilde{y})/2\sqrt{2} & 0 & 0
    \end{pmatrix}, \qquad    Y_{c} \equiv 
    \begin{pmatrix}
    0 & y/\sqrt{2} \\
    \tilde{y}/\sqrt{2} & 0
    \end{pmatrix}.
\end{equation}
The $Z$ couplings are
\begin{equation}
    \Lag_{Z} = \frac{1}{2}Z^{\mu} \bar\sigma_\mu\chi_{n} [J_n^\dag    U_n J_n] \chi_{n} +  Z^{\mu} \chi^{+}_{c} \bar\sigma_\mu [J_+^\dag U_+ J_+] \chi^{+}_{c} + Z^{\mu} \chi^{-}_{c}  \bar\sigma_\mu [ J_-^\dag U_- J_-] \chi^{-}_{c},
\end{equation}
with
\begin{equation}
\begin{gathered}
    U_n \equiv \begin{pmatrix}
    0 & 0 & 0  \\
    0 & 0  & - \sqrt{g^2 + g^{'2}}     \\
    0 & - \sqrt{g^2 + g^{'2}} & 0
    \end{pmatrix}, \\ U_+  \equiv 
    \begin{pmatrix}
    \frac{g^2}{\sqrt{g^2 + g'^2}} \\
    0 & \frac{(g^2 - g'^2)}{2\sqrt{g^2 + g'^2}} 
    \end{pmatrix}, \quad
    U_-  \equiv 
    \begin{pmatrix}
    -\frac{g^2}{\sqrt{g^2 + g'^2}} &0 \\
    0 & -\frac{(g^2 - g'^2)}{2\sqrt{g^2 + g'^2}}
    \end{pmatrix},
\end{gathered}
\end{equation}
while the $W$ couplings are 

\begin{equation}
    \Lag_{W} = W^{\mu +} \chi_{n}  \bar\sigma_\mu [J_n^\dag  D_- J_-] \chi^{-}_{c} +  W^{\mu -} \chi_{n}\bar\sigma_\mu  [J_n^\dag D_+ J_+] \chi^{+}_{c} + \textrm{h.c.} 
\end{equation}
with
\begin{equation}
    D_- \equiv \begin{pmatrix}
    g & 0 \\
    0 & g/2   \\
    0 & -g/2
    \end{pmatrix}, \qquad   D_+ \equiv 
    \begin{pmatrix}
    -g & 0 \\
    0 & g/2   \\
    0 & g/2
    \end{pmatrix}. 
\end{equation}
The charged fermions also couple to the photon with charge $\pm 1$.

\subsection{Constraints}

We treat most of the constraints in the doublet-triplet model similarly to those in the singlet-doublet model. There are two exceptions that we discuss in more detail: the EDM and collider constraints.

\begin{figure}[t]
    \centering
    \includegraphics[width = 120 mm]{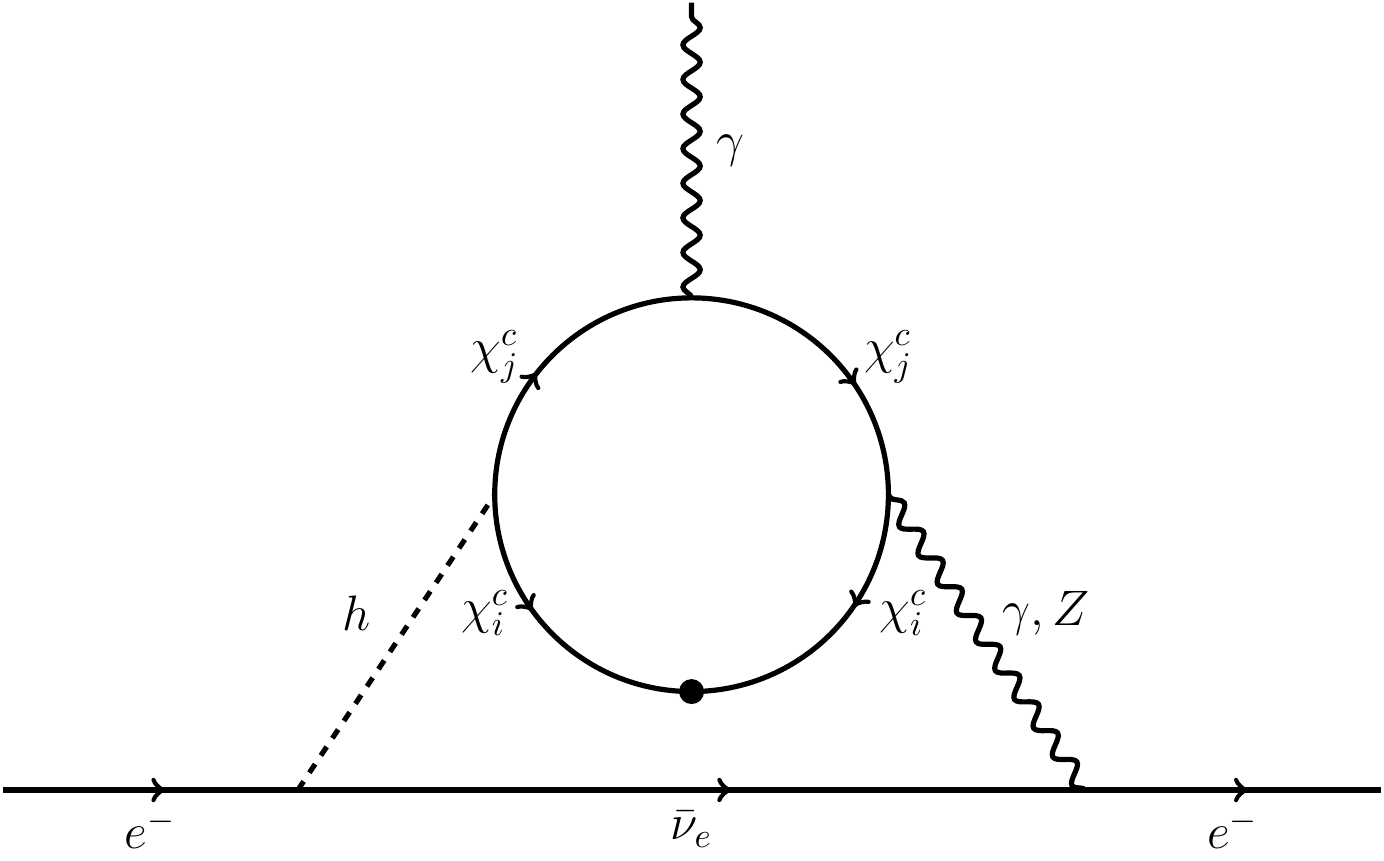}
    \caption{Additional class of Barr-Zee diagrams contributing to the electron EDM. $\chi_{c}$ is the tuple of charged fermions in the mass basis. For the $\gamma h$ diagram, $i = j$, whereas for the $Z h$ diagram, we also have contributions where $i \neq j$.}
\label{fig:BarrZee_Doublet_Triplet}
\end{figure}

The EDM calculation differs from the singlet-doublet case because there are additional diagrams. Like in the singlet-doublet case, the relevant contributing diagrams are all Barr-Zee diagrams \cite{Barr:1990vd}. The diagram with charged $W$ legs, shown in Figure~\ref{fig:BarrZeeW}, that contributed in the singlet-doublet case is still relevant, but for the doublet-triplet model there are two additional relevant Barr-Zee diagrams: $Zh$ and $\gamma h$, shown in Figure \ref{fig:BarrZee_Doublet_Triplet}. There is still no $\gamma Z$ contribution because in that case the same charged fermion runs through the entire loop, leaving no place for $\CP$-violation to enter since the diagonal $Z$ coupling is real. We also neglect the $hh$ diagram since it is suppressed by two factors of the electron Yukawa. We use the general forms of the $Zh$ and $\gamma h$ contributions from \cite{Nakai:2016atk},
\begin{equation}
    d_{e}^{hV} = \frac{1}{16\pi^{2}m_{h}^{2}}\int_{0}^{1}dx\frac{1}{x(1-x)}j\Bigg(\frac{m_{V}^{2}}{m_{h}^{2}}, \frac{\tilde{\Delta}^{V}}{m_{h}^{2}}\Bigg)g_{e}^{V}c_{O}^{V}\frac{m_e}{v},
\end{equation}
where $g_e^V$ is the electron coupling to Z or $\gamma$, $v$ is the Higgs vev, and we define
\begin{equation}
    j(r,s) = \frac{1}{r-s}\Bigg(\frac{r \text{log}r}{r-1} - \frac{s\text{log}s}{s-1}\Bigg).
\end{equation}
$c_{O}^{V}$ and $\tilde{\Delta}^{V}$ are determined by the inner fermion loop which only contains charged fermions for both $\gamma h$ and $\gamma Z$. They are given by
\begin{equation}
\begin{split}
    c_O^{Z} = -\frac{e}{2 \pi^2} \text{Re}\Big( m_c^i x^2 (1 - x)(g_{ij}^Sg_{ji}^{V*} + i g_{ij}^P g_{ji}^{A*}) + (1 - x)^3 m_c^j (g_{ij}^S g_{ji}^{V*} - i g_{ij}^P g_{ji}^{A*})\Big), \\
    \tilde{\Delta}^{Z} = \frac{x m_c^i + (1 - x) m_c^j}{x (1 - x)}, \qquad \quad c_O^{\gamma} = - \frac{e^2 g^P_{jj}}{2 \pi ^2}(1 - x) m_c^j, \qquad \quad \tilde{\Delta}^{\gamma} = \frac{(m_c^{j})^2}{x (1 - x)},
\end{split}
\end{equation}
where
\begin{equation}
    \begin{split}
        g^S &= \frac{1}{2}( J_-^T Y_c J_+ + J_+^\dag Y_c^\dag J_-^*), \qquad g^P = \frac{i}{2}(J_-^T Y_c J_+ - J_+^\dag Y_c^\dag J_-^*), \\
        g^V &= J_-^T U_+ J_-^* + J_+^\dag U_+ J_+,  \qquad \quad g^A =  J_-^T U_+ J_-^* - J_+^\dag U_+ J_+
    \end{split}
\end{equation}
are given in terms of the matrices defined in Section~\ref{sec:doublet_triplet_UV_Model}.
By definition, $\chi_j$ is the fermion which radiates the on-shell external photon, and $g_{ij}^* = (g_{ji})^*$.

A key difference between the singlet-doublet and doublet-triplet cases is that in the latter the mass of the lightest charged fermion is set by similar scales as those that set the mass of the dark matter, and thus generically the lighest charged fermion mass is $\mathcal{O}(100)$ GeV for the doublet-triplet model. This allows us to treat collider constraints differently here than in the singlet-doublet case; we apply generic LEP constraints on charged fermions rather than running the full collider pipeline we considered previously. Specifically, charged fermions lighter than 92.4 GeV are ruled out as long as the mass splitting between the lightest neutral and lightest charged particle is $\geq 100$ MeV \cite{LEPSUSYWG1,LEPSUSYWG2}.\footnote{If the lightest charged state is more than 3 GeV heavier than the lightest neutral state, then there is a stronger bound ruling out charged fermions up to mass 103.5 GeV~\cite{LEPSUSYWG2}. We use the smaller of the two values for simplicity since it is sufficient for our purposes.}

\subsection{Full Exclusion Limits and Discussion}

Unlike in the singlet-doublet case, there is no viable parameter space in this model. In order to show this, we consider three different cases. First, we discuss the case where the magnitude of the couplings is small, for any phase. Then we discuss the case of large coupling and large phase. Finally we discuss the case of large coupling but very small phase.

\begin{figure}
    \centering
     \begin{subfigure}{0.8\linewidth}
    \centering
    \includegraphics[width = \textwidth]{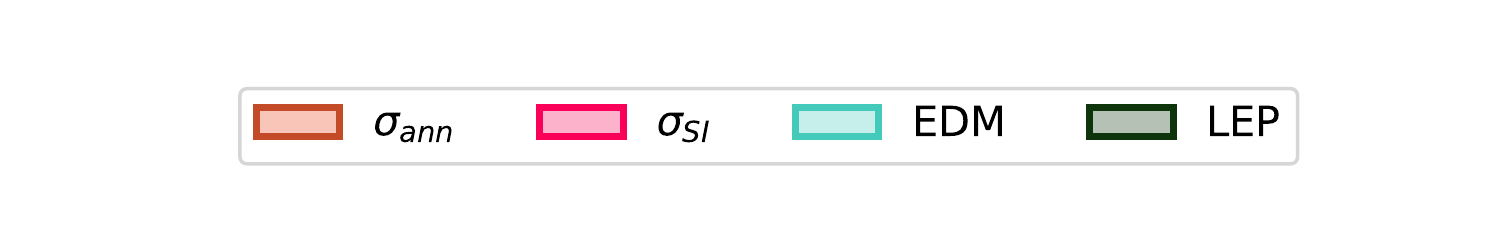} 
    \end{subfigure}
    \begin{subfigure}{0.5\linewidth}
    \centering
    \includegraphics[width = \textwidth]{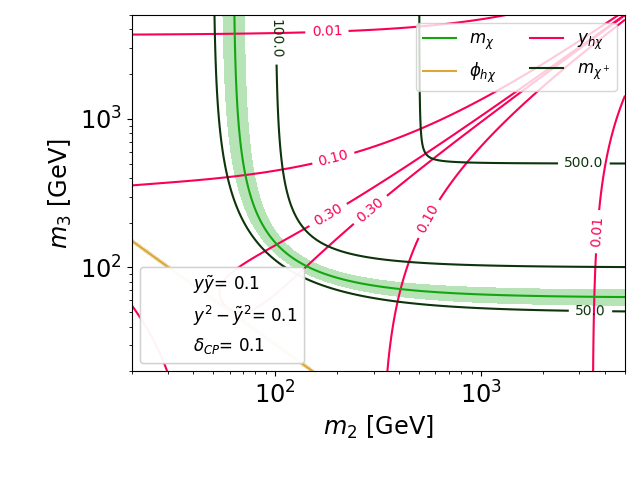} 
    \end{subfigure}%
    \begin{subfigure}{0.5\linewidth}
    \includegraphics[width = \textwidth]{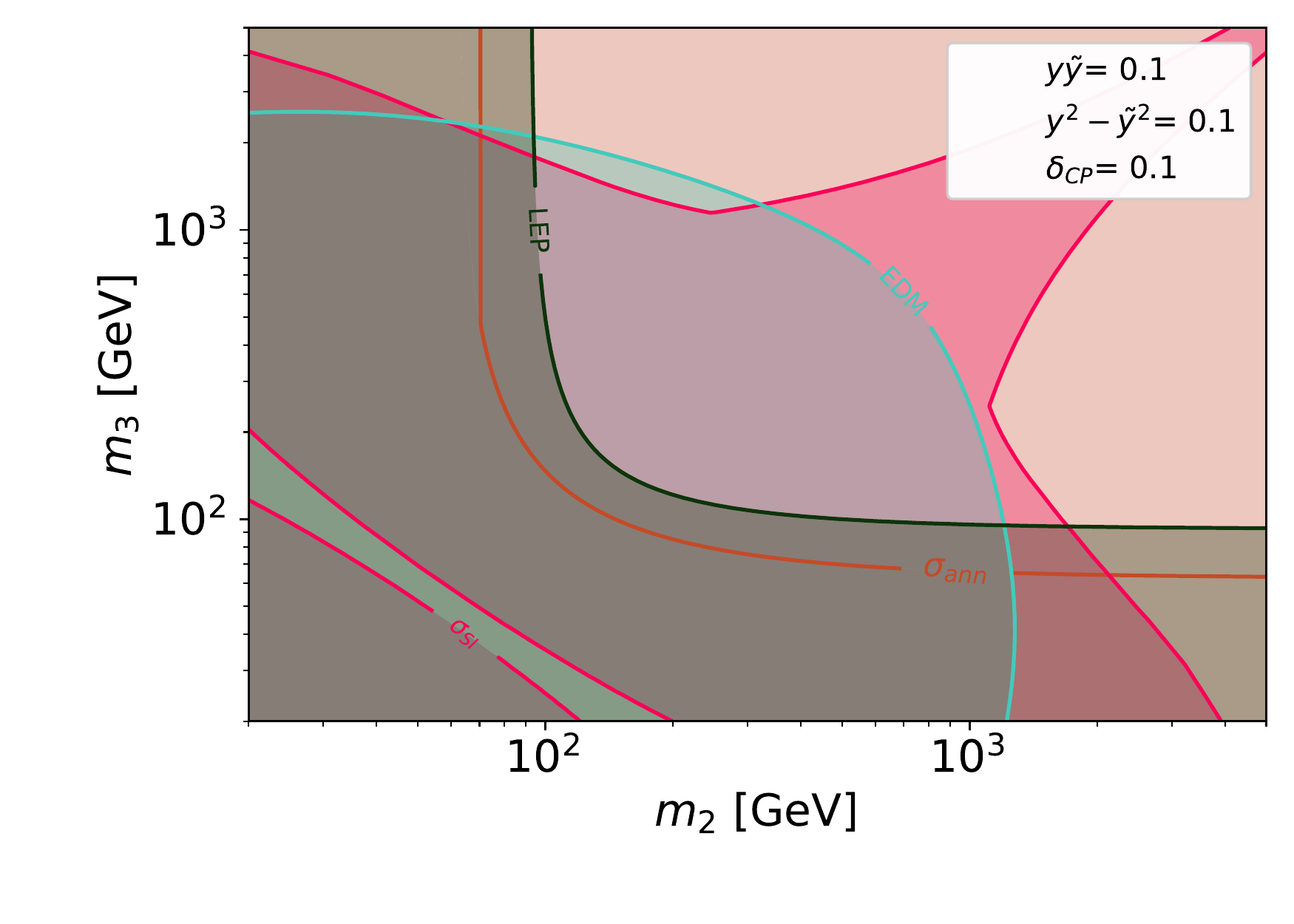}
    \end{subfigure}
    \caption{Example where the magnitude of couplings $y$ and $\tilde{y}$ are small, for different values of $m_2$ and $m_3$. The left plot shows the values of the EFT parameters: dark matter mass, dark matter-Higgs coupling phase, dark matter-Higgs coupling magnitude, and lightest charged fermion mass. Regions around the mass and phase points of interest are shaded: $55 \text{ GeV} \leq m_\chi \leq 70 \text{ GeV}$ and $1.3 \leq \phi_{h\chi} \leq 1.85$. The right shows the annihilation signal and a subset of relevant constraints including EDM~\cite{Andreev:2018ayy}, spin-independent direct detection~\cite{Aprile:2017iyp,Aprile:2018dbl}, and charged fermion constraints from LEP~\cite{LEPSUSYWG1,LEPSUSYWG2}. The annihilation signal appears as a single brown line because a viable annihilation signal is only achievable in a tuned region of parameter space.}
    \label{fig:Doublet_triplet_region1}
\end{figure}

In the first case, parameter space is entirely ruled out by charged fermion constraints, as we can see from Figure~\ref{fig:Doublet_triplet_region1}. On the left, this figure shows the values of several EFT parameters for fixed $y, \tilde{y}$, and $\delta_{CP}$ and various values of $m_2$ and $m_3$. On the right, we show the annihilation signal and a subset of constraints that are sufficient to rule out this region of parameter space.\footnote{The other constraints from the singlet-doublet case still apply here, but we omit them from these plots for clarity.} From these plots, we can see that since the couplings are small, in order to get a sufficient annihilation signal one of $m_2$ or $m_3$ must be $\gtrsim m_h/2$, with the other UV mass larger. Since the magnitude of the couplings is small while the UV masses are large, in this region there will only be a very small splitting between charged and neutral fermions. Therefore, the parameter space here will be entirely ruled out by charged fermion constraints from LEP. This occurs regardless of phase, though EDM constraints are also strong enough to rule this out for larger phases.

\begin{figure}
    \centering
     \begin{subfigure}{0.8\linewidth}
    \centering
    \includegraphics[width = \textwidth]{Plots/LegendDT.pdf} 
    \end{subfigure}
    \begin{subfigure}{0.5\linewidth}
    \centering
    \includegraphics[width = \textwidth]{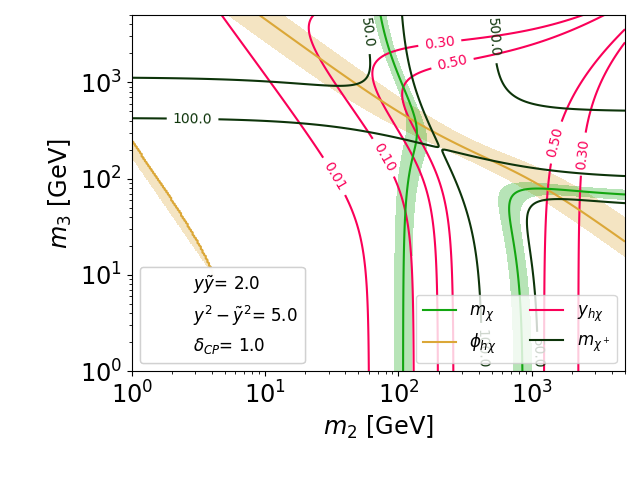} 
    \end{subfigure}%
    \begin{subfigure}{0.5\linewidth}
    \includegraphics[width = \textwidth]{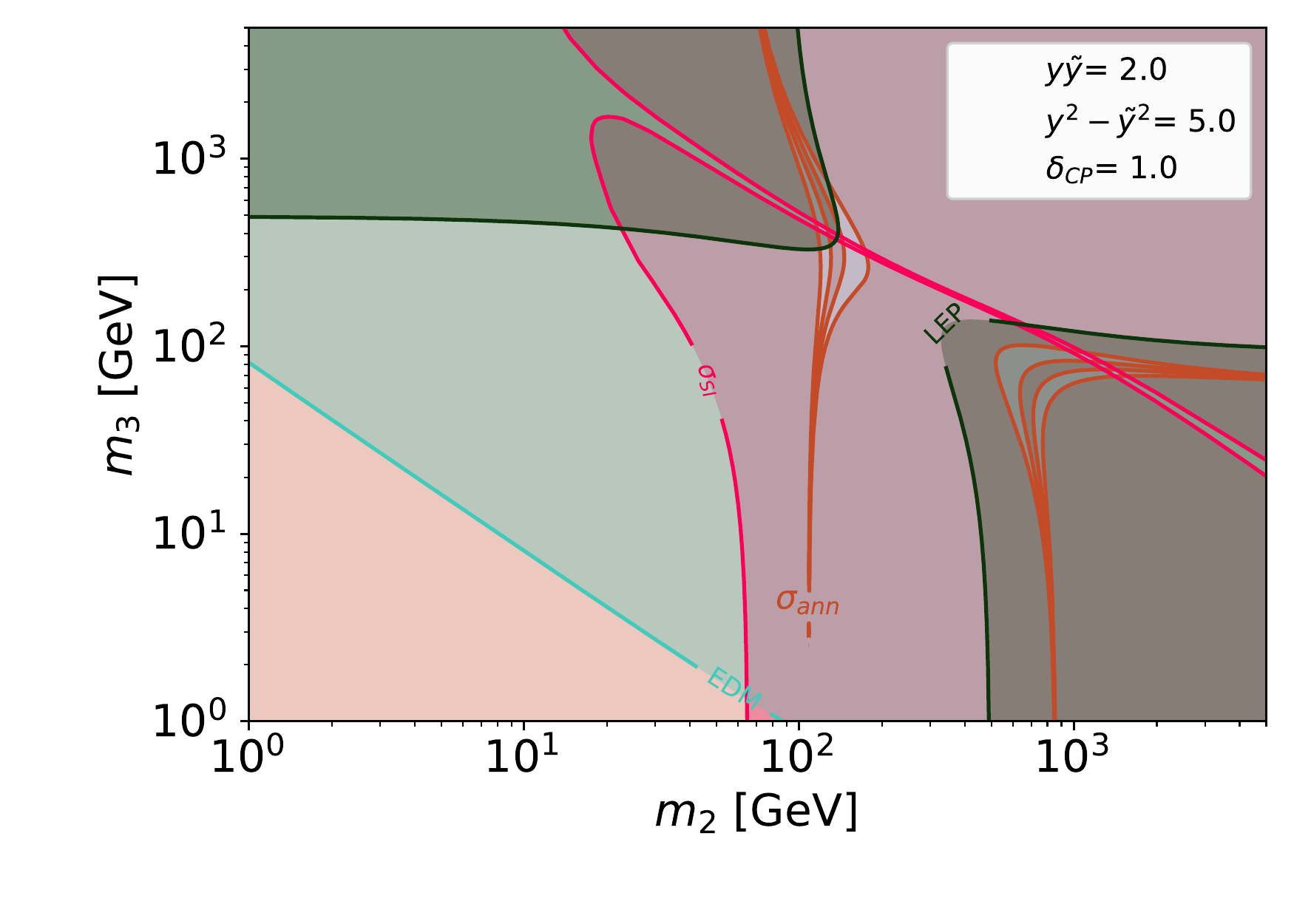}
    \end{subfigure}
    \caption{Here we show plots where the magnitude of couplings $y$ and $\tilde{y}$ are large and $\delta_{CP}$ is also large, for different values of $m_2$ and $m_3$. The left plot shows the values of EFT parameters: dark matter mass, dark matter-Higgs coupling phase, dark matter-Higgs coupling magnitude, and lightest charged fermion mass. Regions around the mass and phase points of interest are shaded: $55 \text{ GeV} \leq m_\chi \leq 70 \text{ GeV}$ and $1.3 \leq \phi_{h\chi} \leq 1.85$. The right shows the annihilation signal and a subset of relevant constraints, and from here we can see that the combination of EDM constraints and spin-independent constraints entirely rule out the parameter space generating a viable annihilation signal.}
    \label{fig:Doublet_triplet_region2}
\end{figure}

In the second case of large coupling and large phase, EDM constraints are typically very strong. The only exceptions are if both $m_2$ and $m_3$ are very large (which can't generate the necessary annihilation signal) or if one of $m_2$ or $m_3$ is very small. This is because in the limit where one of $m_2$ or $m_3$ is exactly zero, the phase becomes unphysical since we can rotate it away. In the limit where $m_2$ is small, the lightest state will have mass even less than $m_2$ and the DM mass won't be in the right mass range to generate the necessary annihilation signal. But in the limit where $m_3$ is small, if the couplings are large enough we can potentially generate the right annihilation signal. However, since the physical phase is small, the EFT phase will also be small, and spin-independent direct detection constraints will always rule out any part of the annihilation signal that isn't constrained by the EDM. This can be seen in Figure~\ref{fig:Doublet_triplet_region2}, which again shows various values of EFT parameters for fixed 
$y, \tilde{y}$, and $\delta_{CP}$ and different $m_2$ and $m_3$ values on the left, and the annihilation signal and a subset of constraints on the right.

\begin{figure}
    \centering
     \begin{subfigure}{0.8\linewidth}
    \centering
    \includegraphics[width = \textwidth]{Plots/LegendDT.pdf} 
    \end{subfigure}
    \begin{subfigure}{0.5\linewidth}
    \centering
    \includegraphics[width = \textwidth]{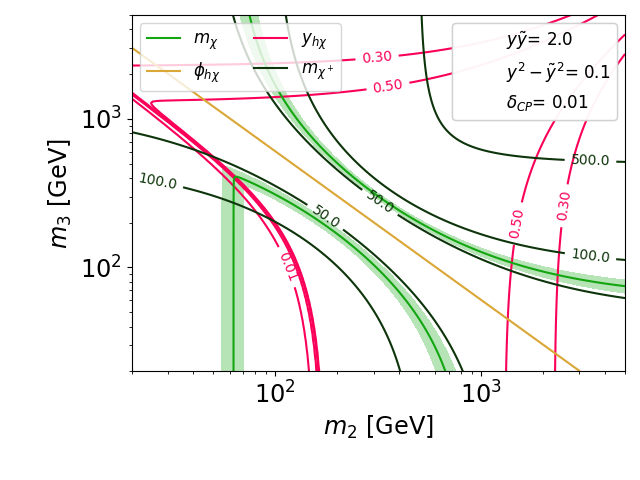} 
    \end{subfigure}%
    \begin{subfigure}{0.5\linewidth}
    \centering
    \includegraphics[width = \textwidth]{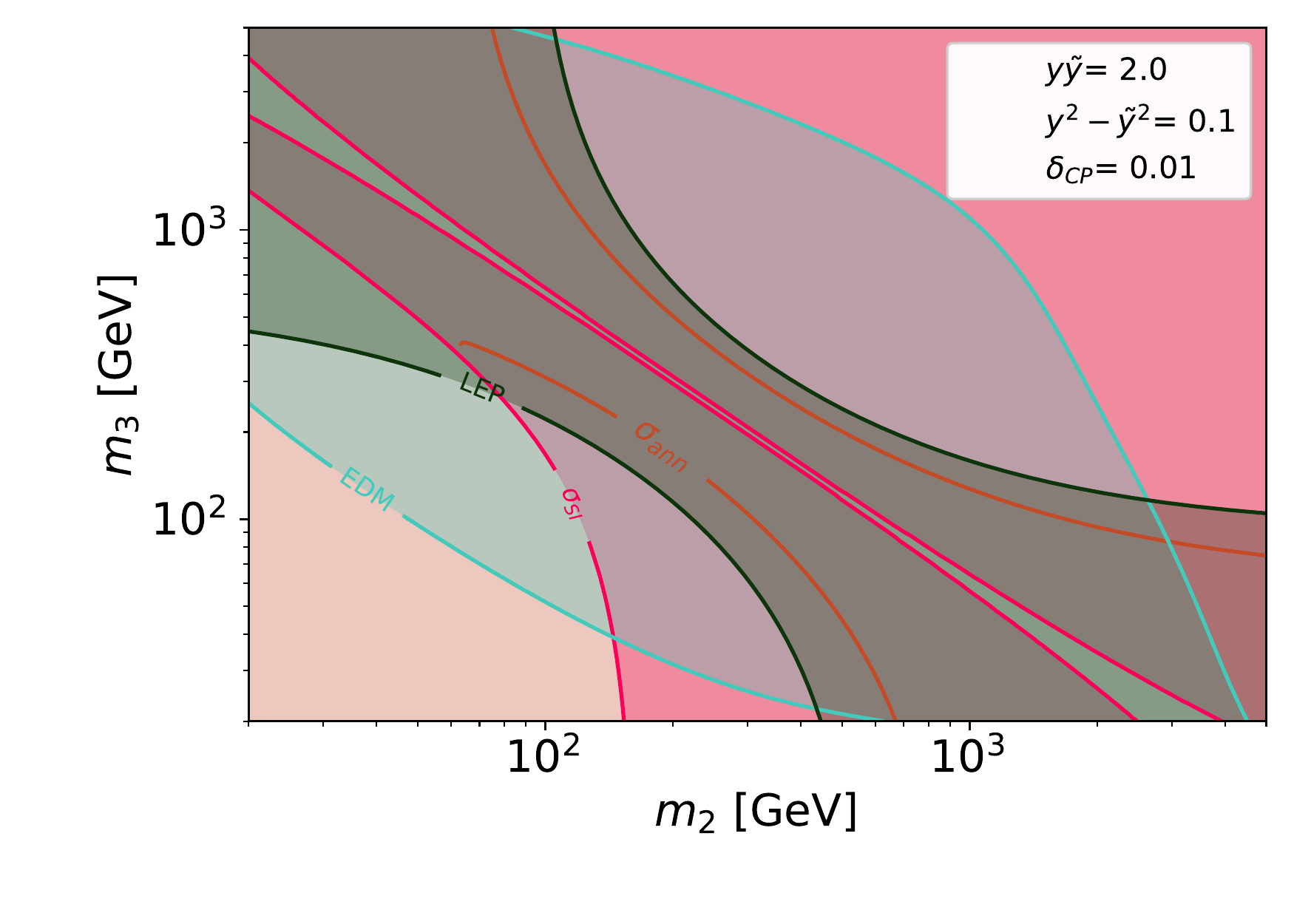} 
    \end{subfigure}
    \begin{subfigure}{0.5\linewidth}
    \centering
    \includegraphics[width = \textwidth]{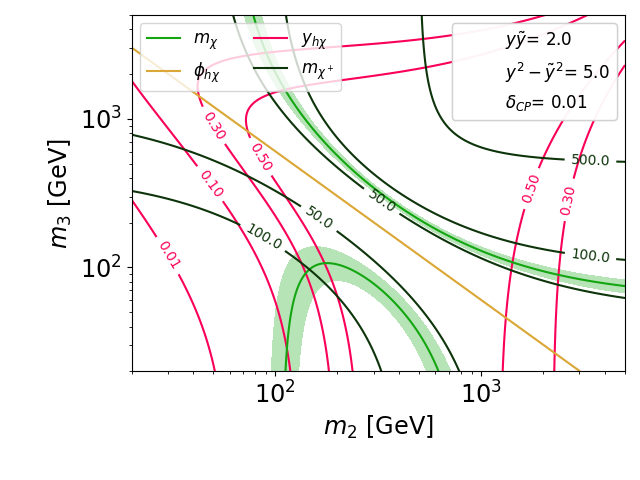} 
    \end{subfigure}%
    \begin{subfigure}{0.5\linewidth}
    \includegraphics[width = \textwidth]{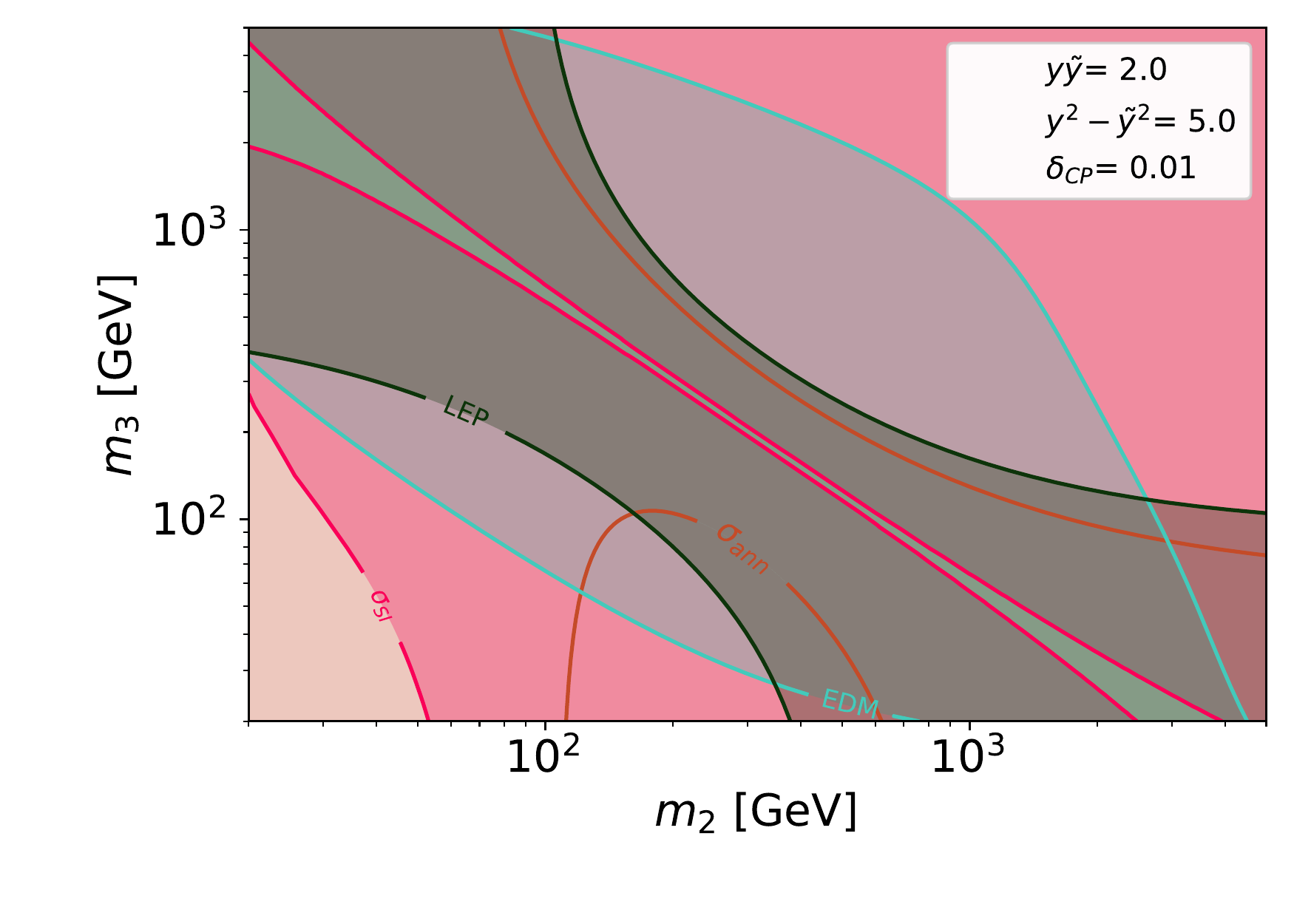}
    \end{subfigure}
    \caption{Two examples where the magnitude of couplings $y$ and $\tilde{y}$ are large and $\delta_{CP}$ is small, for different values of $m_2$ and $m_3$. The top plots show the case where $y$ and $\tilde{y}$ have similar magnitudes; the bottom plots show the case where their magnitudes are very different. As in the other doublet-triplet plots, the left plots show the values of various EFT parameters with shaded regions of interest and the right plots show the annihilation signal and a subset of relevant constraints. The annihilation signal appears as two brown lines on each plot, since the region of allowed masses is so narrow. In both cases, spin-independent constraints rule out the signal. In the case where the couplings are nearly equal, there is also a region where the lightest neutral state decouples, and the dark matter-Higgs coupling is insufficient to generate the annihilation signal despite the dark matter mass being close to $m_h/2$.}
    \label{fig:Doublet_triplet_region3}
\end{figure}

The third case of large magnitude coupling but very small phase is shown in Figure~\ref{fig:Doublet_triplet_region3}. The top plots show the case where $y$ and $\tilde{y}$ are similar in magnitude, while the bottom plots show a large splitting between $y$ and $\tilde{y}$. In both, the EFT coupling is mostly real since the phase is small. There are two different trends depending on the magnitude of the coupling. In both plots, we see regions where the magnitude of the EFT coupling is large, and the annihilation signal is ruled out by spin-independent constraints. In the case of small splitting, we also see a region where the EFT coupling is small (because the lightest state doesn't mix), which is unable to generate the necessary annihilation signal. 

\section{Conclusion}
\label{sec:conclusion}
Given that the GCE is one of the most persistent signals of potential new physics, it is worth cataloging and understanding what could generate it. While there is still substantial debate over the source of the GCE, one promising and well explored possibility is dark matter annihilating to $b\overline{b}$. In this work, we revisit the particular case where dark matter is a Majorana fermion with a $\CP$-violating Higgs coupling, which allows annihilation and spin-independent scattering to be governed by different parameters. Specifically, the leading contribution to annihilation is determined by the imaginary part of the coupling to the Higgs, while spin-independent scattering constraints depend primarily on the real part of the coupling to the Higgs in the mass ranges we are interested in. We study the EFT of this dark matter model for the GCE in detail, and find that while tuning the dark matter mass very close to half the Higgs mass is one potential way to obtain a large enough signal, tuning the phase of the Higgs coupling to make it near imaginary loosens this restriction in the context of the EFT.

We also explore two potential UV completions: singlet-doublet dark matter and doublet-triplet dark matter. In both, the story is more complicated than the EFT because the UV phase and mass are not independent parameters. Although more elaborate supersymmetric realizations of a $\CP$-violating Higgs portal have been discussed in~\cite{Carena:2019pwq}, our goal throughout this paper has been to gain a more detailed qualitative understanding of the mechanism through simpler models. In particular, we have discussed the scaling of the signal and various constraints with the different parameters in the simplified models, as well as quantified how much tuning is necessary to explain the signal without running into constraints. The singlet-doublet dark matter case is particularly interesting because it is a minimal working example of how Majorana dark matter could explain the GCE through the Higgs portal. 

We find that in the minimal singlet-doublet case, there is still viable parameter space when the doublet mass is much larger than the singlet mass. There are two viable regions of  parameter space for the singlet-doublet model. In the case where the UV couplings are small, the tuning of the dark matter mass manifests as a tuning of the singlet mass, but the restriction on both UV and EFT phase is loose. 
When the couplings are larger, the doublet mass is required to be $\gtrsim \mathcal{O}(1)$ TeV. The EFT phase, and often the UV phase as well, must be close to pure imaginary to avoid spin-independent constraints, and the dark matter and singlet masses also must still be relatively close to $m_h/2$ to generate a sufficient annihilation signal (though the allowed region is comparatively much wider). 

Upcoming direct detection and EDM experiments, such as LZ, XENONnT and ACME, will search through significant portions of the remaining parameter space. These two types of probes combine to explore both the limits of minimal and maximal $\CP$-violation, and we expect to definitively rule out doublet masses below the TeV scale in the small coupling case. In the more optimistic case of larger coupling, new experiments will be able to probe doublet masses up to $\mathcal{O}(15)$ TeV or larger for some phases.
In either case, this type of model offers a range of complementary detection avenues that may combine to elucidate the nature of annihilating dark matter. 

In the doublet-triplet case, we do not find any viable parameter space. Spin-independent and EDM constraints restrict the size of the real and imaginary parts of the Higgs coupling, respectively. When the coupling is small in overall magnitude, the annihilation signal requires a dark matter mass near the $m_h/2$ resonance, and the small splitting between the lightest charged and neutral states results in a prohibitively light charged fermion. Hence, the remaining parameter space is ruled out by LEP.

While our results are framed in the context of the GCE, models which include a $\CP$-violating Higgs portal interaction coupling the dark and visible sectors are also compelling for other reasons. These types of interactions could be the key to some of the biggest mysteries of particle physics, including the particle nature of dark matter and also various problems that $\CP$-violation is necessary to solve, such as the matter/antimatter asymmetry of the universe and the strong $\CP$ problem. For example, for some models the addition of a $\CP$ phase around the weak scale could increase the viability of electroweak baryogenesis. While new Higgs boson couplings have the potential to make the hierarchy problem worse, the minimal models we studied can also be realized within the larger framework of SUSY~\cite{Carena:2019pwq} which can ameliorate this issue. These connections could be potential avenues for further exploration, if it turns out that dark matter communicates with the Standard Model through a $\CP$-violating Higgs portal.

\acknowledgments{
We thank Prateek Agrawal and Matthew Reece for useful discussions and feedback on this manuscript. KF is supported by the National Science Foundation Graduate Research Fellowship Program under Grant No. DGE1745303. AP is supported in part by an NSF Graduate Research Fellowship Grant DGE1745303, the DOE Grant DE-SC0013607, and the Alfred P. Sloan Foundation Grant No. G-2019-12504. WLX is supported in part by NSF grants PHY-1620806 and PHY-1915071, the Kavli Foundation grant ``Kavli Dream Team", and
the Moore Foundation Award 8342.}

\appendix

\section{Next-Order Velocity Expansion of the Annihilation Cross Section \label{Section:yhpsi_real}}

As established, to leading order in dark matter velocity, the annihilation signal is set by the pseudoscalar coupling  $\mathrm{Im}[y_{h\chi}]$ (and subdominantly by $g_{Z\chi}$), while spin-independent scattering is set by the scalar coupling $\mathrm{Re}[y_{h\chi}]$. However, we would also like to understand whether we can generate the annihilation signal at all in the limit that $y_{h\chi}$ is real. In this limit, the leading velocity independent term vanishes, and we need to consider terms of higher order in the halo velocity $v$.

For this argument we will neglect the contribution of the $Z$ portal; a $g_{Z\chi}$ consistent with spin-dependent constraints cannot generate a thermal relic annihilation cross section, as it does not have a mass resonance.\footnote{In fact, the $\mathcal{O}(v^2)$ $Z$-coupling term does have a mediator resonance, but enhancement is limited by the significantly larger width of the $Z$ boson.} Thus, for hypothetically viable parameter space it is safe to assume that the $Z$-mediated annihilation is subdominant.   

When $y_{h\chi}$ has vanishing imaginary part, the leading contribution to the spin averaged annihilation amplitude squared is 

\begin{equation}
    |\mathcal{M}|^2_{\chi\chi\to f\bar f} = \frac{4 y_{hf}^2 y_{h \chi}^2 m_\chi^2 (m_\chi^2 - m_f^2)v^2}{ ( m_h^2 -4 m_\chi^2)^2 + m_h^2 \Gamma_h^2} + \mathcal{O}(v^4)
    \label{vel_dep_higgs_ann}.
\end{equation}
This term is suppressed by the non-relativistic speeds of dark matter, for typical values $v^2 \sim 10^{-6}$, and the magnitude of the purely real coupling is stringently constrained by direct detection. Thus, any allowed parameter space would require precise fine-tuning of the dark matter mass. However, the enhancement obtained from the $m_\chi \to m_h/2$ resonance is limited by the finite width of the Higgs, which is $\sim 4$ MeV in the SM~\cite{Sirunyan:2019twz}. Since the branching ratio of $h \to \chi\chi$ near the resonance is vanishingly small due to phase space suppression, we may take 4 MeV as a conservative bound for the Higgs width. Thus, the comparative ratio between annihilation and scattering cross sections, given in Equations~\ref{eqn:Annihilation_Cross_Section} and \ref{eqn:spin_indep_cross_section}, can be bounded by 

\begin{equation}
    \frac{\langle \sigma v\rangle_{\rm ann}}{\sigma_{SI}}\Bigg|_{\mathrm{Im}[y_{h\chi}] =0} \sim   \frac{6\times10^4\;\mathrm{GeV}^2\; m^2_\chi}{(m_h^2 - 4 m_\chi^2)^2 + m_h^2 \Gamma_h^2}  < \frac{10^5\;\mathrm{GeV}^2\; m^2_\chi}{m_h^2 \Gamma_h^2} \approx 10^{9}.
\end{equation}

As the current direct detection limits bound the spin-independent scattering rate at $\leq 10^{-10}$ pb, a model without {\CP}-violation may exhibit an annihilation cross section of at most $\mathcal{O}(0.1)$ pb. We emphasize here that these statements are specifically valid for Majorana fermion dark matter, and dark matter models with a different {\CP}-structure could certainly achieve the required hierarchy between annihilation and scattering with sufficient tuning on this resonance.

\bibliographystyle{JHEP.bst}
\bibliography{main.bib}

\providecommand{\href}[2]{#2}\begingroup\raggedright\begin{thebibliography}{100}

\bibitem{TheFermi-LAT:2015kwa}
{\bf Fermi-LAT} Collaboration, M.~Ajello et~al., {\it {Fermi-LAT Observations
  of High-Energy $\gamma$-Ray Emission Toward the Galactic Center}},  {\em
  Astrophys. J.} {\bf 819} (2016), no.~1 44,
  [\href{http://arxiv.org/abs/1511.02938}{{\tt arXiv:1511.02938}}].

\bibitem{Goodenough:2009gk}
L.~Goodenough and D.~Hooper, {\it {Possible Evidence For Dark Matter
  Annihilation In The Inner Milky Way From The Fermi Gamma Ray Space
  Telescope}},  \href{http://arxiv.org/abs/0910.2998}{{\tt arXiv:0910.2998}}.

\bibitem{Hooper:2010mq}
D.~Hooper and L.~Goodenough, {\it {Dark Matter Annihilation in The Galactic
  Center As Seen by the Fermi Gamma Ray Space Telescope}},  {\em Phys. Lett.}
  {\bf B697} (2011) 412--428, [\href{http://arxiv.org/abs/1010.2752}{{\tt
  arXiv:1010.2752}}].

\bibitem{Hooper:2011ti}
D.~Hooper and T.~Linden, {\it {On The Origin Of The Gamma Rays From The
  Galactic Center}},  {\em Phys. Rev. D} {\bf 84} (2011) 123005,
  [\href{http://arxiv.org/abs/1110.0006}{{\tt arXiv:1110.0006}}].

\bibitem{Gordon:2013vta}
C.~Gordon and O.~Macias, {\it {Dark Matter and Pulsar Model Constraints from
  Galactic Center Fermi-LAT Gamma Ray Observations}},  {\em Phys. Rev. D} {\bf
  88} (2013), no.~8 083521, [\href{http://arxiv.org/abs/1306.5725}{{\tt
  arXiv:1306.5725}}]. [Erratum: Phys.Rev.D 89, 049901 (2014)].

\bibitem{Abazajian:2014fta}
K.~N. Abazajian, N.~Canac, S.~Horiuchi, and M.~Kaplinghat, {\it {Astrophysical
  and Dark Matter Interpretations of Extended Gamma-Ray Emission from the
  Galactic Center}},  {\em Phys. Rev. D} {\bf 90} (2014), no.~2 023526,
  [\href{http://arxiv.org/abs/1402.4090}{{\tt arXiv:1402.4090}}].

\bibitem{Daylan:2014rsa}
T.~Daylan, D.~P. Finkbeiner, D.~Hooper, T.~Linden, S.~K.~N. Portillo, N.~L.
  Rodd, and T.~R. Slatyer, {\it {The characterization of the gamma-ray signal
  from the central Milky Way: A case for annihilating dark matter}},  {\em
  Phys. Dark Univ.} {\bf 12} (2016) 1--23,
  [\href{http://arxiv.org/abs/1402.6703}{{\tt arXiv:1402.6703}}].

\bibitem{Calore:2014xka}
F.~Calore, I.~Cholis, and C.~Weniger, {\it {Background Model Systematics for
  the Fermi GeV Excess}},  {\em JCAP} {\bf 03} (2015) 038,
  [\href{http://arxiv.org/abs/1409.0042}{{\tt arXiv:1409.0042}}].

\bibitem{Aguilar:2016kjl}
{\bf AMS} Collaboration, M.~Aguilar et~al., {\it {Antiproton Flux,
  Antiproton-to-Proton Flux Ratio, and Properties of Elementary Particle Fluxes
  in Primary Cosmic Rays Measured with the Alpha Magnetic Spectrometer on the
  International Space Station}},  {\em Phys. Rev. Lett.} {\bf 117} (2016),
  no.~9 091103.

\bibitem{Cuoco:2017rxb}
A.~Cuoco, J.~Heisig, M.~Korsmeier, and M.~Krämer, {\it {Probing dark matter
  annihilation in the Galaxy with antiprotons and gamma rays}},  {\em JCAP}
  {\bf 10} (2017) 053, [\href{http://arxiv.org/abs/1704.08258}{{\tt
  arXiv:1704.08258}}].

\bibitem{Cuoco:2019kuu}
A.~Cuoco, J.~Heisig, L.~Klamt, M.~Korsmeier, and M.~Krämer, {\it {Scrutinizing
  the evidence for dark matter in cosmic-ray antiprotons}},  {\em Phys. Rev. D}
  {\bf 99} (2019), no.~10 103014, [\href{http://arxiv.org/abs/1903.01472}{{\tt
  arXiv:1903.01472}}].

\bibitem{Cholis:2019ejx}
I.~Cholis, T.~Linden, and D.~Hooper, {\it {A Robust Excess in the Cosmic-Ray
  Antiproton Spectrum: Implications for Annihilating Dark Matter}},  {\em Phys.
  Rev. D} {\bf 99} (2019), no.~10 103026,
  [\href{http://arxiv.org/abs/1903.02549}{{\tt arXiv:1903.02549}}].

\bibitem{Hooper:2019xss}
D.~Hooper, R.~K. Leane, Y.-D. Tsai, S.~Wegsman, and S.~J. Witte, {\it {A
  Systematic Study of Hidden Sector Dark Matter: Application to the Gamma-Ray
  and Antiproton Excesses}},  \href{http://arxiv.org/abs/1912.08821}{{\tt
  arXiv:1912.08821}}.

\bibitem{Boudaud:2019efq}
M.~Boudaud, Y.~Génolini, L.~Derome, J.~Lavalle, D.~Maurin, P.~Salati, and
  P.~D. Serpico, {\it {AMS-02 antiprotons are consistent with a secondary
  astrophysical origin}},  {\em Phys. Rev. Res.} {\bf 2} (2020) 023022,
  [\href{http://arxiv.org/abs/1906.07119}{{\tt arXiv:1906.07119}}].

\bibitem{Heisig:2020nse}
J.~Heisig, M.~Korsmeier, and M.~W. Winkler, {\it {Dark matter or correlated
  errors? Systematics of the AMS-02 antiproton excess}},
  \href{http://arxiv.org/abs/2005.04237}{{\tt arXiv:2005.04237}}.

\bibitem{Cholis:2014noa}
I.~Cholis, D.~Hooper, and T.~Linden, {\it {A New Determination of the Spectra
  and Luminosity Function of Gamma-Ray Millisecond Pulsars}},
  \href{http://arxiv.org/abs/1407.5583}{{\tt arXiv:1407.5583}}.

\bibitem{Cholis:2014lta}
I.~Cholis, D.~Hooper, and T.~Linden, {\it {Challenges in Explaining the
  Galactic Center Gamma-Ray Excess with Millisecond Pulsars}},  {\em JCAP} {\bf
  06} (2015) 043, [\href{http://arxiv.org/abs/1407.5625}{{\tt
  arXiv:1407.5625}}].

\bibitem{Lee:2014mza}
S.~K. Lee, M.~Lisanti, and B.~R. Safdi, {\it {Distinguishing Dark Matter from
  Unresolved Point Sources in the Inner Galaxy with Photon Statistics}},  {\em
  JCAP} {\bf 05} (2015) 056, [\href{http://arxiv.org/abs/1412.6099}{{\tt
  arXiv:1412.6099}}].

\bibitem{Bartels:2015aea}
R.~Bartels, S.~Krishnamurthy, and C.~Weniger, {\it {Strong support for the
  millisecond pulsar origin of the Galactic center GeV excess}},  {\em Phys.
  Rev. Lett.} {\bf 116} (2016), no.~5 051102,
  [\href{http://arxiv.org/abs/1506.05104}{{\tt arXiv:1506.05104}}].

\bibitem{Lee:2015fea}
S.~K. Lee, M.~Lisanti, B.~R. Safdi, T.~R. Slatyer, and W.~Xue, {\it {Evidence
  for Unresolved $\gamma$-Ray Point Sources in the Inner Galaxy}},  {\em Phys.
  Rev. Lett.} {\bf 116} (2016), no.~5 051103,
  [\href{http://arxiv.org/abs/1506.05124}{{\tt arXiv:1506.05124}}].

\bibitem{Macias:2016nev}
O.~Macias, C.~Gordon, R.~M. Crocker, B.~Coleman, D.~Paterson, S.~Horiuchi, and
  M.~Pohl, {\it {Galactic bulge preferred over dark matter for the Galactic
  centre gamma-ray excess}},  {\em Nature Astron.} {\bf 2} (2018), no.~5
  387--392, [\href{http://arxiv.org/abs/1611.06644}{{\tt arXiv:1611.06644}}].

\bibitem{Haggard:2017lyq}
D.~Haggard, C.~Heinke, D.~Hooper, and T.~Linden, {\it {Low Mass X-Ray Binaries
  in the Inner Galaxy: Implications for Millisecond Pulsars and the GeV
  Excess}},  {\em JCAP} {\bf 05} (2017) 056,
  [\href{http://arxiv.org/abs/1701.02726}{{\tt arXiv:1701.02726}}].

\bibitem{Bartels:2017vsx}
R.~Bartels, E.~Storm, C.~Weniger, and F.~Calore, {\it {The Fermi-LAT GeV excess
  as a tracer of stellar mass in the Galactic bulge}},  {\em Nature Astron.}
  {\bf 2} (2018), no.~10 819--828, [\href{http://arxiv.org/abs/1711.04778}{{\tt
  arXiv:1711.04778}}].

\bibitem{Macias:2019omb}
O.~Macias, S.~Horiuchi, M.~Kaplinghat, C.~Gordon, R.~M. Crocker, and D.~M.
  Nataf, {\it {Strong Evidence that the Galactic Bulge is Shining in Gamma
  Rays}},  {\em JCAP} {\bf 09} (2019) 042,
  [\href{http://arxiv.org/abs/1901.03822}{{\tt arXiv:1901.03822}}].

\bibitem{Leane:2019xiy}
R.~K. Leane and T.~R. Slatyer, {\it {Revival of the Dark Matter Hypothesis for
  the Galactic Center Gamma-Ray Excess}},  {\em Phys. Rev. Lett.} {\bf 123}
  (2019), no.~24 241101, [\href{http://arxiv.org/abs/1904.08430}{{\tt
  arXiv:1904.08430}}].

\bibitem{Zhong:2019ycb}
Y.-M. Zhong, S.~D. McDermott, I.~Cholis, and P.~J. Fox, {\it {A New Mask for An
  Old Suspect: Testing the Sensitivity of the Galactic Center Excess to the
  Point Source Mask}},  {\em Phys. Rev. Lett.} {\bf 124} (2020), no.~23 231103,
  [\href{http://arxiv.org/abs/1911.12369}{{\tt arXiv:1911.12369}}].

\bibitem{Leane:2020nmi}
R.~K. Leane and T.~R. Slatyer, {\it {Spurious Point Source Signals in the
  Galactic Center Excess}},  \href{http://arxiv.org/abs/2002.12370}{{\tt
  arXiv:2002.12370}}.

\bibitem{Leane:2020pfc}
R.~K. Leane and T.~R. Slatyer, {\it {The Enigmatic Galactic Center Excess:
  Spurious Point Sources and Signal Mismodeling}},
  \href{http://arxiv.org/abs/2002.12371}{{\tt arXiv:2002.12371}}.

\bibitem{Buschmann:2020adf}
M.~Buschmann, N.~L. Rodd, B.~R. Safdi, L.~J. Chang, S.~Mishra-Sharma,
  M.~Lisanti, and O.~Macias, {\it {Foreground Mismodeling and the Point Source
  Explanation of the Fermi Galactic Center Excess}},
  \href{http://arxiv.org/abs/2002.12373}{{\tt arXiv:2002.12373}}.

\bibitem{Abazajian:2020tww}
K.~N. Abazajian, S.~Horiuchi, M.~Kaplinghat, R.~E. Keeley, and O.~Macias, {\it
  {Strong constraints on thermal relic dark matter from Fermi-LAT observations
  of the Galactic Center}},  \href{http://arxiv.org/abs/2003.10416}{{\tt
  arXiv:2003.10416}}.

\bibitem{List:2020mzd}
F.~List, N.~L. Rodd, G.~F. Lewis, and I.~Bhat, {\it {The GCE in a New Light:
  Disentangling the $\gamma$-ray Sky with Bayesian Graph Convolutional Neural
  Networks}},  \href{http://arxiv.org/abs/2006.12504}{{\tt arXiv:2006.12504}}.

\bibitem{Mishra-Sharma:2020kjb}
S.~Mishra-Sharma and K.~Cranmer, {\it {Semi-parametric $\gamma$-ray modeling
  with Gaussian processes and variational inference}},
  \href{http://arxiv.org/abs/2010.10450}{{\tt arXiv:2010.10450}}.

\bibitem{Karwin:2016tsw}
C.~Karwin, S.~Murgia, T.~M.~P. Tait, T.~A. Porter, and P.~Tanedo, {\it {Dark
  Matter Interpretation of the Fermi-LAT Observation Toward the Galactic
  Center}},  {\em Phys. Rev. D} {\bf 95} (2017), no.~10 103005,
  [\href{http://arxiv.org/abs/1612.05687}{{\tt arXiv:1612.05687}}].

\bibitem{Arcadi:2017kky}
G.~Arcadi, M.~Dutra, P.~Ghosh, M.~Lindner, Y.~Mambrini, M.~Pierre, S.~Profumo,
  and F.~S. Queiroz, {\it {The waning of the WIMP? A review of models,
  searches, and constraints}},  {\em Eur. Phys. J. C} {\bf 78} (2018), no.~3
  203, [\href{http://arxiv.org/abs/1703.07364}{{\tt arXiv:1703.07364}}].

\bibitem{Huang:2013apa}
W.-C. Huang, A.~Urbano, and W.~Xue, {\it {Fermi Bubbles under Dark Matter
  Scrutiny Part II: Particle Physics Analysis}},  {\em JCAP} {\bf 04} (2014)
  020, [\href{http://arxiv.org/abs/1310.7609}{{\tt arXiv:1310.7609}}].

\bibitem{Boehm:2014hva}
C.~Boehm, M.~J. Dolan, C.~McCabe, M.~Spannowsky, and C.~J. Wallace, {\it
  {Extended gamma-ray emission from Coy Dark Matter}},  {\em JCAP} {\bf 05}
  (2014) 009, [\href{http://arxiv.org/abs/1401.6458}{{\tt arXiv:1401.6458}}].

\bibitem{Cheung:2014lqa}
C.~Cheung, M.~Papucci, D.~Sanford, N.~R. Shah, and K.~M. Zurek, {\it {NMSSM
  Interpretation of the Galactic Center Excess}},  {\em Phys. Rev.} {\bf D90}
  (2014), no.~7 075011, [\href{http://arxiv.org/abs/1406.6372}{{\tt
  arXiv:1406.6372}}].

\bibitem{Guo:2014gra}
J.~Guo, J.~Li, T.~Li, and A.~G. Williams, {\it {NMSSM explanations of the
  Galactic center gamma ray excess and promising LHC searches}},  {\em Phys.
  Rev. D} {\bf 91} (2015), no.~9 095003,
  [\href{http://arxiv.org/abs/1409.7864}{{\tt arXiv:1409.7864}}].

\bibitem{Cao:2014efa}
J.~Cao, L.~Shang, P.~Wu, J.~M. Yang, and Y.~Zhang, {\it {Supersymmetry
  explanation of the Fermi Galactic Center excess and its test at LHC run II}},
   {\em Phys. Rev. D} {\bf 91} (2015), no.~5 055005,
  [\href{http://arxiv.org/abs/1410.3239}{{\tt arXiv:1410.3239}}].

\bibitem{Berlin:2015wwa}
A.~Berlin, S.~Gori, T.~Lin, and L.-T. Wang, {\it {Pseudoscalar Portal Dark
  Matter}},  {\em Phys. Rev. D} {\bf 92} (2015) 015005,
  [\href{http://arxiv.org/abs/1502.06000}{{\tt arXiv:1502.06000}}].

\bibitem{Gherghetta:2015ysa}
T.~Gherghetta, B.~von Harling, A.~D. Medina, M.~A. Schmidt, and T.~Trott, {\it
  {SUSY implications from WIMP annihilation into scalars at the Galactic
  Center}},  {\em Phys. Rev.} {\bf D91} (2015) 105004,
  [\href{http://arxiv.org/abs/1502.07173}{{\tt arXiv:1502.07173}}].

\bibitem{Duerr:2015bea}
M.~Duerr, P.~Fileviez~P\'erez, and J.~Smirnov, {\it {Gamma-Ray Excess and the
  Minimal Dark Matter Model}},  {\em JHEP} {\bf 06} (2016) 008,
  [\href{http://arxiv.org/abs/1510.07562}{{\tt arXiv:1510.07562}}].

\bibitem{Carena:2019pwq}
M.~Carena, J.~Osborne, N.~R. Shah, and C.~E.~M. Wagner, {\it {Return of the
  WIMP: Missing energy signals and the Galactic Center excess}},  {\em Phys.
  Rev.} {\bf D100} (2019), no.~5 055002,
  [\href{http://arxiv.org/abs/1905.03768}{{\tt arXiv:1905.03768}}].

\bibitem{Mahbubani:2005pt}
R.~Mahbubani and L.~Senatore, {\it {The Minimal model for dark matter and
  unification}},  {\em Phys. Rev. D} {\bf 73} (2006) 043510,
  [\href{http://arxiv.org/abs/hep-ph/0510064}{{\tt hep-ph/0510064}}].

\bibitem{DEramo:2007anh}
F.~D'Eramo, {\it {Dark matter and Higgs boson physics}},  {\em Phys. Rev. D}
  {\bf 76} (2007) 083522, [\href{http://arxiv.org/abs/0705.4493}{{\tt
  arXiv:0705.4493}}].

\bibitem{Enberg_2007}
R.~Enberg, P.~Fox, L.~Hall, A.~Papaioannou, and M.~Papucci, {\it Lhc and dark
  matter signals of improved naturalness},  {\em Journal of High Energy
  Physics} {\bf 2007} (Nov, 2007) 014–014.

\bibitem{Cohen:2011ec}
T.~Cohen, J.~Kearney, A.~Pierce, and D.~Tucker-Smith, {\it {Singlet-Doublet
  Dark Matter}},  {\em Phys. Rev. D} {\bf 85} (2012) 075003,
  [\href{http://arxiv.org/abs/1109.2604}{{\tt arXiv:1109.2604}}].

\bibitem{Cheung:2013dua}
C.~Cheung and D.~Sanford, {\it {Simplified Models of Mixed Dark Matter}},  {\em
  JCAP} {\bf 02} (2014) 011, [\href{http://arxiv.org/abs/1311.5896}{{\tt
  arXiv:1311.5896}}].

\bibitem{Abe:2014gua}
T.~Abe, R.~Kitano, and R.~Sato, {\it {Discrimination of dark matter models in
  future experiments}},  {\em Phys. Rev. D} {\bf 91} (2015), no.~9 095004,
  [\href{http://arxiv.org/abs/1411.1335}{{\tt arXiv:1411.1335}}]. [Erratum:
  Phys.Rev.D 96, 019902 (2017)].

\bibitem{Calibbi:2015nha}
L.~Calibbi, A.~Mariotti, and P.~Tziveloglou, {\it {Singlet-Doublet Model: Dark
  matter searches and LHC constraints}},  {\em JHEP} {\bf 10} (2015) 116,
  [\href{http://arxiv.org/abs/1505.03867}{{\tt arXiv:1505.03867}}].

\bibitem{Freitas:2015hsa}
A.~Freitas, S.~Westhoff, and J.~Zupan, {\it {Integrating in the Higgs Portal to
  Fermion Dark Matter}},  {\em JHEP} {\bf 09} (2015) 015,
  [\href{http://arxiv.org/abs/1506.04149}{{\tt arXiv:1506.04149}}].

\bibitem{Banerjee:2016hsk}
S.~Banerjee, S.~Matsumoto, K.~Mukaida, and Y.-L.~S. Tsai, {\it {WIMP Dark
  Matter in a Well-Tempered Regime: A case study on Singlet-Doublets Fermionic
  WIMP}},  {\em JHEP} {\bf 11} (2016) 070,
  [\href{http://arxiv.org/abs/1603.07387}{{\tt arXiv:1603.07387}}].

\bibitem{Cai_2017}
C.~Cai, Z.-H. Yu, and H.-H. Zhang, {\it Cepc precision of electroweak oblique
  parameters and weakly interacting dark matter: The fermionic case},  {\em
  Nuclear Physics B} {\bf 921} (Aug, 2017) 181–210.

\bibitem{Lopez-Honorez:2017ora}
L.~Lopez~Honorez, M.~H.~G. Tytgat, P.~Tziveloglou, and B.~Zaldivar, {\it {On
  Minimal Dark Matter coupled to the Higgs}},  {\em JHEP} {\bf 04} (2018) 011,
  [\href{http://arxiv.org/abs/1711.08619}{{\tt arXiv:1711.08619}}].

\bibitem{Dedes:2014hga}
A.~Dedes and D.~Karamitros, {\it {Doublet-Triplet Fermionic Dark Matter}},
  {\em Phys. Rev. D} {\bf 89} (2014), no.~11 115002,
  [\href{http://arxiv.org/abs/1403.7744}{{\tt arXiv:1403.7744}}].

\bibitem{Zeldovich:1965gev}
Y.~Zeldovich, {\em {Survey of Modern Cosmology}}, vol.~3, pp.~241--379.
\newblock 1965.

\bibitem{Chiu:1966kg}
H.-Y. Chiu, {\it {Symmetry between particle and anti-particle populations in
  the universe}},  {\em Phys. Rev. Lett.} {\bf 17} (1966) 712.

\bibitem{Lee:1977ua}
B.~W. Lee and S.~Weinberg, {\it {Cosmological Lower Bound on Heavy Neutrino
  Masses}},  {\em Phys. Rev. Lett.} {\bf 39} (1977) 165--168.

\bibitem{Hut:1977zn}
P.~Hut, {\it {Limits on Masses and Number of Neutral Weakly Interacting
  Particles}},  {\em Phys. Lett. B} {\bf 69} (1977) 85.

\bibitem{Wolfram:1978gp}
S.~Wolfram, {\it {Abundances of Stable Particles Produced in the Early
  Universe}},  {\em Phys. Lett. B} {\bf 82} (1979) 65--68.

\bibitem{Steigman:1979kw}
G.~Steigman, {\it {Cosmology Confronts Particle Physics}},  {\em Ann. Rev.
  Nucl. Part. Sci.} {\bf 29} (1979) 313--338.

\bibitem{Scherrer:1985zt}
R.~J. Scherrer and M.~S. Turner, {\it {On the Relic, Cosmic Abundance of Stable
  Weakly Interacting Massive Particles}},  {\em Phys. Rev. D} {\bf 33} (1986)
  1585. [Erratum: Phys.Rev.D 34, 3263 (1986)].

\bibitem{Bertstein:1985}
J.~Bernstein, L.~S. Brown, and G.~Feinberg, {\it Cosmological heavy-neutrino
  problem},  {\em Phys. Rev. D} {\bf 32} (Dec, 1985) 3261--3267.

\bibitem{Srednicki:1988ce}
M.~Srednicki, R.~Watkins, and K.~A. Olive, {\it {Calculations of Relic
  Densities in the Early Universe}},  {\em Nucl. Phys. B} {\bf 310} (1988) 693.

\bibitem{Griest:1990kh}
K.~Griest and D.~Seckel, {\it {Three exceptions in the calculation of relic
  abundances}},  {\em Phys. Rev. D} {\bf 43} (1991) 3191--3203.

\bibitem{Gondolo:1990dk}
P.~Gondolo and G.~Gelmini, {\it {Cosmic abundances of stable particles:
  Improved analysis}},  {\em Nucl. Phys. B} {\bf 360} (1991) 145--179.

\bibitem{Steigman:2012nb}
G.~Steigman, B.~Dasgupta, and J.~F. Beacom, {\it {Precise Relic WIMP Abundance
  and its Impact on Searches for Dark Matter Annihilation}},  {\em Phys. Rev.
  D} {\bf 86} (2012) 023506, [\href{http://arxiv.org/abs/1204.3622}{{\tt
  arXiv:1204.3622}}].

\bibitem{Binder:2017rgn}
T.~Binder, T.~Bringmann, M.~Gustafsson, and A.~Hryczuk, {\it {Early kinetic
  decoupling of dark matter: when the standard way of calculating the thermal
  relic density fails}},  {\em Phys. Rev. D} {\bf 96} (2017), no.~11 115010,
  [\href{http://arxiv.org/abs/1706.07433}{{\tt arXiv:1706.07433}}]. [Erratum:
  Phys.Rev.D 101, 099901 (2020)].

\bibitem{Abe:2020obo}
T.~Abe, {\it {The effect of the early kinetic decoupling in a fermionic dark
  matter model}},  \href{http://arxiv.org/abs/2004.10041}{{\tt
  arXiv:2004.10041}}.

\bibitem{DiMauro:2021raz}
M.~Di~Mauro, {\it {The characteristics of the Galactic center excess measured
  with 11 years of Fermi-LAT data}},
  \href{http://arxiv.org/abs/2101.04694}{{\tt arXiv:2101.04694}}.

\bibitem{Lin:2019uvt}
T.~Lin, {\it {Dark matter models and direct detection}},  {\em PoS} {\bf 333}
  (2019) 009, [\href{http://arxiv.org/abs/1904.07915}{{\tt arXiv:1904.07915}}].

\bibitem{DelNobile:2013sia}
M.~Cirelli, E.~Del~Nobile, and P.~Panci, {\it {Tools for model-independent
  bounds in direct dark matter searches}},  {\em JCAP} {\bf 10} (2013) 019,
  [\href{http://arxiv.org/abs/1307.5955}{{\tt arXiv:1307.5955}}].

\bibitem{Hill:2014yxa}
R.~J. Hill and M.~P. Solon, {\it {Standard Model anatomy of WIMP dark matter
  direct detection II: QCD analysis and hadronic matrix elements}},  {\em Phys.
  Rev. D} {\bf 91} (2015) 043505, [\href{http://arxiv.org/abs/1409.8290}{{\tt
  arXiv:1409.8290}}].

\bibitem{Bishara:2017pfq}
F.~Bishara, J.~Brod, B.~Grinstein, and J.~Zupan, {\it {From quarks to nucleons
  in dark matter direct detection}},  {\em JHEP} {\bf 11} (2017) 059,
  [\href{http://arxiv.org/abs/1707.06998}{{\tt arXiv:1707.06998}}].

\bibitem{Ellis:2018dmb}
J.~Ellis, N.~Nagata, and K.~A. Olive, {\it {Uncertainties in WIMP Dark Matter
  Scattering Revisited}},  {\em Eur. Phys. J. C} {\bf 78} (2018), no.~7 569,
  [\href{http://arxiv.org/abs/1805.09795}{{\tt arXiv:1805.09795}}].

\bibitem{Aprile:2017iyp}
{\bf XENON} Collaboration, E.~Aprile et~al., {\it {First Dark Matter Search
  Results from the XENON1T Experiment}},  {\em Phys. Rev. Lett.} {\bf 119}
  (2017), no.~18 181301, [\href{http://arxiv.org/abs/1705.06655}{{\tt
  arXiv:1705.06655}}].

\bibitem{Aprile:2018dbl}
{\bf XENON} Collaboration, E.~Aprile et~al., {\it {Dark Matter Search Results
  from a One Ton-Year Exposure of XENON1T}},  {\em Phys. Rev. Lett.} {\bf 121}
  (2018), no.~11 111302, [\href{http://arxiv.org/abs/1805.12562}{{\tt
  arXiv:1805.12562}}].

\bibitem{Akerib:2018lyp}
{\bf LUX-ZEPLIN} Collaboration, D.~Akerib et~al., {\it {Projected WIMP
  sensitivity of the LUX-ZEPLIN dark matter experiment}},  {\em Phys. Rev. D}
  {\bf 101} (2020), no.~5 052002, [\href{http://arxiv.org/abs/1802.06039}{{\tt
  arXiv:1802.06039}}].

\bibitem{Aprile:2019dbj}
{\bf XENON} Collaboration, E.~Aprile et~al., {\it {Constraining the
  spin-dependent WIMP-nucleon cross sections with XENON1T}},  {\em Phys. Rev.
  Lett.} {\bf 122} (2019), no.~14 141301,
  [\href{http://arxiv.org/abs/1902.03234}{{\tt arXiv:1902.03234}}].

\bibitem{PhysRevLett.118.251301}
{\bf PICO Collaboration} Collaboration, C.~Amole et~al., {\it Dark matter
  search results from the $\mathrm{PICO}\text{\ensuremath{-}}60{\text{
  }\mathrm{c}}_{3}{\mathrm{f}}_{8}$ bubble chamber},  {\em Phys. Rev. Lett.}
  {\bf 118} (Jun, 2017) 251301.

\bibitem{PhysRevD.100.022001}
{\bf PICO Collaboration} Collaboration, C.~Amole et~al., {\it Dark matter
  search results from the complete exposure of the pico-60
  ${\mathrm{c}}_{3}{\mathrm{f}}_{8}$ bubble chamber},  {\em Phys. Rev. D} {\bf
  100} (Jul, 2019) 022001.

\bibitem{Aprile:2020vtw}
{\bf XENON} Collaboration, E.~Aprile et~al., {\it {Projected WIMP Sensitivity
  of the XENONnT Dark Matter Experiment}},
  \href{http://arxiv.org/abs/2007.08796}{{\tt arXiv:2007.08796}}.

\bibitem{Aartsen:2016zhm}
{\bf IceCube} Collaboration, M.~Aartsen et~al., {\it {Search for annihilating
  dark matter in the Sun with 3 years of IceCube data}},  {\em Eur. Phys. J. C}
  {\bf 77} (2017), no.~3 146, [\href{http://arxiv.org/abs/1612.05949}{{\tt
  arXiv:1612.05949}}]. [Erratum: Eur.Phys.J.C 79, 214 (2019)].

\bibitem{Akerib:2016lao}
{\bf LUX} Collaboration, D.~Akerib et~al., {\it {Results on the Spin-Dependent
  Scattering of Weakly Interacting Massive Particles on Nucleons from the Run 3
  Data of the LUX Experiment}},  {\em Phys. Rev. Lett.} {\bf 116} (2016),
  no.~16 161302, [\href{http://arxiv.org/abs/1602.03489}{{\tt
  arXiv:1602.03489}}].

\bibitem{Akerib:2016vxi}
{\bf LUX} Collaboration, D.~Akerib et~al., {\it {Results from a search for dark
  matter in the complete LUX exposure}},  {\em Phys. Rev. Lett.} {\bf 118}
  (2017), no.~2 021303, [\href{http://arxiv.org/abs/1608.07648}{{\tt
  arXiv:1608.07648}}].

\bibitem{Cui:2017nnn}
{\bf PandaX-II} Collaboration, X.~Cui et~al., {\it {Dark Matter Results From
  54-Ton-Day Exposure of PandaX-II Experiment}},  {\em Phys. Rev. Lett.} {\bf
  119} (2017), no.~18 181302, [\href{http://arxiv.org/abs/1708.06917}{{\tt
  arXiv:1708.06917}}].

\bibitem{Carena:2018nlf}
M.~Carena, J.~Osborne, N.~R. Shah, and C.~E.~M. Wagner, {\it {Supersymmetry and
  LHC Missing Energy Signals}},  {\em Phys. Rev.} {\bf D98} (2018), no.~11
  115010, [\href{http://arxiv.org/abs/1809.11082}{{\tt arXiv:1809.11082}}].

\bibitem{LEPSUSYWG1}
{\bf LEPSUSYWG, ALEPH, DELPHI, L3 and OPAL} Collaboration, {\it {Combined lep
  chargino results, up to 208 gev for large m0}},
  \href{http://arxiv.org/abs/http://lepsusy.web.cern.ch/lepsusy/www/inos\_moriond01/charginos\_pub.html}{{\tt
  http://lepsusy.web.cern.ch/lepsusy/www/inos\_moriond01/charginos\_pub.html}}.

\bibitem{LEPSUSYWG2}
{\bf LEPSUSYWG, ALEPH, DELPHI, L3 and OPAL} Collaboration, {\it {Combined lep
  chargino results, up to 208 gev for low dm}},
  \href{http://arxiv.org/abs/http://lepsusy.web.cern.ch/lepsusy/www/inoslowdmsummer02/charginolowdm\_pub.html}{{\tt
  http://lepsusy.web.cern.ch/lepsusy/www/inoslowdmsummer02/charginolowdm\_pub.html}}.

\bibitem{Andreev:2018ayy}
{\bf ACME} Collaboration, V.~Andreev et~al., {\it {Improved limit on the
  electric dipole moment of the electron}},  {\em Nature} {\bf 562} (2018),
  no.~7727 355--360.

\bibitem{Barr:1990vd}
S.~M. Barr and A.~Zee, {\it {Electric Dipole Moment of the Electron and of the
  Neutron}},  {\em Phys. Rev. Lett.} {\bf 65} (1990) 21--24. [Erratum: Phys.
  Rev. Lett.65,2920(1990)].

\bibitem{Atwood:1990cm}
D.~Atwood, C.~P. Burgess, C.~Hamazaou, B.~Irwin, and J.~A. Robinson, {\it {One
  loop P and T odd W+- electromagnetic moments}},  {\em Phys. Rev.} {\bf D42}
  (1990) 3770--3777.

\bibitem{Peskin:1991sw}
M.~E. Peskin and T.~Takeuchi, {\it {Estimation of oblique electroweak
  corrections}},  {\em Phys. Rev. D} {\bf 46} (1992) 381--409.

\bibitem{Cacciapaglia:2006pk}
G.~Cacciapaglia, C.~Csaki, G.~Marandella, and A.~Strumia, {\it {The Minimal Set
  of Electroweak Precision Parameters}},  {\em Phys. Rev. D} {\bf 74} (2006)
  033011, [\href{http://arxiv.org/abs/hep-ph/0604111}{{\tt hep-ph/0604111}}].

\bibitem{RPP2020}
{\bf Particle Data Group} Collaboration, P.~Zyla et~al., {\it {Review of
  Particle Physics}},  {\em Progress of Theoretical and Experimental Physics}
  {\bf 2020} (08, 2020)
  [\href{http://arxiv.org/abs/https://academic.oup.com/ptep/article-pdf/2020/8/083C01/33653179/ptaa104.pdf}{{\tt
  https://academic.oup.com/ptep/article-pdf/2020/8/083C01/33653179/ptaa104.pdf}}].
  083C01.

\bibitem{Barbieri:2004qk}
R.~Barbieri, A.~Pomarol, R.~Rattazzi, and A.~Strumia, {\it {Electroweak
  symmetry breaking after LEP-1 and LEP-2}},  {\em Nucl. Phys. B} {\bf 703}
  (2004) 127--146, [\href{http://arxiv.org/abs/hep-ph/0405040}{{\tt
  hep-ph/0405040}}].

\bibitem{Khosa:2020zar}
C.~K. Khosa, S.~Kraml, A.~Lessa, P.~Neuhuber, and W.~Waltenberger, {\it
  {SModelS database update v1.2.3}},  {\em {to appear in LHEP. }} (2020)
  [\href{http://arxiv.org/abs/2005.00555}{{\tt arXiv:2005.00555}}].

\bibitem{Ambrogi:2018ujg}
F.~Ambrogi et~al., {\it {SModelS v1.2: long-lived particles, combination of
  signal regions, and other novelties}},
  \href{http://arxiv.org/abs/1811.10624}{{\tt arXiv:1811.10624}}.

\bibitem{Dutta:2018ioj}
J.~Dutta, S.~Kraml, A.~Lessa, and W.~Waltenberger, {\it {SModelS extension with
  the CMS supersymmetry search results from Run 2}},  {\em LHEP} {\bf 1}
  (2018), no.~1 5--12, [\href{http://arxiv.org/abs/1803.02204}{{\tt
  arXiv:1803.02204}}].

\bibitem{Ambrogi:2017neo}
F.~Ambrogi, S.~Kraml, S.~Kulkarni, U.~Laa, A.~Lessa, V.~Magerl, J.~Sonneveld,
  M.~Traub, and W.~Waltenberger, {\it {SModelS v1.1 user manual}},
  \href{http://arxiv.org/abs/1701.06586}{{\tt arXiv:1701.06586}}.

\bibitem{Kraml:2013mwa}
S.~Kraml, S.~Kulkarni, U.~Laa, A.~Lessa, W.~Magerl, D.~Proschofsky, and
  W.~Waltenberger, {\it {SModelS: a tool for interpreting simplified-model
  results from the LHC and its application to supersymmetry}},  {\em
  Eur.Phys.J.} {\bf C74} (2014) 2868,
  [\href{http://arxiv.org/abs/1312.4175}{{\tt arXiv:1312.4175}}].

\bibitem{ATL-PHYS-PUB-2019-029}
{ATLAS Collaboration}, {\it {Reproducing searches for new physics with the
  ATLAS experiment through publication of full statistical likelihoods}},
  Tech. Rep. ATL-PHYS-PUB-2019-029, CERN, Geneva, Aug, 2019.
\newblock \url{https://cds.cern.ch/record/2684863}.

\bibitem{Skands:2003cj}
P.~Z. Skands, B.~Allanach, H.~Baer, C.~Balazs, G.~Belanger, et~al., {\it {SUSY
  Les Houches accord: Interfacing SUSY spectrum calculators, decay packages,
  and event generators}},  {\em JHEP} {\bf 0407} (2004) 036,
  [\href{http://arxiv.org/abs/hep-ph/0311123}{{\tt hep-ph/0311123}}].

\bibitem{Alwall:2006yp}
J.~Alwall, A.~Ballestrero, P.~Bartalini, S.~Belov, E.~Boos, et~al., {\it {A
  Standard format for Les Houches event files}},  {\em Comput.Phys.Commun.}
  {\bf 176} (2007) 300--304, [\href{http://arxiv.org/abs/hep-ph/0609017}{{\tt
  hep-ph/0609017}}].

\bibitem{Buckley:2013jua}
A.~Buckley, {\it {PySLHA: a Pythonic interface to SUSY Les Houches Accord
  data}},  \href{http://arxiv.org/abs/1305.4194}{{\tt arXiv:1305.4194}}.

\bibitem{Staub:2008uz}
F.~Staub, {\it {SARAH}},  \href{http://arxiv.org/abs/0806.0538}{{\tt
  arXiv:0806.0538}}.

\bibitem{Staub:2013tta}
F.~Staub, {\it {SARAH 4 : A tool for (not only SUSY) model builders}},  {\em
  Comput. Phys. Commun.} {\bf 185} (2014) 1773--1790,
  [\href{http://arxiv.org/abs/1309.7223}{{\tt arXiv:1309.7223}}].

\bibitem{Staub:2015kfa}
F.~Staub, {\it {Exploring new models in all detail with SARAH}},  {\em Adv.
  High Energy Phys.} {\bf 2015} (2015) 840780,
  [\href{http://arxiv.org/abs/1503.04200}{{\tt arXiv:1503.04200}}].

\bibitem{Porod:2003um}
W.~Porod, {\it {SPheno, a program for calculating supersymmetric spectra, SUSY
  particle decays and SUSY particle production at e+ e- colliders}},  {\em
  Comput. Phys. Commun.} {\bf 153} (2003) 275--315,
  [\href{http://arxiv.org/abs/hep-ph/0301101}{{\tt hep-ph/0301101}}].

\bibitem{Porod:2011nf}
W.~Porod and F.~Staub, {\it {SPheno 3.1: Extensions including flavour,
  CP-phases and models beyond the MSSM}},  {\em Comput. Phys. Commun.} {\bf
  183} (2012) 2458--2469, [\href{http://arxiv.org/abs/1104.1573}{{\tt
  arXiv:1104.1573}}].

\bibitem{Alwall:2014hca}
J.~Alwall, R.~Frederix, S.~Frixione, V.~Hirschi, F.~Maltoni, O.~Mattelaer,
  H.~S. Shao, T.~Stelzer, P.~Torrielli, and M.~Zaro, {\it {The automated
  computation of tree-level and next-to-leading order differential cross
  sections, and their matching to parton shower simulations}},  {\em JHEP} {\bf
  07} (2014) 079, [\href{http://arxiv.org/abs/1405.0301}{{\tt
  arXiv:1405.0301}}].

\bibitem{Alwall:2011uj}
J.~Alwall, M.~Herquet, F.~Maltoni, O.~Mattelaer, and T.~Stelzer, {\it {MadGraph
  5 : Going Beyond}},  {\em JHEP} {\bf 06} (2011) 128,
  [\href{http://arxiv.org/abs/1106.0522}{{\tt arXiv:1106.0522}}].

\bibitem{Sirunyan:2019twz}
{\bf CMS} Collaboration, A.~M. Sirunyan et~al., {\it {Measurements of the Higgs
  boson width and anomalous $HVV$ couplings from on-shell and off-shell
  production in the four-lepton final state}},  {\em Phys. Rev. D} {\bf 99}
  (2019), no.~11 112003, [\href{http://arxiv.org/abs/1901.00174}{{\tt
  arXiv:1901.00174}}].

\bibitem{Nakai:2016atk}
Y.~Nakai and M.~Reece, {\it {Electric Dipole Moments in Natural
  Supersymmetry}},  {\em JHEP} {\bf 08} (2017) 031,
  [\href{http://arxiv.org/abs/1612.08090}{{\tt arXiv:1612.08090}}].

\end{thebibliography}\endgroup
\end{document}